\documentclass[aps,preprintnumbers,amsmath,amssymb,nofootinbib,eqsecnum,preprintnumbers]{revtex4}
\DeclareUnicodeCharacter{2217}{\textasteriskcentered}
\usepackage[a4paper,margin=1.25in]{geometry}
\usepackage{amsmath}
\usepackage{hyperref}
\usepackage{ragged2e}
\usepackage{mathrsfs}
\usepackage{bigints}
\usepackage[utf8]{inputenc}
\usepackage{subcaption}
\usepackage{amsfonts,amssymb,amscd,amsthm,amsfonts,epsfig,enumerate,yfonts}
\usepackage{physics}
\usepackage[font=small,skip=0pt]{caption}
\usepackage{float}
\usepackage{multirow}
\usepackage{tikz}
\usepackage{ragged2e}
\usetikzlibrary{positioning}
\usepackage[font=small,skip=7pt]{caption}
\begin{document}
\title{Shadow geometry of Kerr MOG naked singularity and analysis of accretion disk luminosity}
\author{Saira Yasmin$^{1}$}
\email{sairayasmeen555@gmail.com}
\author{Mubasher Jamil$^{1,2}$}
\email{mjamil@sns.nust.edu.pk}
\affiliation{$^1$School of Natural Sciences, National University of Sciences and Technology (NUST), H-12, Islamabad 44000, Pakistan}
\affiliation{$^2$Research Center of Astrophysics and Cosmology, Khazar University, Baku, AZ 1096, 41 Mehseti Street, Azerbaijan}
\date{\today}

\begin{abstract}
\begin{justify}
\textbf{Abstract:} Naked singularities are hypothetical astrophysical entities featuring gravitational singularities without event horizons. In this study, we analyze the shadow properties of Kerr Modified Gravity (Kerr MOG) naked singularities (KMNSs). We show that the KMNS shadow can appear closed, open, or even vanish, depending on the dimensionless spin parameter $a$, the modified gravity parameter $\alpha$, and the observer's inclination angle. We identify the critical conditions under which the KMNS shadow develops a gap, a unique feature not present in BH shadows.  We analyze the properties of a thin accretion disk surrounding a KMNS, within the framework of MOG characterized by the parameter \( \alpha \). The study includes a detailed examination of the spacetime geometry and the equations of motion for test particles. In addition, we adopt a simplified model for the disk’s radiative flux, temperature distribution, and spectral luminosity. Our analysis primarily focuses on the flux distribution of the accretion disk around KMNS with identical mass but varying spin and MOG deformation parameters. This allows us to explore how modifications in rotation and the MOG parameter \( \alpha \) influence the radiative properties of the disk. Further, these observational signatures may serve as effective tools for clearly distinguishing KMNS from standard Kerr naked singularities (KNSs), where the MOG parameter \( \alpha = 0 \).
\end{justify}
\end{abstract}
\maketitle
\section{Introduction}
\begin{justify}
General relativity ($\mathrm{GR}$), proposed by Einstein, remains an exceptionally effective theory of gravity. A spacetime singularity would necessarily develop, according to the  $\mathrm{GR}$, when masses large enough collapse under the pull of their own gravity. In 1969, Roger Penrose presented the Cosmic Censorship Conjecture (CCC), which emphasizes that horizonless strong spacetime singularities are not feasible \cite{penrose1965gravitational}. However, several studies focusing on the continuous gravitational collapse of the inhomogeneous matter cloud demonstrate that spacetime singularities formed during gravitational collapse can be observable by an outside observer \cite{patel2022light}. In $\mathrm{GR}$, such a situation can also emerge where the continual gravitational collapse could bring about an equilibrium static spacetime with a central observable or naked singularity, such as Joshi-Malafarina-Narayan (JMN) spacetime \cite{Joshi2011zm}. 
The naked singularity can occur as the final state of continuous gravitational collapse of an inhomogeneous matter cloud. Information about the distinguished spacetime structure near a naked singularity may be retrieved by photon trajectories \cite{janis1968reality}. Therefore, to predict the physical signature of a naked singularity, a detailed study of null geodesics near a naked singularity is very significant.  A valuable approach for classifying the spacetime structure of a naked singularity is to investigate the shadow of a naked singularity \cite{joshi2020shadow}. The analysis of the shadow also provides an initial way of figuring out the properties of the naked singularity. The shadow of a black hole (BH) has attracted significant attention within the astrophysical community, particularly following the first image of the region around the supermassive BH at the center of the M87* galaxy, captured by the Event Horizon Telescope (EHT) \cite{akiyama2019first}. This interest further increased with the release of the image of the region around the BH at the center of our own galaxy, SgrA* \cite{akiyama2022first}. The EHT collaboration analyzed the shadow image of Sgr A*, concluding that the central object is the finest possible nonspinning BH imitators \cite{akiyama2022first}. Initial efforts to model the shadow of BHs in asymptotically flat spacetimes focused on simple configurations, including non-rotating Schwarzschild BHs~\cite{1968AmSci..56....1W}, Schwarzschild BHs surrounded by accretion disks~\cite{1974IAUS...64..132B}, and rotating Kerr BHs~\cite{luminet2019image}. Since then, the study of BH shadows has significantly evolved to encompass a wide range of spacetime geometries within the framework of $\mathrm{GR}$.
 Researchers have modeled these shadows using diverse spacetime geometries, with and without the presence of matter, applying different methods and even exploring alternative gravity theories \cite{cunningham1973optical, Jafarzade2025xcn,khaijde2024zqq,universe8020102,PhysRevD.101.044035,PhysRevD.100.sh}.
\end{justify}
\indent Moffat proposed a scalar–tensor–vector gravity (STVG) theory \cite{moffat2006scalar}, commonly referred to as modified gravity theory (MOG). The MOG theory was developed to resolve the discrepancies between $\mathrm{GR}$ and various large-scale astronomical observations which include the dynamics of galaxies, flat rotation curves, the mass distribution in systems such as the Bullet cluster, and different gravitational lensing phenomenon \cite{Rehman:2025rxw}, all without requiring dark matter. MOG has also been investigated as a possible explanation for the accelerated expansion of the universe \cite{brownstein2007bullet} as well as local particle periodic motions near MOG BH \cite{Shabbir:2025kqh}.
 Additionally, the MOG theory provides a description of structure formation in the early universe, the matter power spectrum, and the acoustic power spectrum of the cosmic microwave background (CMB) \cite{moffat2015black}.
This theory introduces new fields into $\mathrm{GR}$, strengthening the gravitational field. Its action consists of the usual Einstein-Hilbert term associated with the metric tensor \(g_{\mu\nu}\), a massive vector field \(\phi_{\mu}\), and three scalar fields that represent the running values of the gravitational constant \(G\), the coupling constant \(\Omega\) (which determines the strength of the coupling between matter and the vector field), and the vector field’s mass \(\mu\) (which adjusts the range of the coupling) \cite{moffat2006scalar}. The scalar field \( G = G_N (1 + \alpha) \) represents the strength of the gravitational attraction, where \( G_N \) is Newton’s gravitational constant and \( \alpha \) is a dimensionless parameter of the theory \cite{yasmin2025shadow}.
Einstein’s $\mathrm{GR}$, although immensely successful, presents a fascinating paradox involving spacetime singularities, specifically the emergence of closed timelike curves (CTCs), which permit time to loop backward \cite{hawking1992chronolgy}. These trajectories violate the normal sequence of time (chronology), resulting in a breakdown of determinism. The paradox posed by CTCs is distinct from that of curvature singularities. While singularities signal a breakdown in the laws of physics, CTCs imply a failure of predictability. Nevertheless, both lead to challenges in formulating well-defined initial conditions. The region of spacetime that permits CTCs, commonly referred to as the causality-violating region, results in an ill-posed Cauchy problem, and its boundary is known as the Cauchy horizon  \cite{1995lwetbookV}. Various attempts have been made to avoid the formation of CTCs through modifications to known spacetime solutions \cite{vaishak2019chronolgy}, but a complete theoretical resolution remains elusive \cite{bajowald2005black,hossenfelder2010model}. The concept of time travel via CTCs introduces additional paradoxes, such as the well-known grandfather paradox and other logical inconsistencies.
However, recent work by Tobar and Costa suggests that paradox-free (or consistency-free) time travel could be possible within CTC frameworks \cite{germain2020reversible}. They propose that multiple, in-equivalent but self-consistent histories may coexist, enabling consistent dynamics in such spacetimes. Similarly, Nolan explored the motion of gyroscopes along CTCs and analyzed the behavior of T-periodic spin vectors. Their results, across various CTC admitting spacetimes, were interpreted through the lens of the consistency principle~\cite{ikeda2021black}. CTCs naturally arise in many stationary, axially symmetric rotating solutions of general relativity. The first such example was G\"{o}del’s rotating cosmological model~\cite{lum2021closed}, presented in 1949. 
Significant effort has been directed toward finding alternatives to the KNS spacetime that avoid the causality-violating regions, often referred to as time machines \cite{calvani1978time}, which are predicted to emerge near the ring singularity \cite{bcarter1973equi}. One such class of alternatives includes string-inspired solutions known as superspinars, which model rapidly rotating compact objects. In these models, the external spacetime remains described by the standard Kerr NS geometry, while the interior is governed by hypothetical string-theoretic effects \cite{GIMON2009astro,Babar2016dyna}. This approach effectively hides the causality-violating region behind a regular inner structure, with the boundary of the inner solution expected at or beyond \( r \geq 0 \).  
The study of mass accretion around rotating BHs in the context of $\mathrm{GR}$ was first undertaken in~\cite{novikov1973black,PhysRevD.92.043008}. By adopting an equatorial approximation to the stationary and axisymmetric spacetime of rotating BHs, steady-state thin disk models were developed, extending the theory of non-relativistic accretion~\cite{shakura1973black}. In these models, hydrodynamical equilibrium is sustained through efficient cooling mechanisms driven by radiation transport, with the accreting matter following a Keplerian rotation. The radiation emitted from the disk surface was also examined, assuming blackbody radiation emerging from the disk in thermodynamic equilibrium. The radiation properties of thin accretion disks were further explored in~\cite{thorne1974disk}, where the effects of photon capture by the BH on its spin evolution were also discussed. These studies also computed the efficiency with which BHs convert rest mass into outgoing radiation during the accretion process. More recently, the emissivity properties of accretion disks have been studied for exotic central objects, including wormholes, non-rotating and rotating quark, boson, or fermion stars, brane-world BHs, $f(R)$ gravity models, and Horava-Lifshitz gravity~\cite{harko2011thin}. In all these cases, it was shown that distinct signatures could emerge in the electromagnetic spectrum, thus providing a means to directly test different physical models using astrophysical observations of the emission spectra from accretion disks.
{In the literature, studies on shadows and accretion images in modified gravity can also be found in \cite{Wang2018prk, zheng2025shadows, yasmin2025shadow,Jafarzade2025nbe}}.

This paper also investigates the properties of thin accretion disks around rotating naked singularities, specifically comparing the KMNS and the KNS (with $\alpha = 0$). We conduct a comparative analysis of the physical properties of these disks, including the emitted energy flux, the temperature distribution across the disk surface, and the spectrum of emitted equilibrium radiation. Due to the distinct differences in the external geometry, the thermodynamic and electromagnetic properties of the disks, such as energy flux, temperature distribution, and radiation spectrum, differ significantly between these two types of compact objects. These differences may provide unique observational signatures, potentially enabling the distinction between naked singularities and BHs, at least in principle. It should be noted that the proposed method for detecting naked singularities through the study of accretion disks is indirect and should be complemented by direct observational techniques. These may include the examination of the "surface" of the compact object candidates and/or the study of the lensing properties of the central object.

In this paper, we systematically study the projection of the unstable spherical photon orbits around KMNSs at infinity, which is referred as the ``shadow." We analyze the geometry of the naked singularity and compute the KMNS shadow analytically by separating the Hamilton-Jacobi equation and examining the radial effective potential. We demonstrate that the shadow of the KMNS can be closed, open, or even vanish, depending on the spin parameter and the inclination angle of the observer. Additionally, we present the critical parameters at which the shadow undergoes a topological transition, which is discussed in Section~ \ref{secshad}. In Section~\ref{secIIA}, we examine the unstable spherical photon orbits around KMNS spacetimes, whereas Section~\ref{secIIB} examines the spherical photon orbits and the resulting shadow structure. In Section~\ref{secIII}, we derive the main physical parameters for test particles, including the specific energy, specific angular momentum, and angular velocity. Additionally, the properties of standard thin accretion disks are reviewed in this section. The energy flux, temperature distribution, and radiation spectrum from thin disks around KMNSs are also discussed. We discuss and conclude our results in Section~\ref{secconc}.

\section{Analytical Shadow of Kerr MOG Naked Singularity} \label{secshad}

The Kerr MOG metric in the Boyer-Lindquist coordinates $(t,r,\theta,\phi)$ is given by \cite{moffat2015black}
\begin{align}\label{kmogmetric}
d s^{2}= & -\frac{\Delta-a^{2} \sin ^{2} \theta}{\rho^{2}} d t^{2}+\sin ^{2} \theta\left(\frac{\left(r^{2}+a^{2}\right)^{2}-\Delta a^{2} \sin ^{2} \theta}{\rho^{2}}\right) d \phi^{2}   \nonumber \\
& -2 a \sin ^{2} \theta\left(\frac{r^{2}+a^{2}-\Delta}{\rho^{2}}\right) d t d \phi+\frac{\rho^{2}}{\Delta} d r^{2}+\rho^{2} d \theta^{2} ,
\end{align}
\begin{align}
\Delta &= r^{2} - 2 G \mathscr{M} r + a^{2} + \alpha G_{\mathrm{N}} G \mathscr{M}^{2}, \\
\rho^{2} &= r^{2} + a^{2} \cos^{2} \theta
\end{align}
where $G =G_N(1+\alpha)$, is an enhanced gravitational constant with the contribution of Newton's gravitational constant $G_N$ and the deformation rate $\alpha$.
For $\alpha=0$, it will retrieve the Kerr metric. The Newtonian mass $\mathscr{M}$ and the ADM mass $\mathcal{M}$ are related as
$\mathcal{M}=(1+\alpha) \mathscr{M}$. The Kerr MOG metric has two horizons known as the outer horizon \( (r_+) \) and the inner horizon \( (r_-) \), obtained as roots of the equation $\Delta=0$, 
\begin{align}
 r^2-2 \mathcal{M} r +a^2+\frac{\alpha}{1+\alpha }\mathcal{M}^2=0.
\end{align}
The two horizons are also referred to as the event horizon and the Cauchy horizon, respectively. They are given by \cite{lee2017innermost}
\begin{equation}
r_{\pm} = G_{\mathrm{N}} \mathcal{M} \pm \sqrt{\frac{G_{\mathrm{N}}^2 \mathcal{M}^2}{1+\alpha} - a^2}.
\end{equation}
 For $ a=\frac{a}{G_{\mathcal{N}}\mathcal{M}}> 1$, it is referred to a KNS while for $ a=\frac{a}{G_{\mathcal{N}}\mathcal{M}}<1$ there exists a BH solution. Further $ a=\frac{a}{G_{\mathcal{N}}\mathcal{M}}=1$, yields an extremal BH solution. Meanwhile, for the Kerr MOG BH, these limits can be simplified in the following way. If we take $\alpha=1$ then these limits become $a>\frac{1}{\sqrt{2}}$ for KMNS and $a<\frac{1}{\sqrt{2}}$ for Kerr MOG BH solution and $a=\frac{1}{\sqrt{2}}$, for the extremal limit. Similarly, these limits are also found with $\alpha=2$ 
  \cite{pradhan2020distinguishing}. 
  
  The Hamiltonian of a photon propagation in the Kerr MOG spacetime can be defined as \cite{wang2024ring}
\begin{equation}
\mathcal{H}(x, p)=\frac{1}{2} g^{\mu \nu}(x) p_{\mu} p_{\nu}=\frac{1}{2 \rho^{2}}\left(p_{\theta}^{2}+ p_{r}^{2} \Delta+\mathcal{V}_{\mathrm{eff}}\right)=0, 
\end{equation}
where the effective potential $\mathcal{V}_{\text {eff }}$ is defined as
\begin{equation}
\mathcal{V}_{\mathrm{eff}}=
-\frac{1}{\Delta} \left( a \frac{L_z}{\tilde{E}} - \left( r^2 + a^2 \right) \right)^2 + \left( \frac{1}{\sin \theta} \frac{L_z}{ \tilde{E}} - a \sin^2 \theta \right)^2.
\end{equation}
A significant physical property of the KMNS is the issue of causality violations. The fundamental laws of physics are based on the principle of causality, which asserts that cause precedes effect. A  CTC is a closed path in spacetime along which the tangent is always timelike. The existence of CTCs implies a violation of causality, as an observer traveling along such a path could return to an event that occurred before her own departure.
To address this paradox, Hawking proposed the Chronology Protection Conjecture, which asserts that the laws of physics prevent the formation of CTCs \cite{dutta2024role}. However, causality violations are known to occur in certain BHs, including the Kerr and Kerr–Newman BHs \cite{gott1991closed}. In general, causality violations and CTCs can occur when \( g_{\phi\phi} > 0 \) \cite{azreg2014generating}.

\begin{figure}[!htb]
    \centering
    \subfloat[a = 1.01]{
        \includegraphics[width=0.45\textwidth]{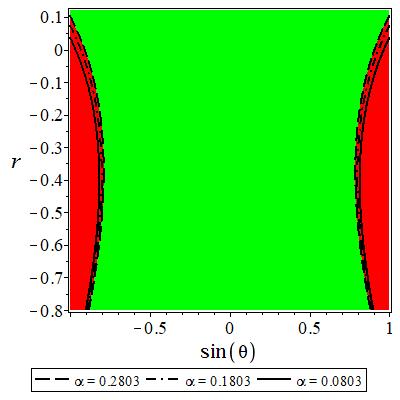}
    }
    \subfloat[a = 1.38]{
        \includegraphics[width=0.45\textwidth]{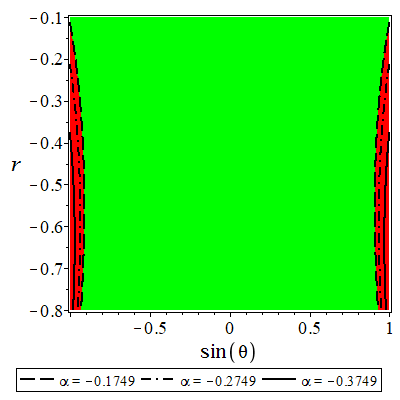}
    }\\[1.5ex]
    \subfloat[a = 1.64]{
        \includegraphics[width=0.45\textwidth]{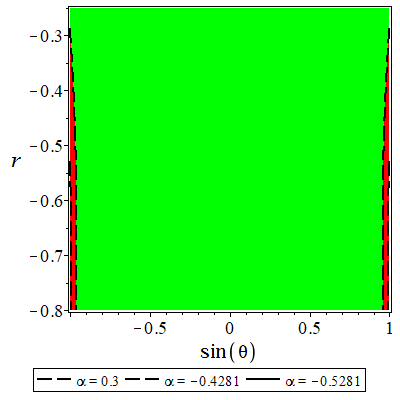}
    }
    \subfloat[a = 1.03]{
        \includegraphics[width=0.45\textwidth]{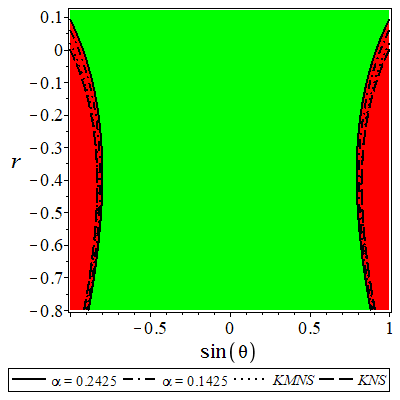}
    }
    \caption{The causality violation/preservation regions of the KMNS are analyzed here. We have fixed \( \mathcal{M} = 1 \) and considered different values of the spin parameter \( a \) along with the corresponding values of \( \alpha \).
}
    \label{Figcas}
\end{figure}

In the context of the KMNS, we analyzed causality violations using the plots shown in Fig .~\ref{Figcas}, which display the relationship between \( r \) and \( \sin(\theta) \). According to (\ref{kmogmetric}), the sign of \( g_{\phi\phi} \) is determined by the parameters \( \alpha \) and \( a \).
Fig ~\ref{Figcas}(a)  shows the causality violation/preservation regions for the KMNS at a specific spin parameter \( a = 1.01 \). The green region indicates areas where causality is preserved, implying the absence of CTCs. In contrast, the red region ($g_{\phi \phi}>0$) represents areas where causality violations occur, and CTCs are present. These are the regions where spacetime curvature allows for time travel or loops, violating causality. The shift and size of the green and red regions depend on the corresponding value of \( \alpha \) in this plot. As \( \alpha \) decreases, the red region (CTCs) reduces in size, and the green region (causality preservation) expands.
 Fig ~\ref{Figcas}(b) shows how the causality violation regions change. The green and red regions alter in size depending on the value of \( \alpha \). As \( \alpha \) increases, the green region expands, indicating a greater area where causality is preserved. In contrast, when \( \alpha \) decreases, the causality violation region (red region) extends further, showing a larger region where CTCs occur. This suggests that for higher \( \alpha \), causality violations are confined to a smaller region.
 Fig ~\ref{Figcas}(c)  shows the behavior for \( a = 1.64 \) with values of \( \alpha  \). With higher spin, the structure of the causality violation regions changes significantly. As the spin parameter \( a \) increases, the red region (representing CTCs) extends further outward, and the green region (causality preservation) shrinks. This behavior suggests that higher spins, along with specific choices of \( \alpha \), lead to an increased extent of CTC formation, making causality violations more widespread.  Fig ~\ref{Figcas}(d) shows the causality violation regions for both the KNS and the KMNS.  The comparison reveals how the causality violation regions differ between the two. In the KMNS, the red region (representing causality violations) is larger, indicating more expansive CTCs. In contrast, for the KNS, the causality violation region is smaller, and the overall structure is more constrained. As $\alpha$ decreases, the CTC region shrinks for both cases, but the Kerr MOG solution consistently exhibits a larger region of CTCs compared to the KNS.

\begin{table}[]
\caption{\label{taba}Corresponding to different values of the spin parameter $a$, the deformation parameter $\alpha$ range is shown for the KMNS. Here the value of the mass parameter is $\mathcal{M}=1$.}
\centering
\begin{tabular}{lll}
\hline
No. & \quad \quad  $a/\mathcal{M}$ & \quad \quad Range of $\alpha$ \\
\hline
1 &  \quad \quad 1.1 &  \quad \quad $\alpha >- 0.17355 $\\
2 &  \quad \quad 1.2 &   \quad \quad $\alpha >- 0.30555$\\
3 &  \quad \quad 1.3 &  \quad \quad $\alpha >- 0.40828$\\
4 &  \quad \quad 1.4 &  \quad \quad $\alpha >- 0.48979$\\
5 &  \quad \quad 1.5 &  \quad \quad $\alpha >- 0.55556$\\
6 &  \quad \quad 1.6 &  \quad \quad $\alpha >- 0.609375$\\
7 &  \quad \quad 1.7 &   \quad \quad$\alpha >- 0.653979$\\
8 &  \quad \quad 1.8 &  \quad \quad $\alpha >- 0.691358$\\
9 &  \quad \quad 1.9 &  \quad \quad $\alpha >- 0.722991$\\
\hline 
\end{tabular}
\end{table}
\subsection{Unstable spherical photon orbits around KMNs} \label{secIIA}
 The Kerr MOG metric exhibits both axisymmetry and time translation symmetry. In these coordinates, the metric coefficients are independent of the \(t\) and \(\phi\) coordinates, which leads to the conservation of the associated conjugate momenta, \(p_\text{t}\) and \(p_\phi\) respectively \cite{yasmin2025shadow}. These conserved quantities correspond to the constants of motion associated with the spacetime's Killing vectors
\begin{align*}
\mathcal{K}^{\mu}=(1,0,0,0), \quad \mathrm{R}^{\mu}=(0,0,0,1).
\end{align*}
Accordingly, the geodesic motion in Kerr MOG geometry is characterized by  two constants of motion, which are 
 \begin{align}\label{E L}
    \tilde{E}&=-\mathcal{K}^{\mu} p_\mu=-p_{\text{t}}, \nonumber \\
\mathrm{L}_\text{z}&=\mathrm{R}^{\mu} p_\mu=p_\phi.  
\end{align}
  The conserved quantities can be obtained from the symmetries of the spacetime.  By projecting the Killing vectors along the covariant four-momentum vector $p_\mu$, we find conserved energy $\tilde{E}$ and conserved angular momentum in the $\phi$ direction $\mathrm{L}_z$, given in (\ref{E L}). To examine the spherical photon orbits and the shadow of KMNS, we use the Hamilton-Jacobi equation \cite{nguyen2023shadow}
\begin{align} \label{Hjacobi}
    \frac{\partial S}{\partial\lambda}=\frac{1}{2} g^{\mu \nu}\frac{\partial S}{\partial x^\mu}\frac{\partial S}{\partial x^\nu},
\end{align}
where $S$ is the action as the function of affine parameter $\lambda$ and coordinates $x^\mu$. 

The solution to (\ref{Hjacobi})  can be separated into different components depending on each coordinate, we use ansatz \cite{perlick2022calculating}
\begin{align}\label{ansatz}
  S&=-\tilde{E}t+S_r(r)+S_\theta(\theta)+L_\text{z}\phi.
\end{align}
Then from the above equations, we can have equations of motion for null geodesic corresponding to each coordinate as follows
\begin{align}
    \rho^2 \Delta \dot{t}&= \Xi \tilde{E}-a(2\mathcal{M}r-\frac{\alpha}{1+\alpha} \mathcal{M}^2)\mathrm{L_z},
\end{align}
    \begin{align}
    \rho^2 \Delta \dot{\phi}= a(2\mathcal{M}r-\frac{\alpha}{1+\alpha} \mathcal{M}^2) \tilde{E}+(\rho^2-2\mathcal{M}r+\frac{\alpha}{1+\alpha}  \mathcal{M}^2)\mathrm{L}_z\csc^2\theta,
\end{align}
\begin{align}
    \rho^4 \dot{\theta}^2= \mathcal{C}-\left(\frac{\mathrm{L}_z^2}{\sin^2 \theta}-\tilde{E^2} a^2\right)\cos^2 \theta,
\end{align}
\begin{align}\label{radial effective KNS1}
    \rho^4 \dot{r}^2= -\mathcal{C} \Delta+(r^2+a^2)\tilde{E}^2-2a( r^2 +a^2-\Delta) \tilde{E}\mathrm{L_z}+a^2\mathrm{L}_z^2-a^2 \tilde{E}^2 \Delta-\mathrm{L_z}^{2}\Delta.
\end{align}
where $ \Xi =(a^2+r^2)^2-\Delta a^2 \sin^2\theta$. The radial equation of motion (\ref{radial effective KNS1}) can be rewritten as
\begin{align}
   \left(\frac{\rho^2}{\mathrm{E}}\right)^2 \dot{r}^2= -\Delta \zeta+(r^2+a^2)^2-2a(r^2+a^2-\Delta)\varphi+a^2\varphi^2-a^2\Delta-\varphi^2 \Delta.
\end{align}
Here, we define the impact parameters as \(\xi = \frac{\mathrm{L}_z}{\tilde{E}}\) and \(\zeta = \frac{\mathcal{C}}{\tilde{E}^2}\), where \(\mathcal{C}\) is Carter's constant, representing a third conserved quantity that arises from the separability of the $\mathrm{HJ}$ equation. $\mathcal{C}$ is related to geodesics in the latitudinal direction.
 Now, the effective potential derived from the radial equation of motion is as follows
\begin{align}
  \mathcal{R}(r)& =-\Delta \zeta+(r^2+a^2)^2-2a(r^2+a^2-\Delta)\xi+a^2\xi^2-a^2\Delta-\xi^2 \Delta.
\end{align}
 \subsection{Spherical photon orbits and shadow} \label{secIIB}
 \begin{justify}
 We obtain two sets of solutions by solving the following conditions for spherical photon orbits with constant radius \( r = r_\text{p} \), given as
\begin{equation}
    \mathcal{R}(r) = 0, \quad \text{and} \quad \frac{d}{dr}\mathcal{R}(r) = 0.
\end{equation}
However, only one of these solutions is physically meaningful and applies to our calculations given by
 \begin{align}\label{impactkm}
    \xi &= \frac{a^2 \mathcal{M} + a^2 r_\text{p} - 3\mathcal{M} r_{\text{p}}^2 + r_\text{p}^3 + a^2 \mathcal{M} \alpha}{a(\mathcal{M} - r_\text{p})(1+\alpha)} 
    + \frac{a^2 r_\text{p} \alpha+ 2 \mathcal{M}^2 r_\text{p} \alpha - 3\mathcal{M} r_{\text{p}}^2 \alpha + r_\text{p}^3 \alpha}{a(\mathcal{M} - r_\text{p})(1+\alpha)} \nonumber \\
    \zeta &= \frac{r_\text{p}}{-a^2 \mathcal{M} + a^2 r_\text{p} - a^2 \mathcal{M} \alpha + r_\text{p} a^2 \alpha} \Bigg( 
   \frac{ -4 a^2 \mathcal{M} (1 + \alpha) \left(r_\text{p} - \mathcal{M} \alpha + r_\text{p} \alpha \right)}{a(\mathcal{M} - r_\text{p})(1+\alpha)} \nonumber \\[1ex]
    &\quad + \frac{\Big(2 \mathcal{M}^2 \alpha + \alpha \mathcal{M}^2 - 3 \mathcal{M} \left(r_\text{p} (1 + \alpha) + r_\text{p}^2 (1 + \alpha)\right) \Big)^2 }{a(\mathcal{M} - r_\text{p})(1+\alpha)}
    \Bigg).
\end{align}

At the equatorial plane, since $\zeta=0$, the critical orbits of the photon ($r_\text{ph}$) can be determined by the roots of (\ref{impactkm}) in KMNS spacetime.
The impact parameters $\xi$ and $\zeta$ are significant in finding the shape of the shadow. We can find stable and unstable photon orbits around the KMNS, as before in \cite{patel2022light} for the KNS. In the context of KMNS, a single photon sphere for retrograde motion presents a specific critical impact parameter that describes the boundary between captured light and light that escapes by the singularity that we can see in Fig.\ref{figab} .
Orbits with positive values of $\xi$ represent the prograde equatorial motion of photons, while those with negative values correspond to retrograde equatorial orbits.
For the KMNS with parameter $\alpha = 0$ ( Kerr case), $\zeta$ vanishes at radius $r_{ph} \approx 4.3$, as shown in the first plot. The corresponding circular photon orbit is retrograde, since $\xi$ is negative at $r_{ph}$. This indicates that no photon sphere exists for prograde equatorial photon motion in this naked singularity spacetime.
When the MOG parameter is set to $\alpha = -0.40$, the photon sphere radius shifts inward to $r_{ph} \approx 3.6$, as seen in the second plot. Again, $\xi$ is negative at this radius, confirming the photon sphere corresponds exclusively to retrograde null geodesics. This confirms that even in the KMNS regime with negative $\alpha$, prograde photon spheres do not form.
\begin{figure}[H] 
     \centering
      {{\includegraphics[height=5cm,width=7cm]{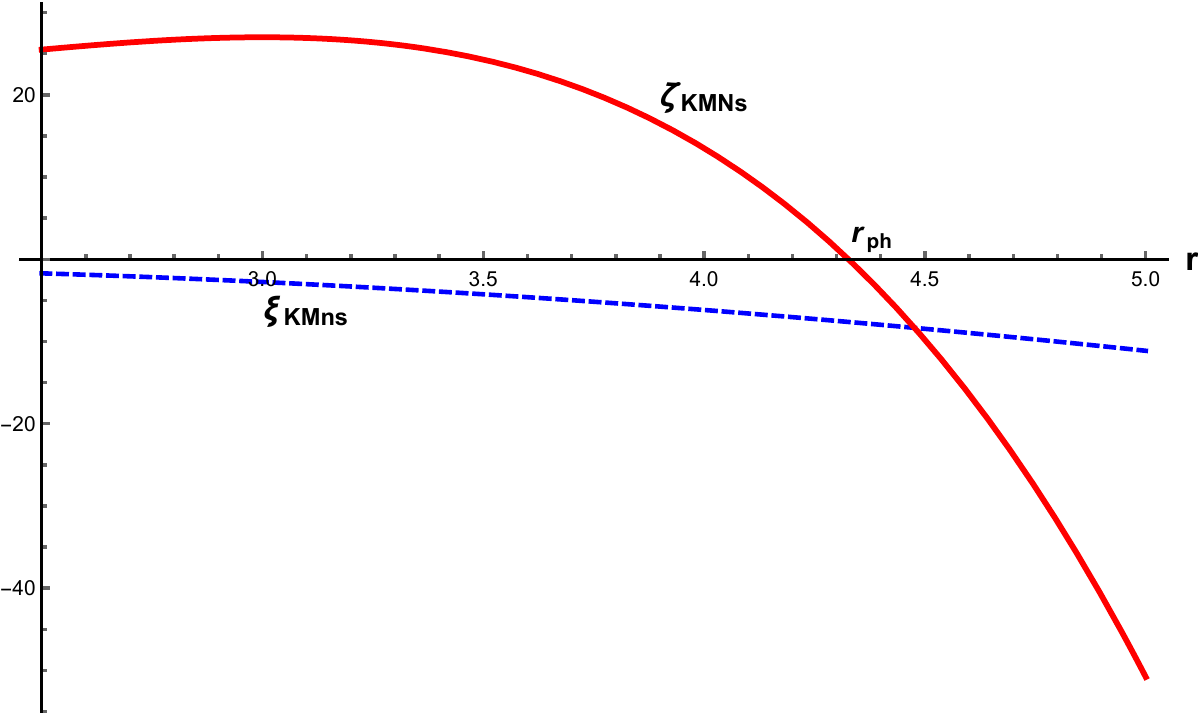}}}
       {{\includegraphics[height=5cm,width=7cm]{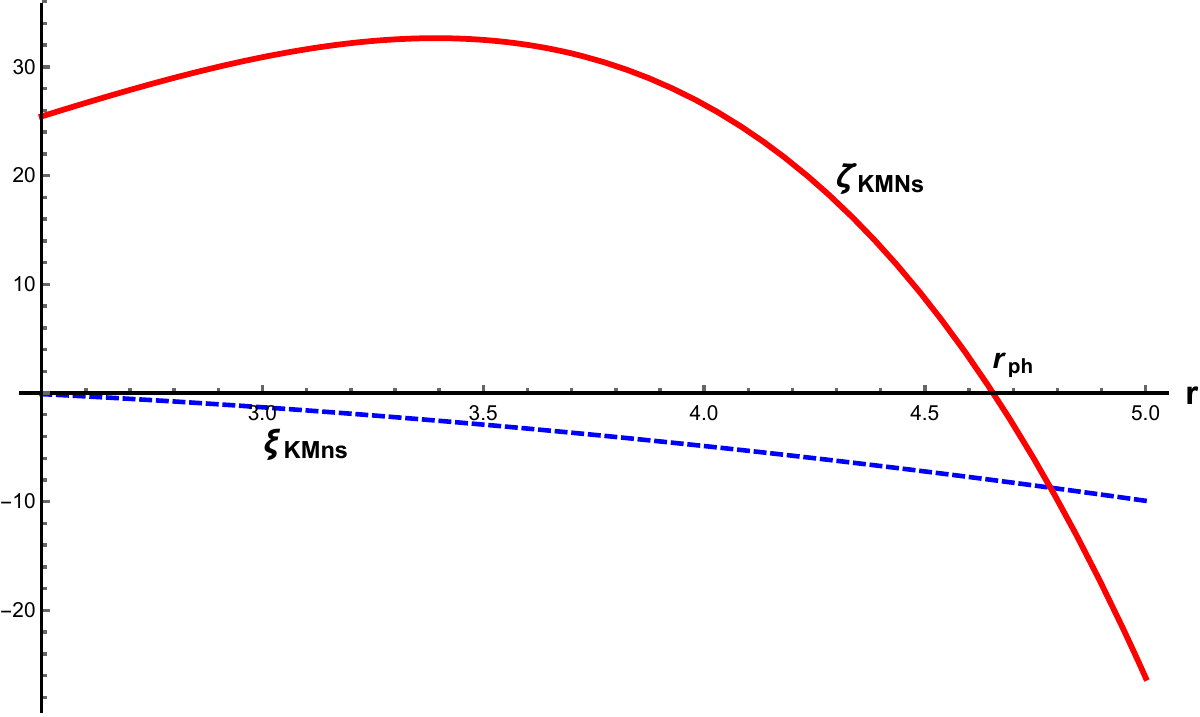}}}
    
       \caption{\label{figab} This figure demonstrates the behavior of $\zeta$ and $\xi$ for the KMNS as functions of the radial coordinate $r$, with parameters $a = 1.38$ and $\mathcal{M} = 1$. The left panel corresponds to the MOG case with $\alpha = -0.40$, while the right panel shows the GR limit with $\alpha = 0$. The plots illustrate how the presence of the MOG parameter $\alpha$ alters the radial profiles of $\zeta$ and $\xi$.
    }
    \end{figure}
\noindent
The radius $r_{ph}$ corresponds to an unstable circular photon orbit, where photons with impact parameters equal to the critical value $\xi(r_{ph})$ are trapped in the photon sphere. Photons with impact parameters smaller than this critical value fall into the naked singularity, while those with larger impact parameters scatter to infinity and can be observed. This unstable photon sphere governs the shadow boundary of the naked singularity, creating a dark region in the observer’s sky surrounded by bright photons scattered from outside the photon sphere.
Thus, the impact parameter plots for KMNSs reveal the absence of prograde photon spheres and illustrate how the Modified Gravity parameter $\alpha$ shifts the photon sphere radius and affects the nature of photon orbits, ultimately influencing the observable shadow size and shape.
\noindent For KMNS, the prograde equatorial orbit does not exist. The retrograde equatorial orbit separates between orbits that terminate at the singularity and those that diminish to infinity. Off the equatorial plane, all orbits that approach the KMNS can escape or unbound. Unstable photon orbits surrounding KMNS exist in the range $r_{ms}<r_\text{p}<r_{ph}$ where $r_{ms}$ is a marginally stable radius and $r_{ph}$ is the equatorial retrograde circular radius that we can get numerically by putting $\zeta=0$ also can be seen in Fig.\ref{figab}. The marginally stable radius $r_{ms}$  can be calculated by  $\frac{d^2}{dr^2}\mathcal{R}(r)=0$, which yields
 \begin{align}\label{rms}
  r_\text{ms}= 
\frac{\mathcal{M} + \mathcal{M}\alpha}{1 + \alpha} + 
\frac{
\left(
a^2 \mathcal{M} - \mathcal{M}^3 + 3 a^2 \mathcal{M} \alpha - 2 \mathcal{M}^3 \alpha + 3 a^2 \mathcal{M} \alpha^2 - \mathcal{M}^3 \alpha^2 + a^2 \mathcal{M} \alpha^3
\right)^{1/3}
}{1 + \alpha} 
\end{align}
While stable photon orbits with radius \( r < r_{\mathrm{ms}} \) are typically hidden inside the event horizon for the Kerr BH, they acquire physical significance for the KNS due to the absence of an event horizon. Nevertheless, bound photon orbits cannot be directly observed by distant observers, and thus, they do not influence the observational signatures of the KNS \cite{nguyen2023shadow}. This consideration similarly applies to the KMNS. 
       \end{justify}
      \begin{figure}[!htb]
    \centering

    \subfloat[a = 1.02, $\alpha$ = 0 ]{
        \includegraphics[width=0.47\textwidth]{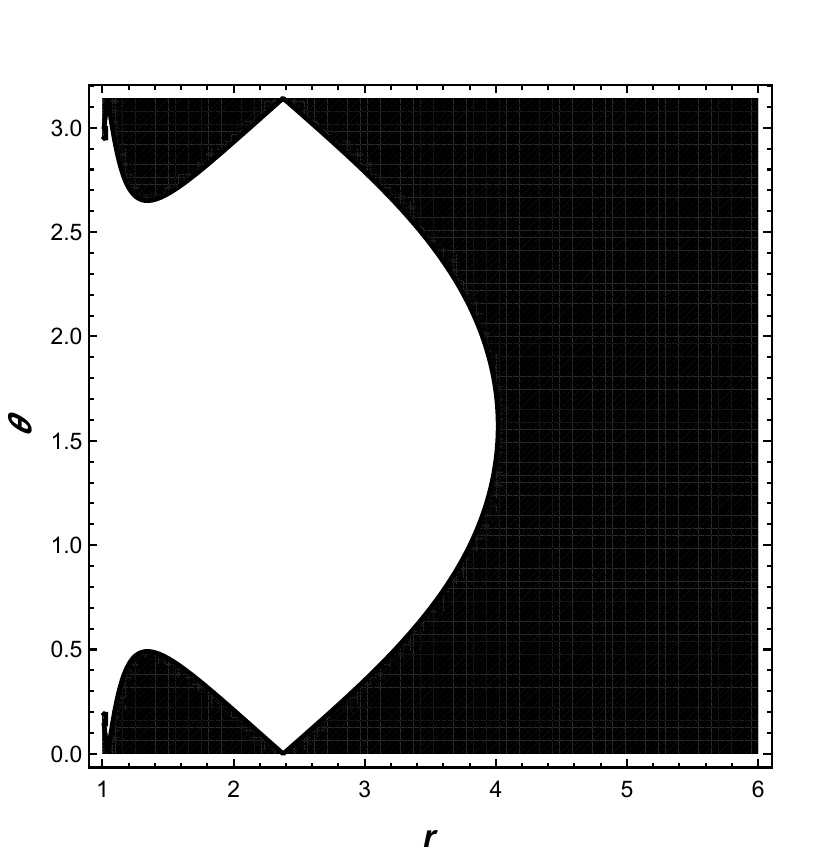}
    }
    \subfloat[a = 1.02, $\alpha$ = 0.06116 ]{
        \includegraphics[width=0.47\textwidth]{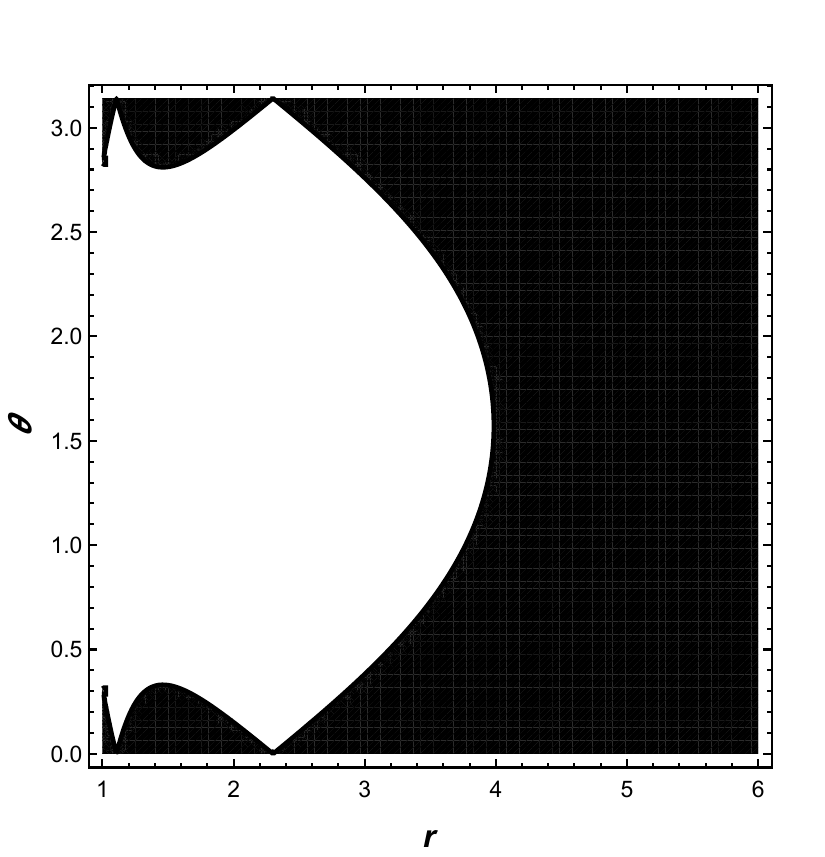}
   }\\[1.5ex]

    \subfloat[a = 1.02, $\alpha$ =0.16116]{
        \includegraphics[width=0.47\textwidth]{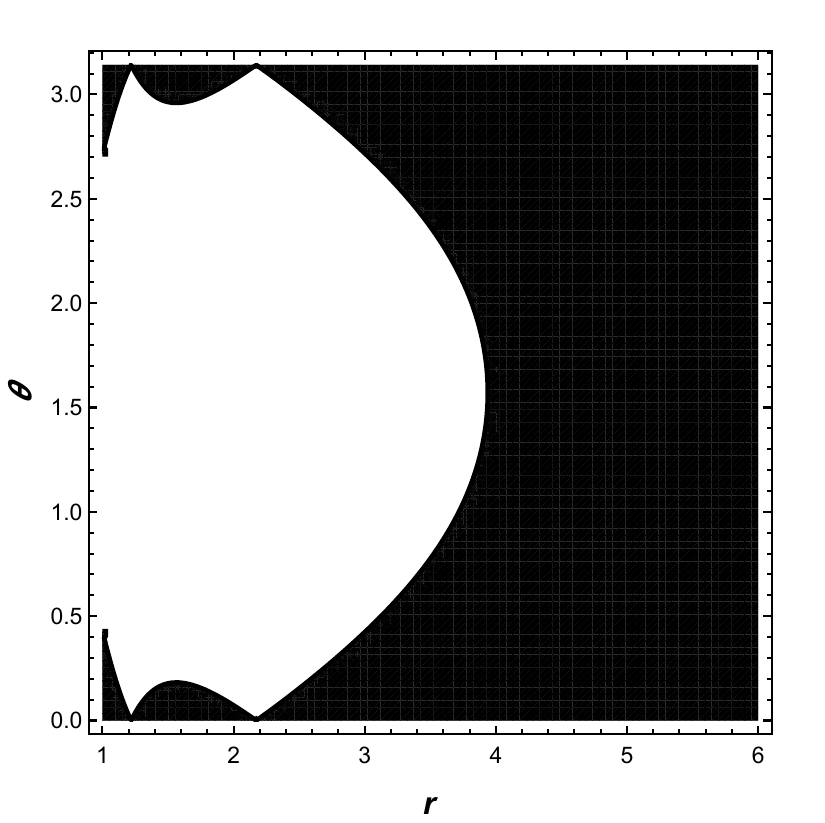}
    }
    \subfloat[a = 1.02, $\alpha$ = 0.26116]{
        \includegraphics[width=0.47\textwidth]{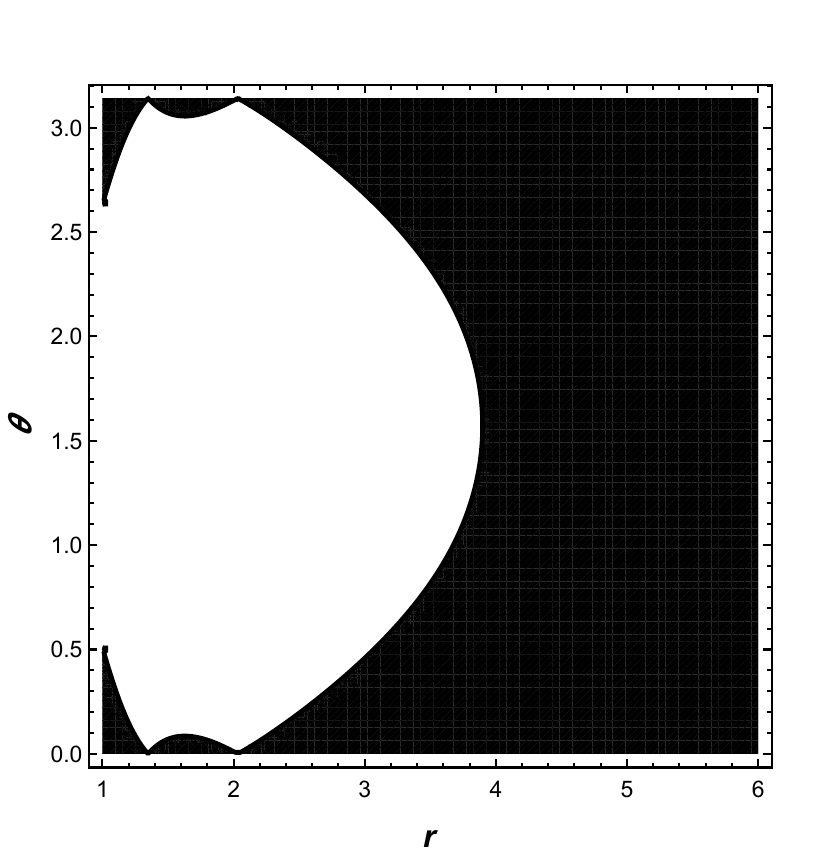}
    }
    \caption{ \label{Figforbid} This figure illustrates the forbidden region for photon motion in the Kerr-MOG spacetime for the parameter values $a = 1.02$ and $\mathcal{M} = 1$. The black-shaded region represents the zone where the effective potential $\mathcal{V}_{\text{eff}} > 0$, indicating that photon propagation is dynamically forbidden in these areas. 
}  
\end{figure}
  We can obtain the apparent shape of the shadow seen by the asymptotic observer if we consider the celestial coordinates $x$ and $y$, which are the coordinates of the asymptotic observer's sky. The general expression to find celestial coordinates $x$ and $y$ are \cite{nguyen2023shadow},
\begin{align}\label{alpha}
    x=\lim_{r_0 \to \infty}\left(-r_0^2 \sin \mathrm{i}\frac{d\Psi}{dr} |_{r_{0},\mathrm{i}} \right)=-\xi \csc \mathrm{i},
\end{align}
\begin{align} \label{beta}
    y=\lim_{r_0 \to \infty}\left(r_0^2 \frac{d\Phi}{dr} |_{r_{0},\mathrm{i}} \right)=\pm\left(\zeta +a^2\cos^2\mathrm{i}-\xi^2 \cot^2 \mathrm{i}\right)^\frac{1}{2}.
\end{align}  
Fig.~\ref{Figforbid} illustrates how the forbidden region for photon motion in Kerr MOG spacetime evolves with increasing MOG deformation parameter $\alpha$, while keeping spin fixed at $a = 1.02$. Fig.~\ref{Figforbid}(a), with $\alpha = 0$, corresponds to the standard KNS and exhibits the largest forbidden region. As $\alpha$ increases from Fig.~\ref{Figforbid}(b) to \ref{Figforbid}(d), the black-shaded region contracts significantly in both radial and angular extent. This behavior indicates that higher values of $\alpha$ reduce the effective potential barrier, allowing more photon trajectories to propagate and escape from the vicinity of the singularity. Consequently, the deformation parameter $\alpha$ plays a crucial role in modifying the photon orbit structure, influencing the shape and size of the shadow observable by a distant observer.
\begin{figure}[!htb]
    \centering
    \subfloat[]{
        \includegraphics[width=0.47\textwidth]{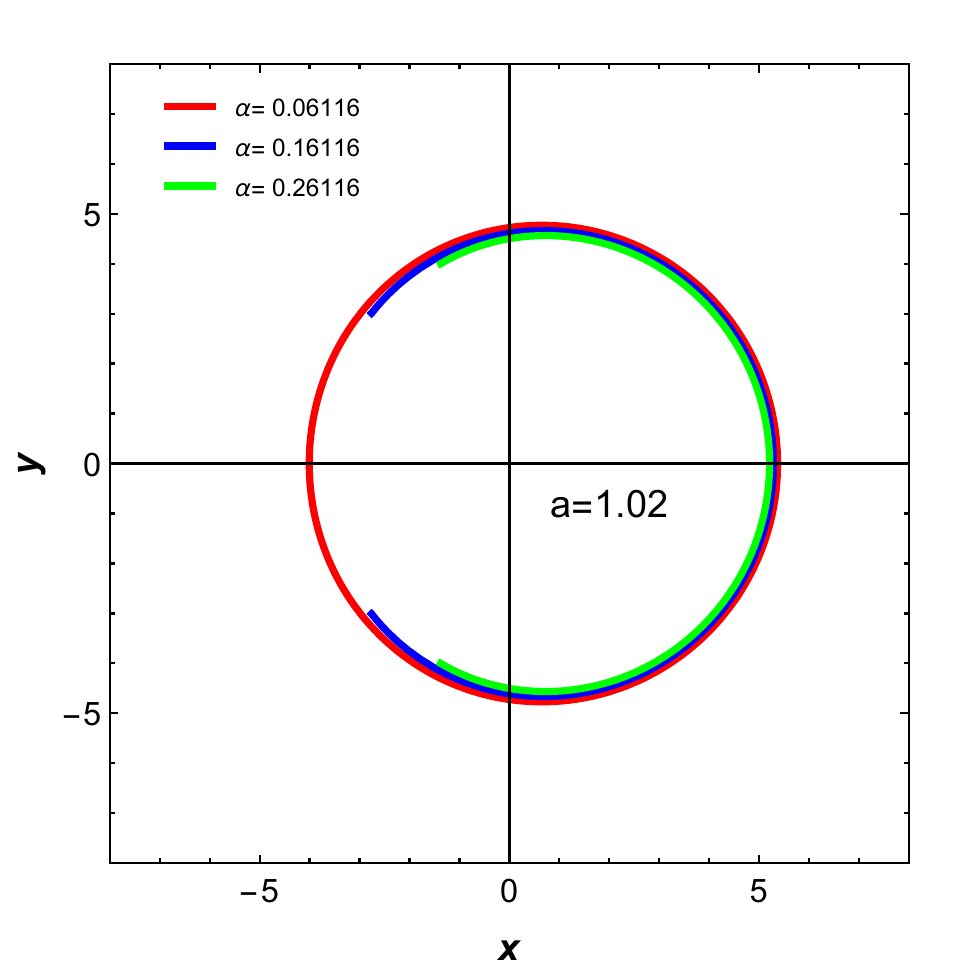}
    }
    \subfloat[]{
        \includegraphics[width=0.47\textwidth]{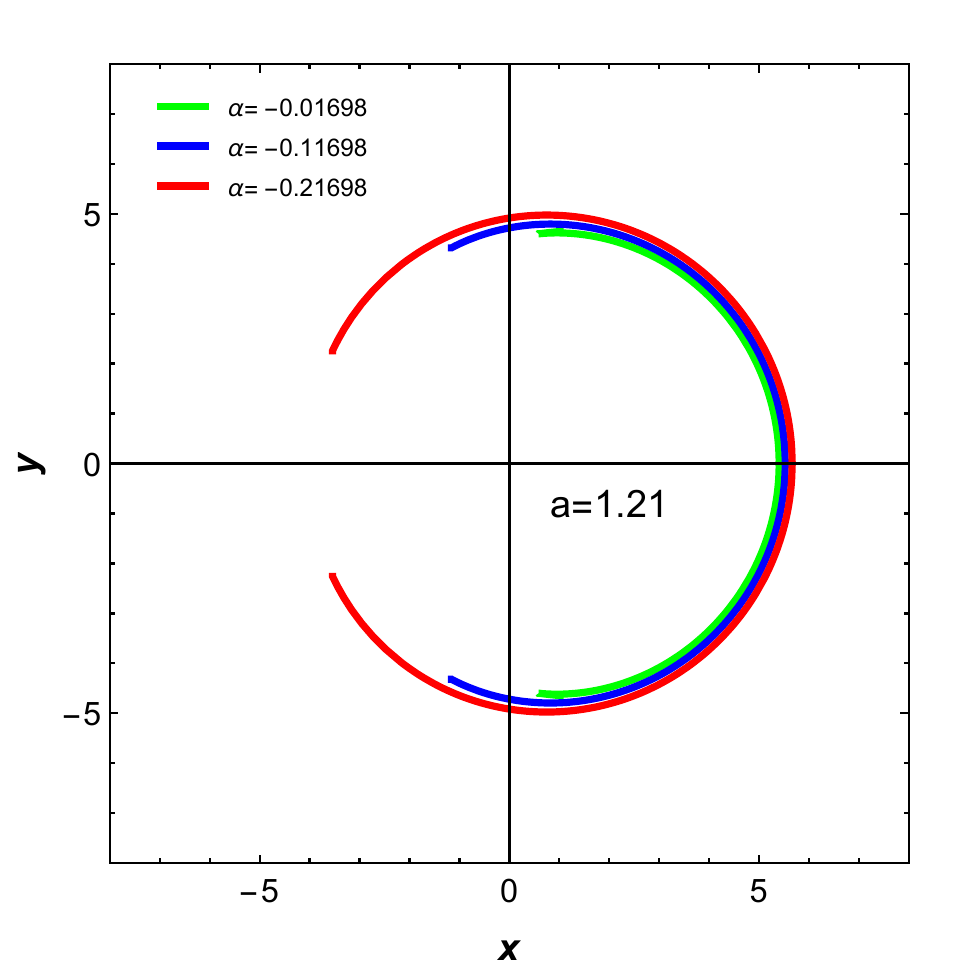}
    }\\[1.5ex]
    \subfloat[]{
        \includegraphics[width=0.47\textwidth]{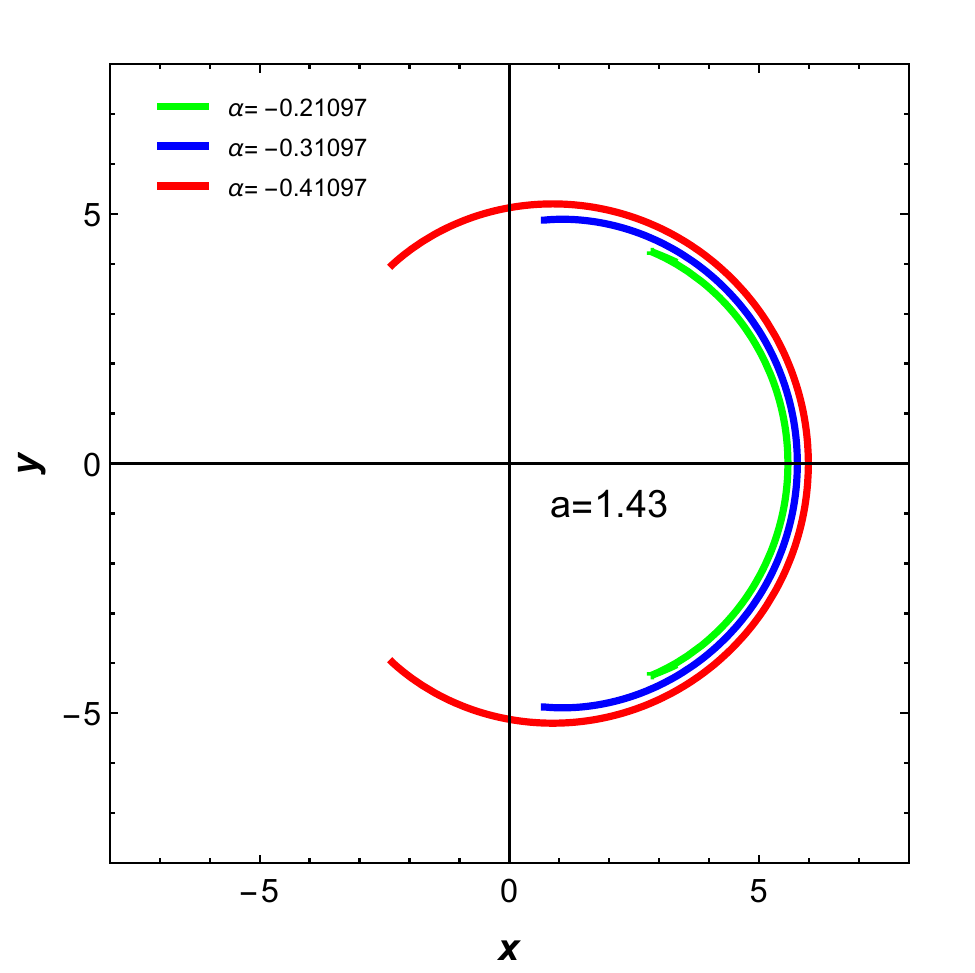}
    }
    \subfloat[]{
        \includegraphics[width=0.47\textwidth]{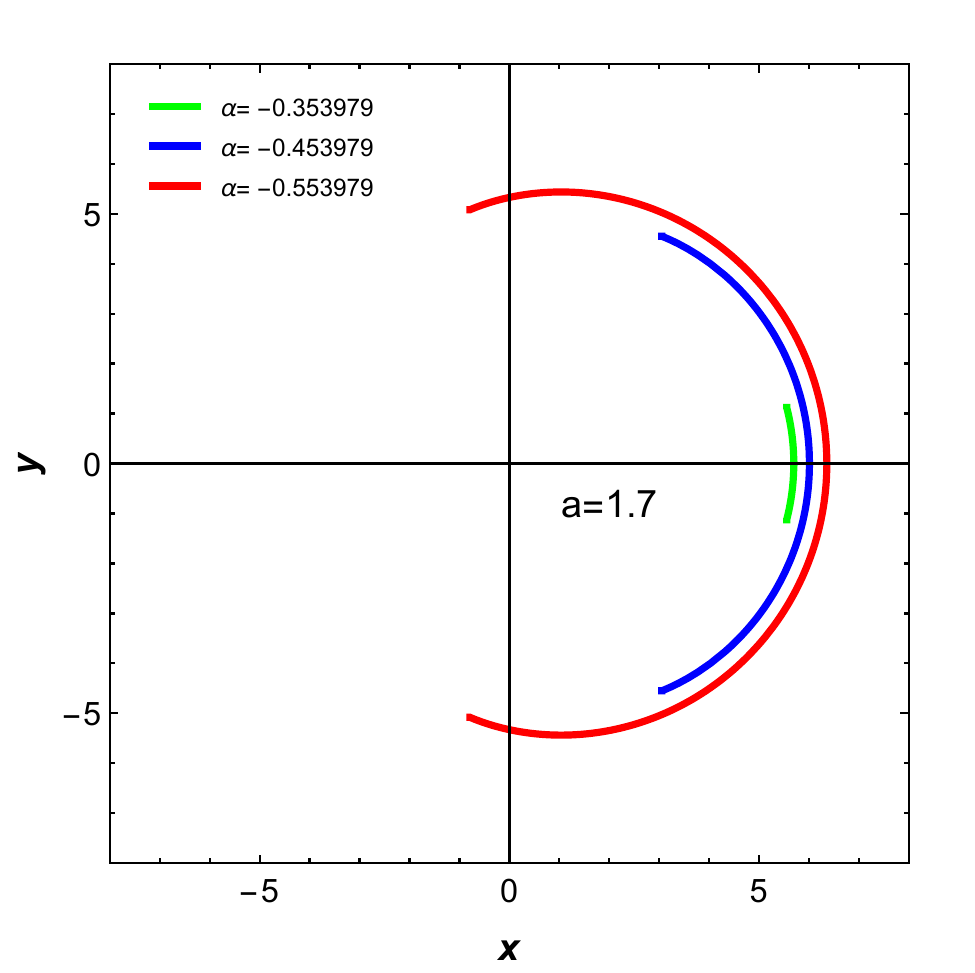}
    }
    \caption{Shadows of a KMNS with observational inclinations angles $"\mathrm{i}"$, spin $"a"$ and MOG parameter $\alpha$.
    The plots from left to right are associated with  $\mathrm{i} = 15^{\circ}$ and $\mathcal{M}=1$. As $\alpha$ increases, the shadow size contracts noticeably due to enhanced MOG corrections, especially at higher spin. The low inclination angle reduces asymmetry, highlighting the radial effect of $\alpha$ on the photon region.
}
    \label{Figshadowa}
\end{figure}
As shown in Fig.~\ref{Figshadowa}, we observe the behavior of shadow contours cast by a KMNS at a low inclination angle of $\mathrm{i} = 15^\circ$, for increasing spin values and varying MOG parameter $\alpha$. For lower spins such as $a = 1.02$ and $a = 1.21$, the shadow retains a nearly circular shape, and its radius decreases smoothly as $\alpha$ increases, indicating a contraction of the photon capture region due to the enhanced gravitational modification. However, in the case of higher spins, particularly $a = 1.43$ and more evidently at $a = 1.7$, a distinct transition occurs. The shadow corresponding to the largest $\alpha$ (red curve) becomes fragmented and eventually disappears at $a = 1.701$, suggesting the absence of closed unstable photon orbits for that configuration.
  \begin{figure}[!htb]
    \centering
    \subfloat[]{
        \includegraphics[width=0.47\textwidth, height=5cm, keepaspectratio]{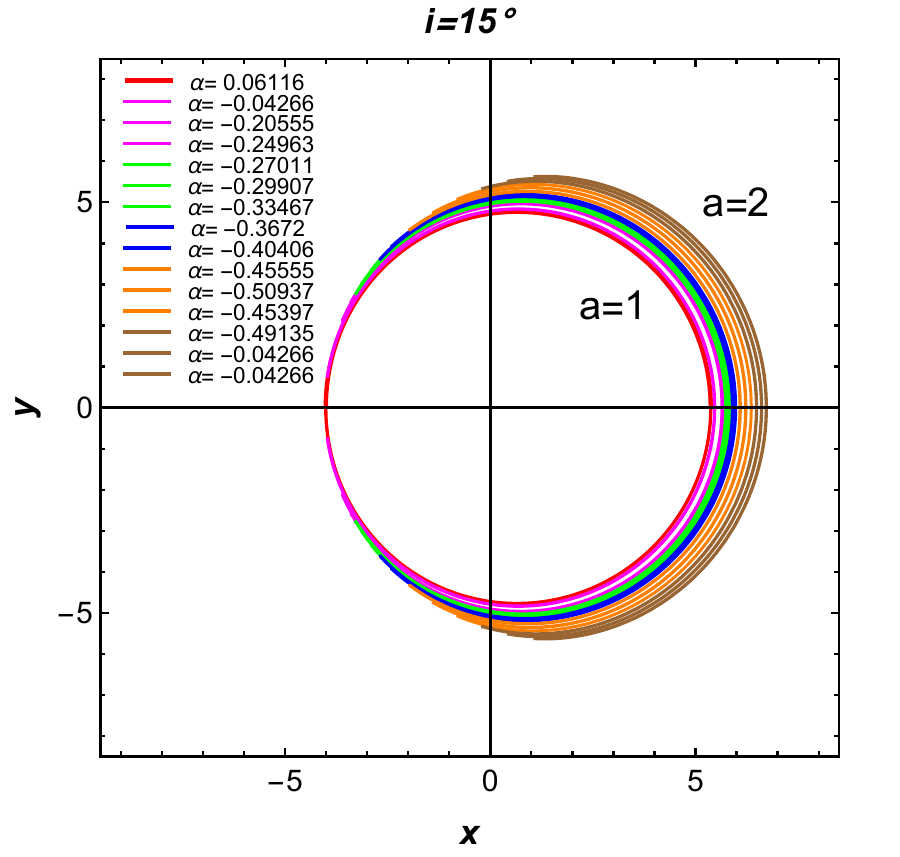}
    }
    \subfloat[]{
        \includegraphics[width=0.47\textwidth, height=5cm, keepaspectratio]{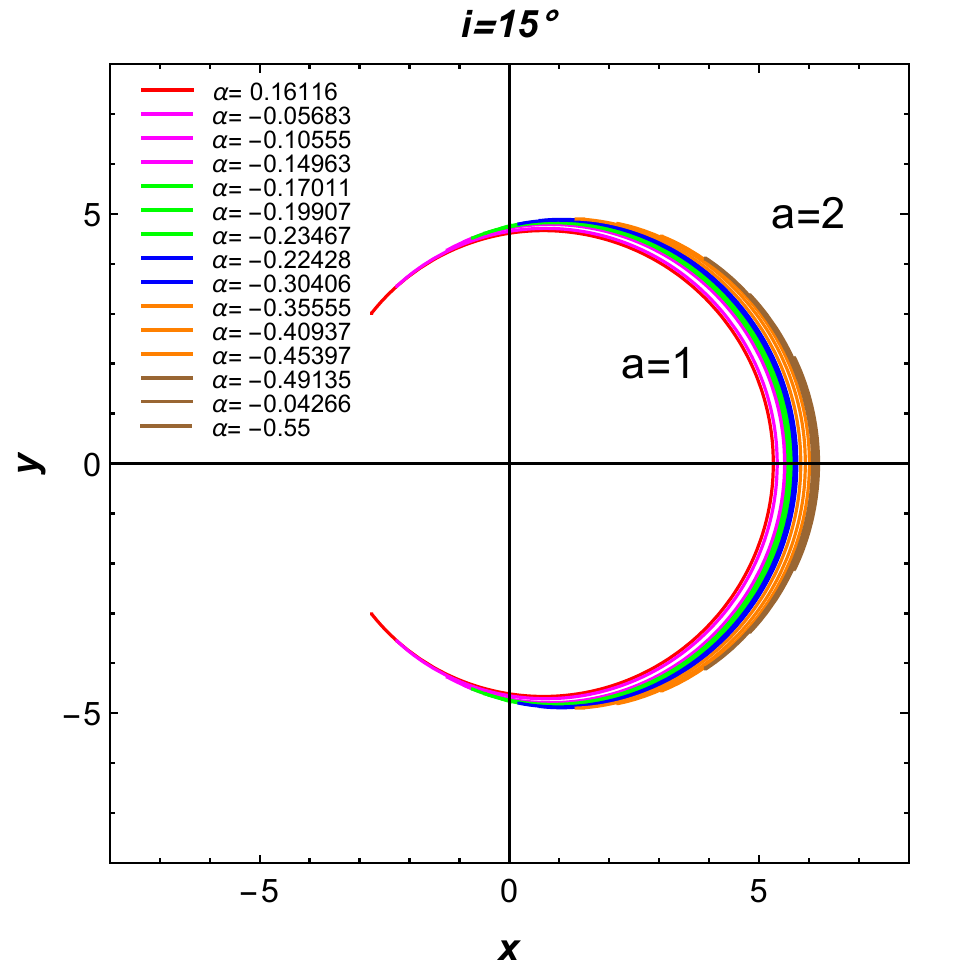}
    }\\[1.5ex]
    \subfloat[]{
        \includegraphics[width=0.47\textwidth, height=5cm, keepaspectratio]{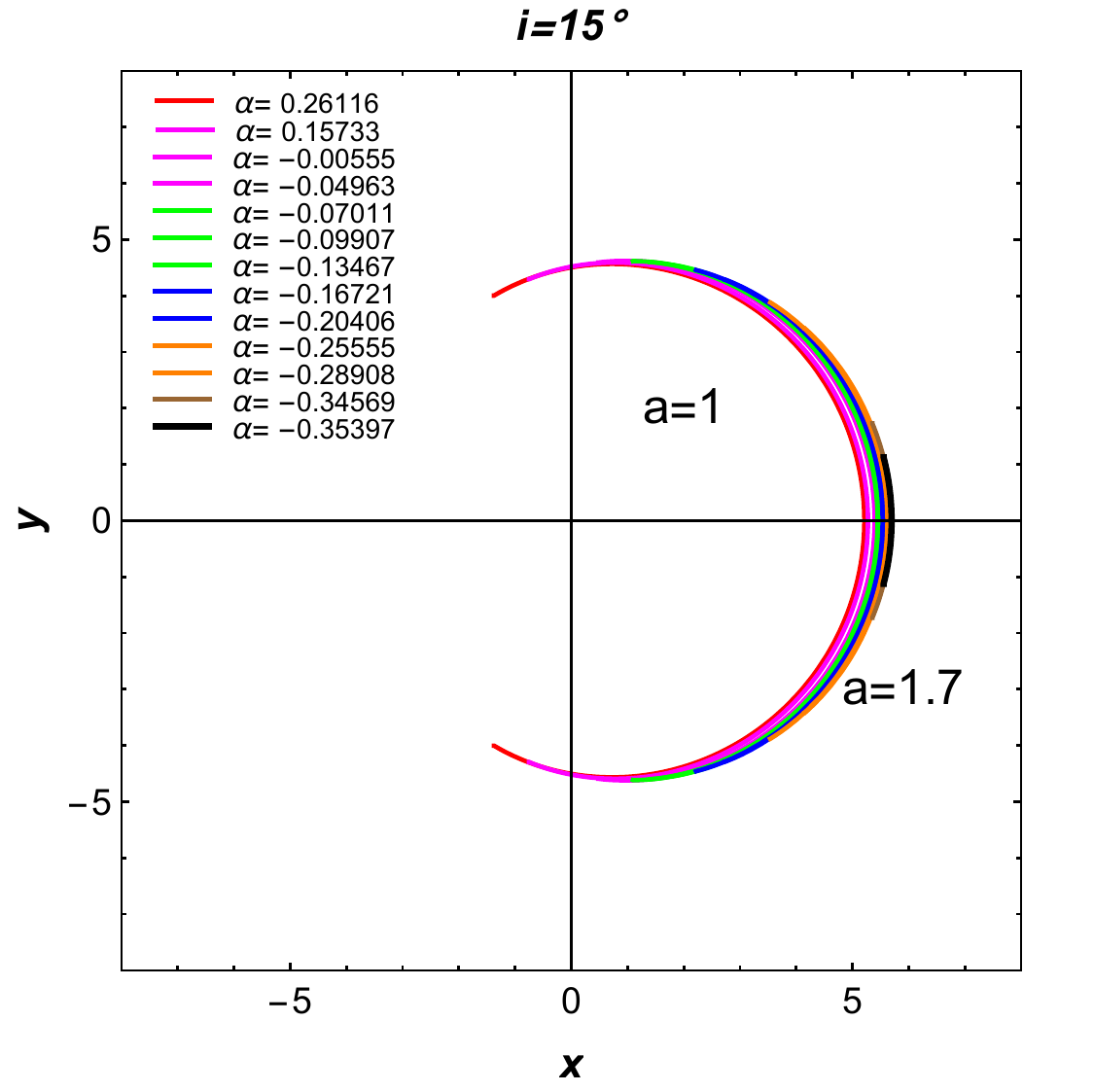}
    }
    \subfloat[]{
        \includegraphics[width=0.47\textwidth, height=5cm, keepaspectratio]{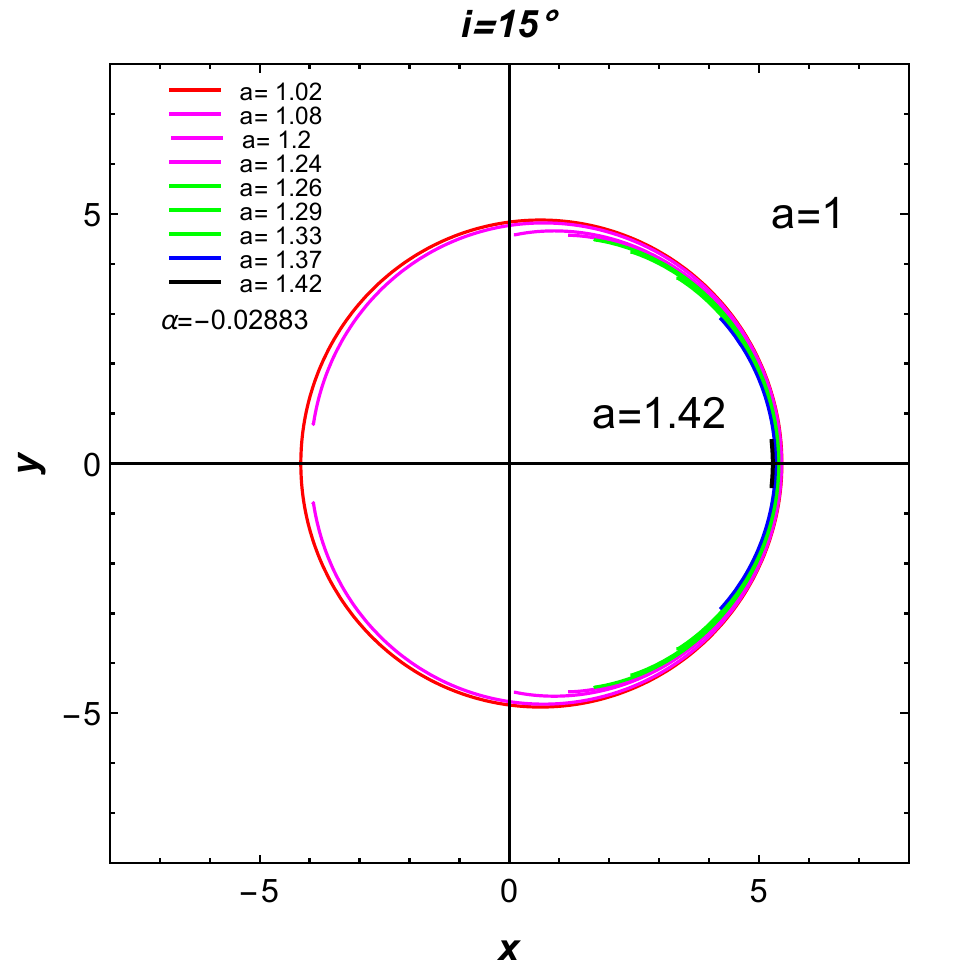}
    }
    \caption{\justifying
Shadows of a KMNS at an inclination angle of $i = 15^\circ$ for various spin parameters $a$ and MOG parameters $\alpha$, with $\mathcal{M} = 1$. 
{Panels (a) and (b) show the evolution of the shadow as the spin parameter $a$ increases from 1 to 2, with corresponding values of $\alpha$. Panel (c) shows the shadow as the spin parameter $a$ increases from 1 to 1.7 with corresponding values of $\alpha$.
} Panel (d) illustrates the effect of increasing spin from $a = 1$ to $a = 1.42$ for fixed $\alpha =-0.02833$. The shadow becomes increasingly deformed and ultimately disappears beyond $a = 1.42$.
}
    \label{Figshadowb}
\end{figure}
In Fig.~\ref{Figshadowb}, we study the shadows of KMNS at an inclination angle of $i = 15^\circ$, using three different values of the MOG parameter $\alpha$ for each spin. As shown in \ref{Figshadowb}(a) to  \ref{Figshadowb}(c), the shadow opens its gap progressively with increasing spin $a$ for each corresponding value of $\alpha$. Moreover, for larger values of $\alpha$, the shadow begins to open its gap at lower spin values compared to smaller $\alpha$, which is clearly visible in \ref{Figshadowb}(a) and \ref{Figshadowb}(b). In \ref{Figshadowb}(c), the shadow completely disappears at $a = 1.71$, indicating the absence of closed photon orbits. {In Fig.~\ref{Figshadowb}(d), the MOG parameter $\alpha = -0.02833$ is fixed, the shadow gradually opens gaps as the spin increases, eventually disappearing at $a = 1.42$.
} As seen in  Fig.~\ref{Figshadowb}, illustrate that larger $\alpha$ values cause the shadow to open and vanish earlier, highlighting the spin and $\alpha$ dependence in the shadow topology of KMNSs. However, in Fig. \ref{Figshadowc}, the shadow of the KMNS at a higher inclination angle does not disappear at lower spins, unlike what is observed in Figure~\ref{Figshadowb}.
 \begin{figure}[!htb]
    \centering
    \subfloat[]{
        \includegraphics[width=0.4\textwidth, height=5cm, keepaspectratio]{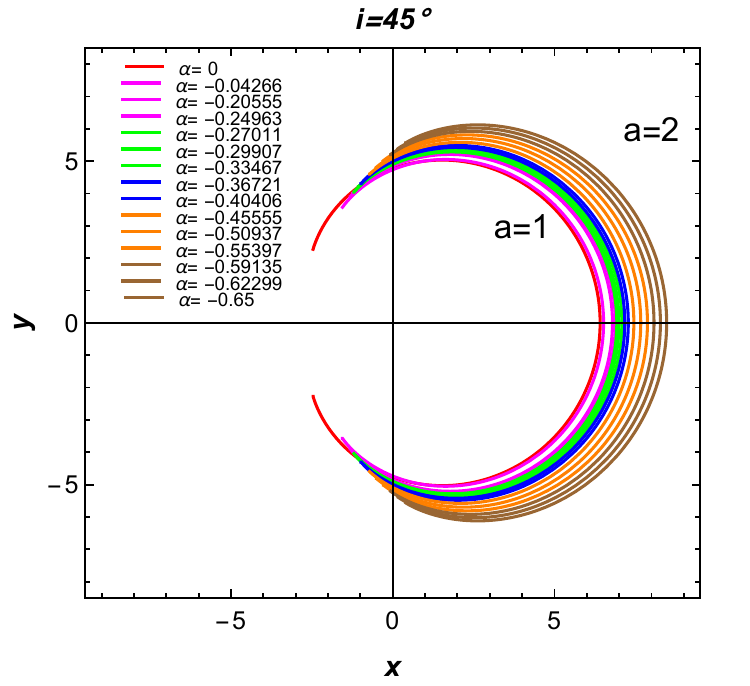}
    }
    \subfloat[]{
        \includegraphics[width=0.355\textwidth, height=5cm, keepaspectratio]{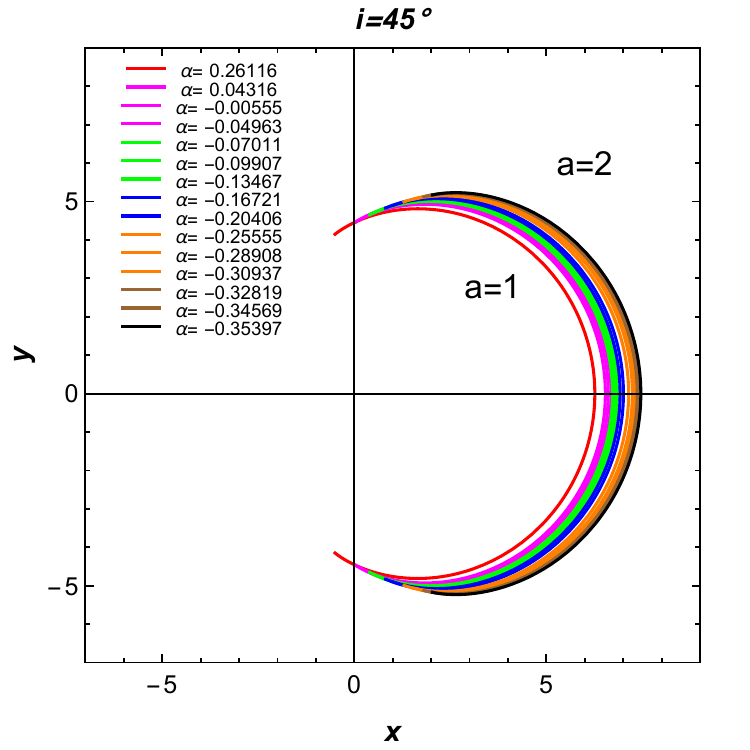}
    }
    \subfloat[]{
        \includegraphics[width=0.35\textwidth, height=5cm, keepaspectratio]{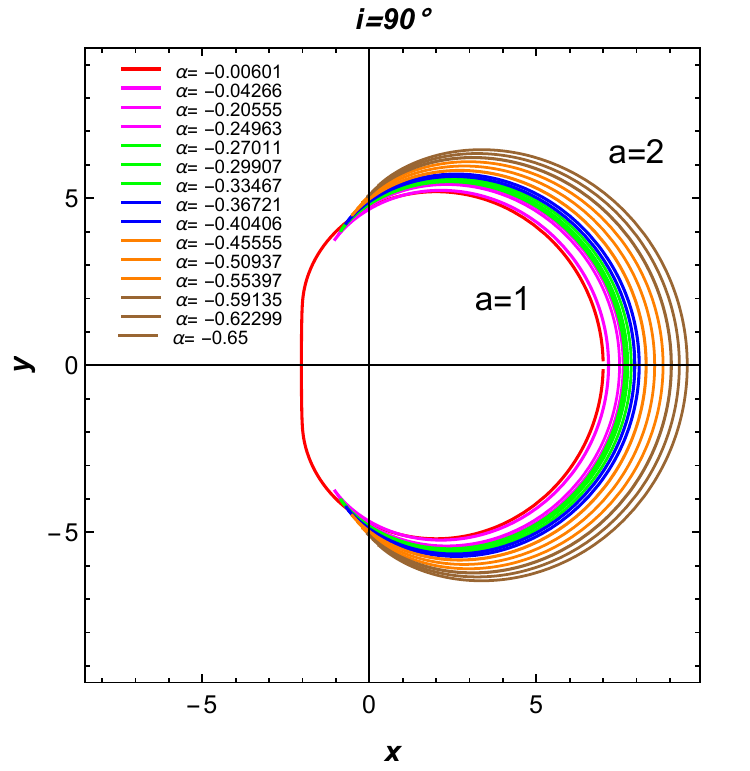}
    }
    \caption{ \justifying
        Shadows of KMNS for various spin parameters $a$, MOG parameter $\alpha$, and observational inclination angles $i$. {Panels (a) and (b) correspond to $i = 45^\circ$, while panel (c) corresponds to $i = 90^\circ$, with $\mathcal{M} = 1$.}
  } 
 \label{Figshadowc}
\end{figure}

\begin{figure}[!htb]
    \centering
    \subfloat[$\alpha=0.06116$]{
        \includegraphics[width=0.35\textwidth]{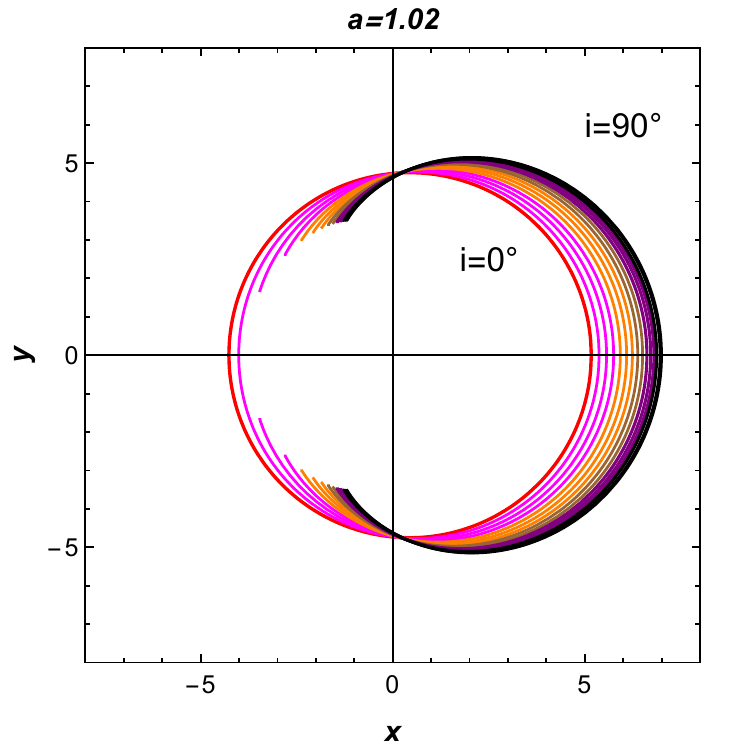}
    }
    \subfloat[$\alpha=0.16116$]{
        \includegraphics[width=0.35\textwidth]{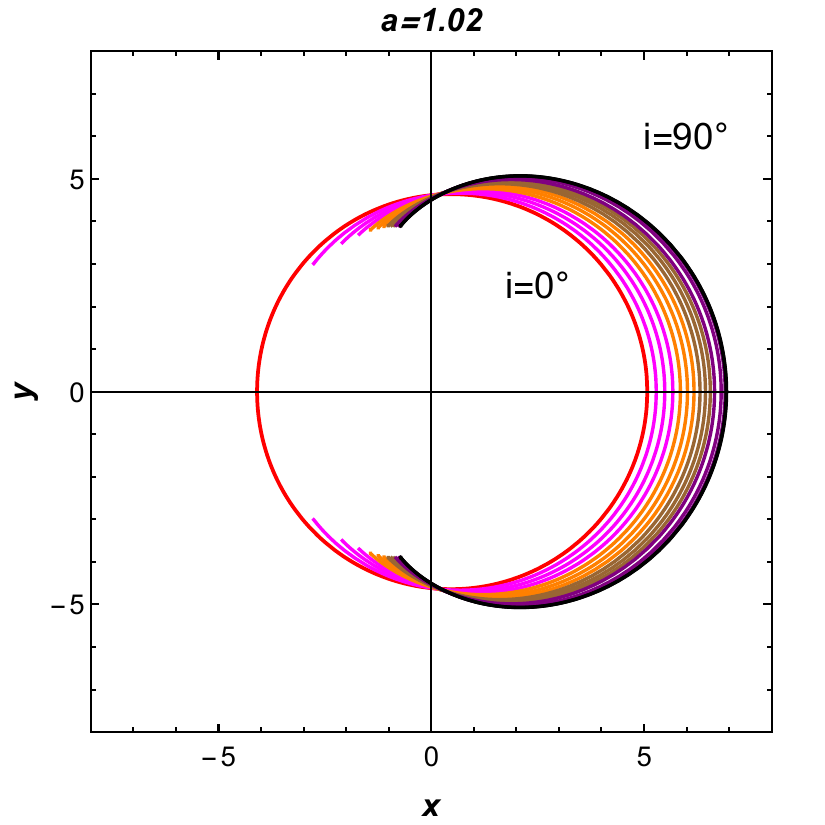}
    }
    \subfloat[$\alpha=0.26116$]{
        \includegraphics[width=0.35\textwidth]{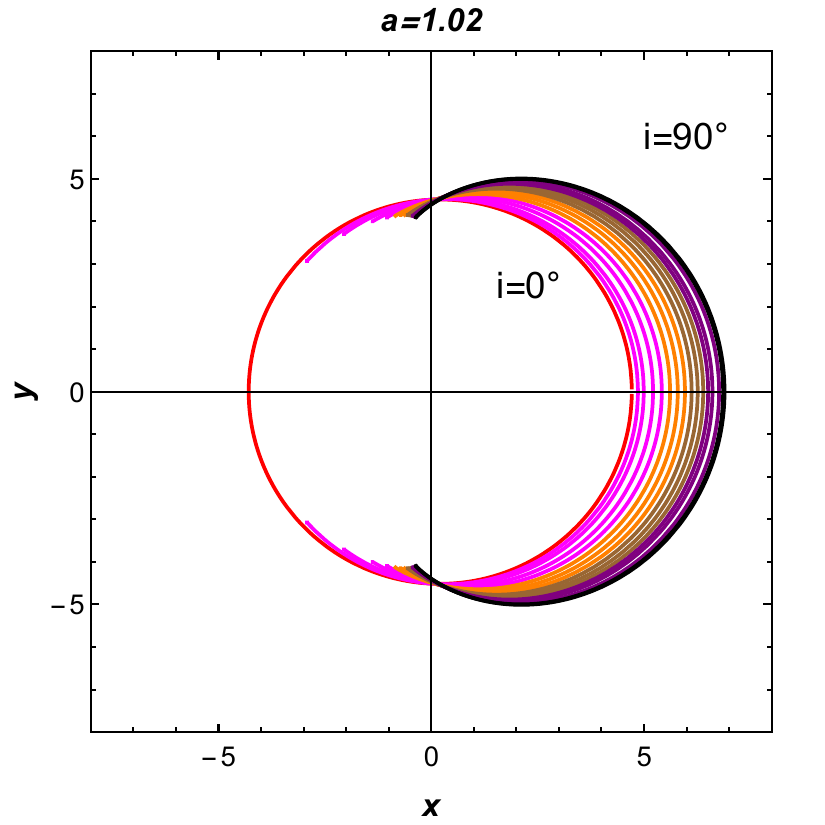}
    }\\[1.5ex]
      \subfloat[$\alpha=-0.23467$]{
        \includegraphics[width=0.45\textwidth]{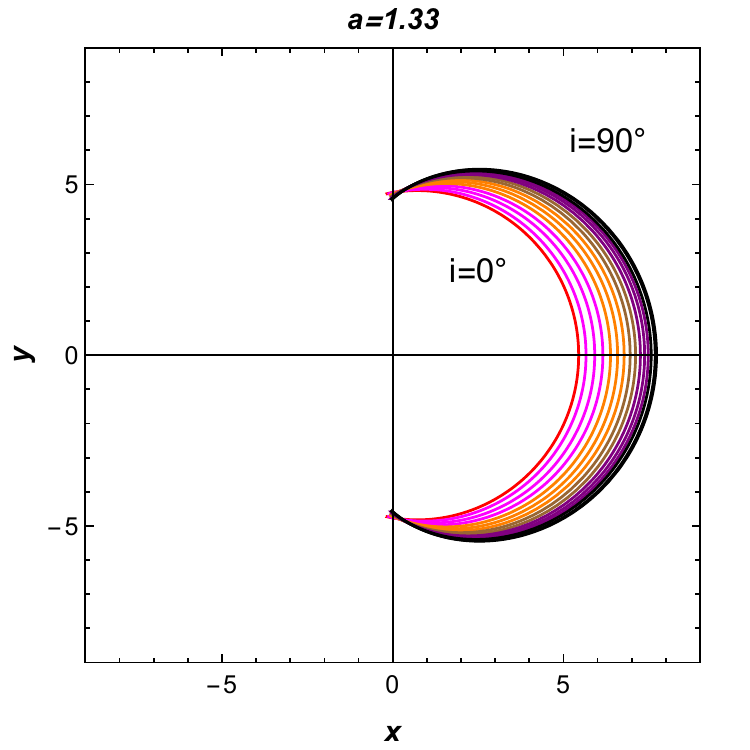}
    }
      \subfloat[$\alpha=-0.35555$]{
        \includegraphics[width=0.45\textwidth]{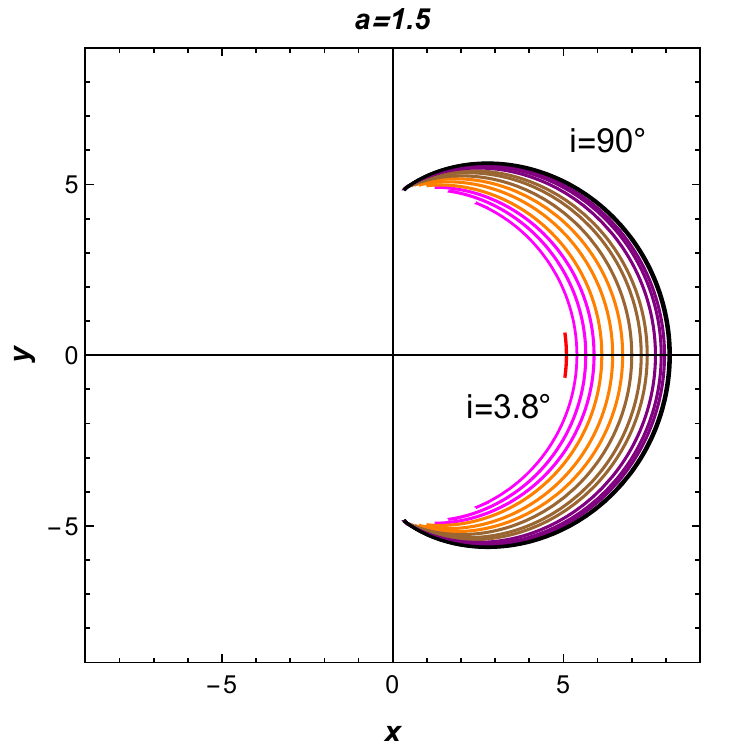}
    }
    \caption{\justifying
        Shadows of KMNS with spin ranging from $a = 1.02$ to $a = 1.5$, $\mathcal{M} = 1$, and varying MOG parameter $\alpha$, observed at different inclination angles $i$. In each plot, colors represent different inclination angles, ranging from red ($i = 0^\circ$) to black ($i = 90^\circ$). Plot (e) starts from $i = 3.8^\circ$ up to $i = 90^\circ$.}
    \label{Figshadowd}
\end{figure}

In Fig.~\ref{Figshadowd}, we present the shadows of KMNS for spin values ranging from $a = 1.02$ to $a = 1.5$, observed at various inclination angles. The top row corresponds to fixed spin $a = 1.02$ with different values of $\alpha$. As seen in the plots, increasing $\alpha$ at fixed spin leads to the shadow opening wider gaps, especially at higher inclination angles. This demonstrates that for larger $\alpha$, the shadow tends to open earlier compared to lower $\alpha$ at the same spin. The bottom row illustrates the shadow morphology for spins $a = 1.33$ and $a = 1.5$ with fixed and varying $\alpha$. {In Fig.~\ref{Figshadowd}, the transition from closed to open or vanishing shadows occurs at lower spin values for smaller inclination angles and shifts to higher spins as the inclination increases, for corresponding values of $\alpha$. This comparison confirms the combined influence of spin, inclination, and $\alpha$ in determining the shadow topology of KMNSs.
}
\newpage
\section{Thin Acceration Disk} \label{secIII} 
In this section, we calculate the fundamental physical quantities associated with the accretion disk around a KMNS. These quantities include the specific energy $\mathrm{E}$, specific angular momentum $L$, angular velocity $\Omega$, and the flux of emitted radiation $\mathrm{F}$ for particles in stable circular orbits. For simplicity, in our analysis, we set \( \mathcal{M} = M \), where \( M \) represents the body’s total mass. The behavior of these parameters follows from the disk's structure equations, which are based on the conservation of mass, energy, and angular momentum. Each kinematic quantity depends on the orbital radius and can be determined using the general formalism developed in references~\cite{Gyulchev2019imageo}. 
The specific energy \( E \), specific angular momentum \( L \), and angular velocity \( \Omega \) of particles moving in circular orbits are given by \cite{Gyulchev2019imageo}
\begin{align}
E &= \frac{ -g_{tt} - g_{t\phi} \, \Omega }{ \sqrt{ -g_{tt} - 2g_{t\phi} \, \Omega - g_{\phi\phi} \, \Omega^2 } }, \label{eq:E} \\
L &= \frac{ g_{t\phi} + g_{\phi\phi} \, \Omega }{ \sqrt{ -g_{tt} - 2g_{t\phi} \, \Omega - g_{\phi\phi} \, \Omega^2 } }, \label{eq:L} \\
\Omega &= \frac{d\phi}{dt} = \frac{ -g_{t\phi,r} + \sqrt{ (g_{t\phi,r})^2 - g_{tt,r} \, g_{\phi\phi,r} } }{ g_{\phi\phi,r} }. \label{eq:Omega}
\end{align}
The marginally stable orbit marks the innermost boundary of stable circular orbits in a Keplerian rotational profile. This marginally stable radius is obtained by \ref{rms}. A stationary observer, defined by motion along a worldline with constant \( r \) and \( \theta \), and rotating at a fixed angular velocity \( \Omega \), possesses a four-velocity vector of the form $
u^\mu \propto \left( \frac{\partial}{\partial t} \right)^\mu + \Omega \left( \frac{\partial}{\partial \phi} \right)^\mu.
$
To remain within the future light cone, this vector must satisfy the condition \cite{kazempour2022analysis}
\begin{equation}
\left[ \left( \frac{\partial}{\partial t} \right)^\mu + \Omega \left( \frac{\partial}{\partial \phi} \right)^\mu \right]^2 = g_{tt} + 2\Omega g_{t\phi} + \Omega^2 g_{\phi\phi} \leq 0. \label{eq:cone_condition}
\end{equation}
 \begin{figure}[!htb]
    \centering
    \subfloat[$a=1.01$]{
        \includegraphics[width=0.47\textwidth]{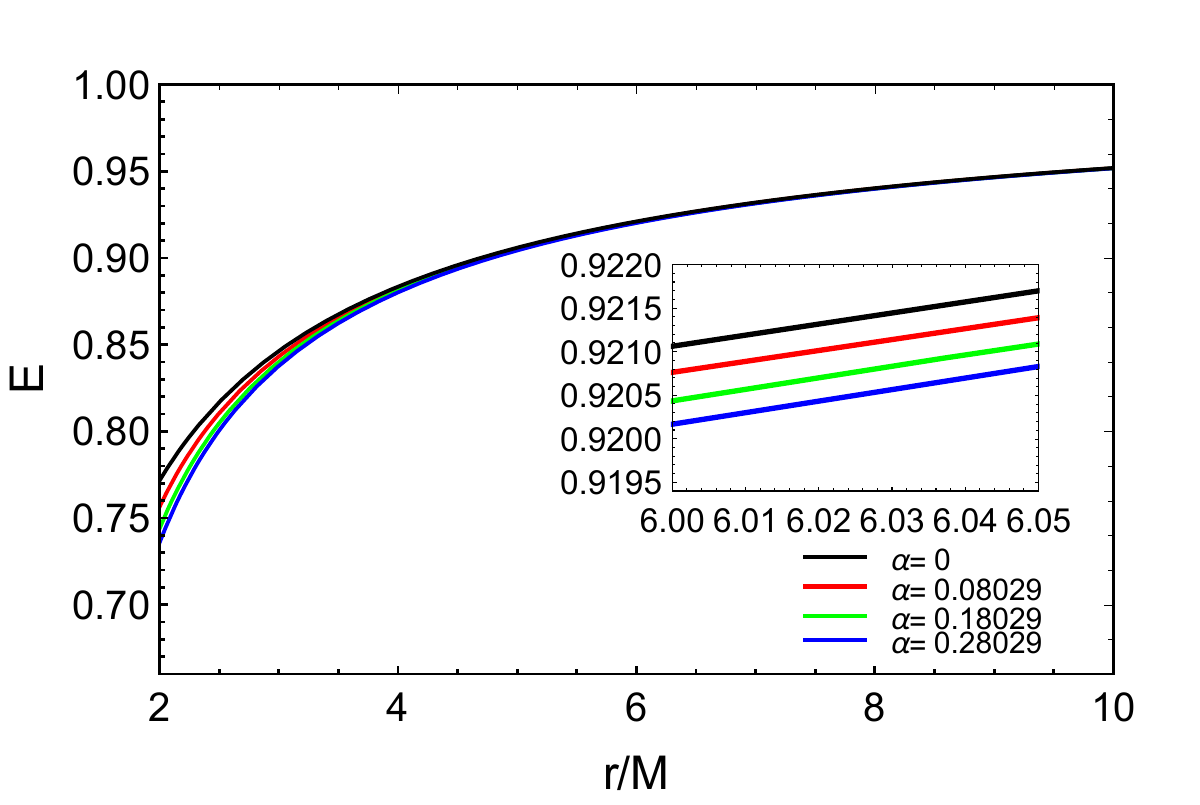}
    }
    \subfloat[$a=1.38$]{
        \includegraphics[width=0.47\textwidth]{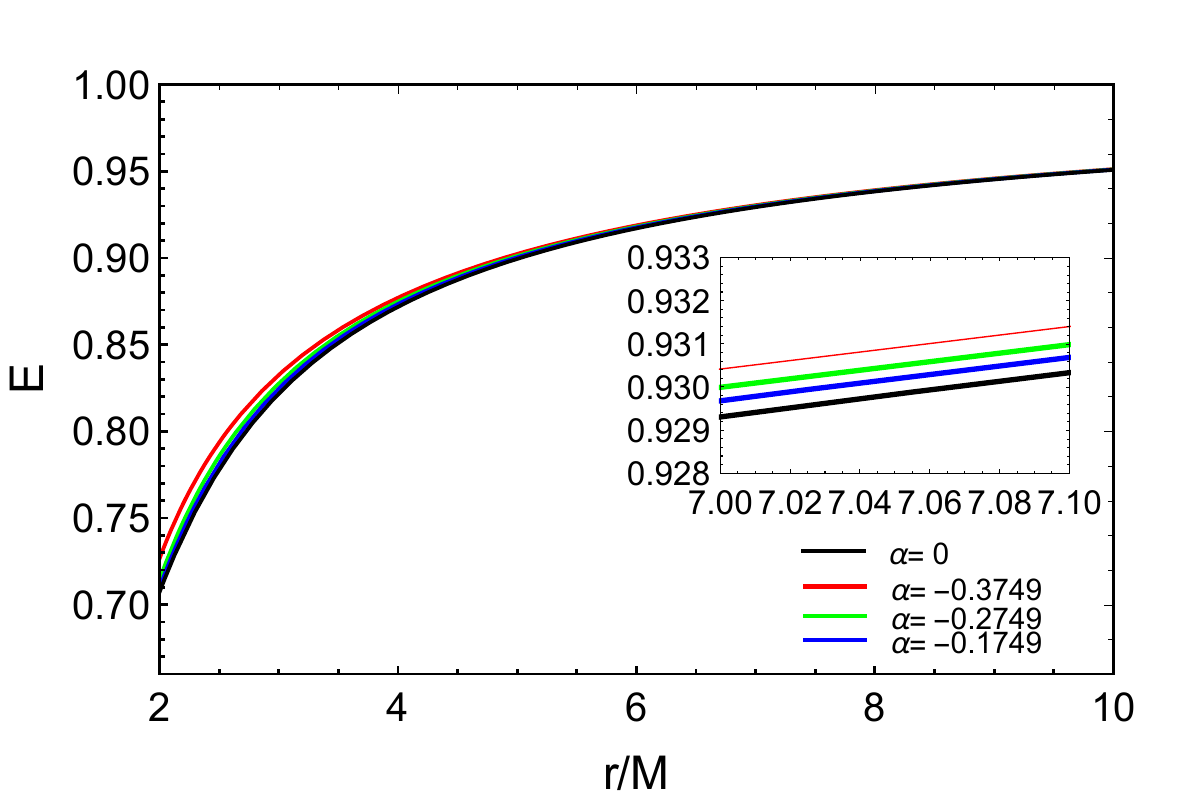}
    }\\[1.5ex]
    \subfloat[$a=1.02,a=1.16, a=1.3, a=1.5, a=1.7, a=2$]{
        \includegraphics[width=0.47\textwidth]{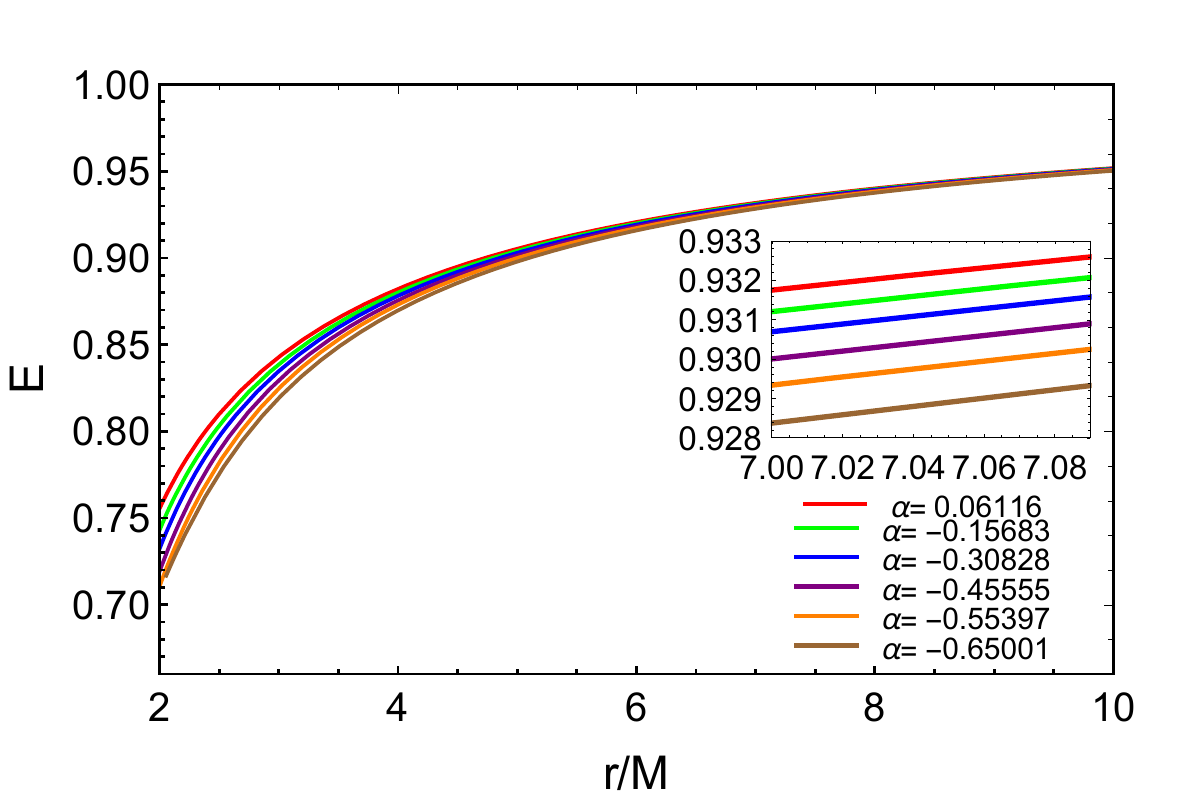}
    }
    \subfloat[$\alpha=-0.02883$]{
        \includegraphics[width=0.47\textwidth]{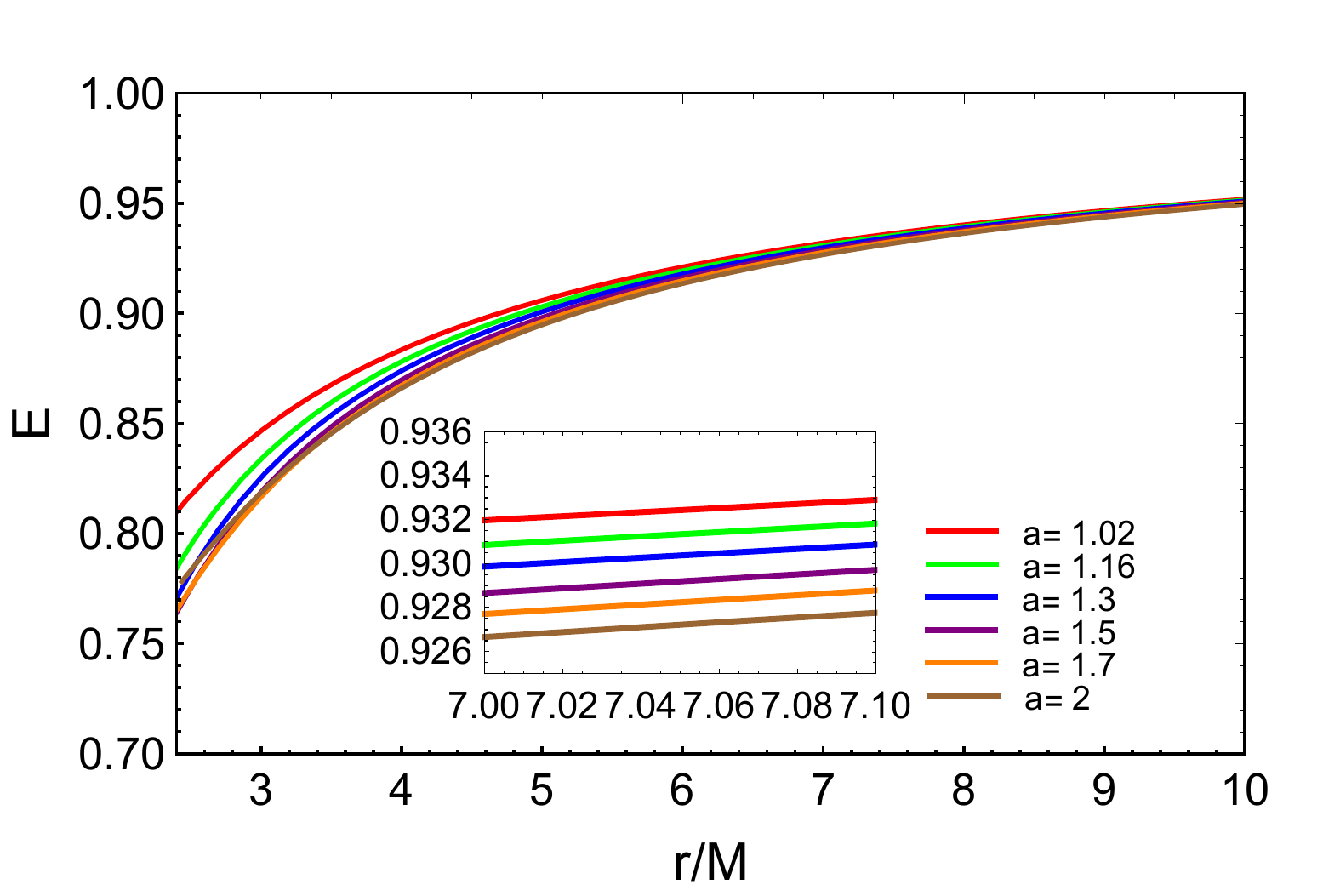}
    }
   \caption{ \justifying \label{Figenergy} Specific energy \(E\) of circular orbits versus radial coordinate \(r/{M}\). (a)-(b) Fixed spin \(a = 1.01, 1.38\) with varying \(\alpha\), including KNS (\(\alpha=0\)). (c) Varying \(a\) with corresponding \(\alpha\) where different colors represent various spin values: red ($a = 1.02$), green ($a = 1.16$), blue ($a = 1.3$), purple ($a = 1.5$), orange ($a = 1.7$), and brown ($a = 2$). (d) Varying \(a\) with fixed \(\alpha =- 0.02883\). Higher \(\alpha\) increases energy, higher spin reduces it, showing competing effects of MOG and frame-dragging.}
\end{figure}
This inequality, which also appears in (\ref{eq:E}) and (\ref{eq:L}), imposes a constraint on the allowed angular velocity \( \Omega \) of stationary observers $\Omega_{\text{min}} < \Omega < \Omega_{\text{max}}$ with the minimum and maximum bounds given by
\begin{align}
\Omega_{\text{min}} &= \omega - \sqrt{ \omega^2 - \frac{g_{tt}}{g_{\phi\phi}} }, \label{eq:omega_min} \\
\Omega_{\text{max}} &= \omega + \sqrt{ \omega^2 - \frac{g_{tt}}{g_{\phi\phi}} }, \label{eq:omega_max}
\end{align}
\begin{figure}[!htb]
    \centering
    \subfloat[a = 1.01]{
        \includegraphics[width=0.47\textwidth]{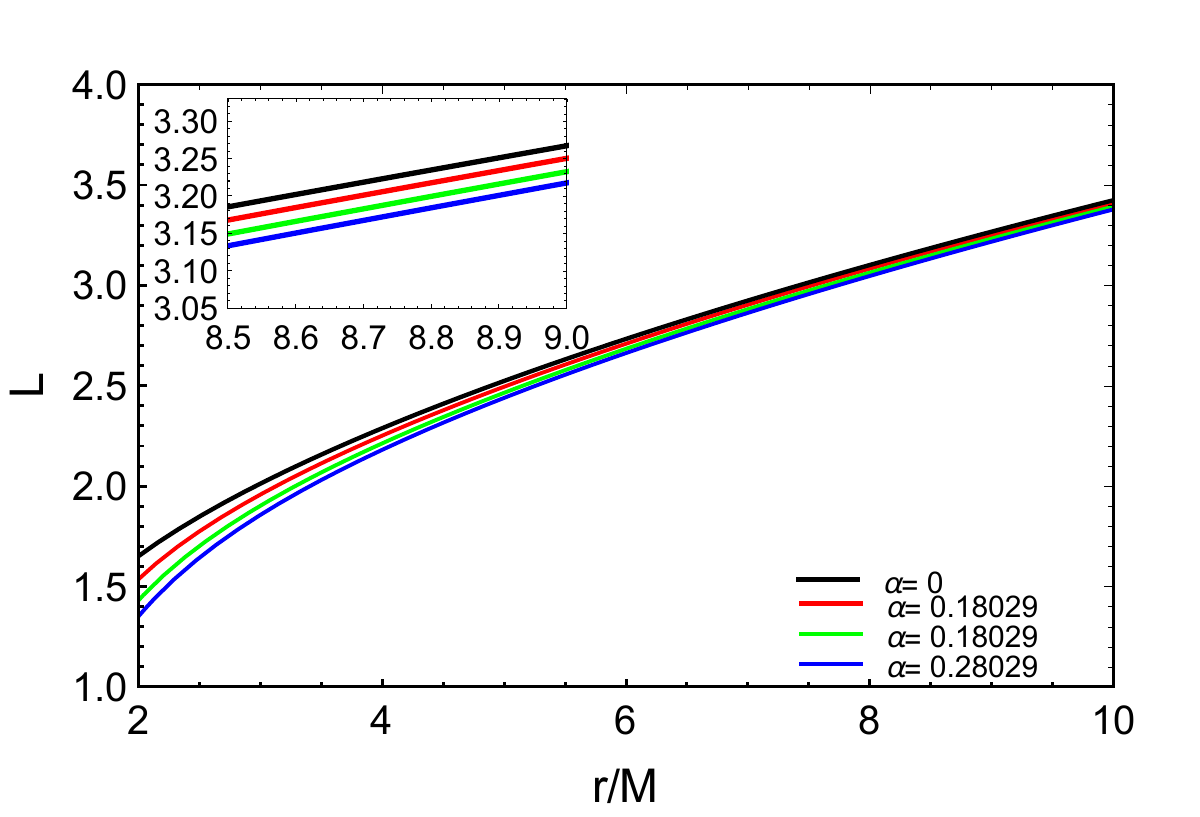}
    }
    \subfloat[a = 1.38]{
        \includegraphics[width=0.47\textwidth]{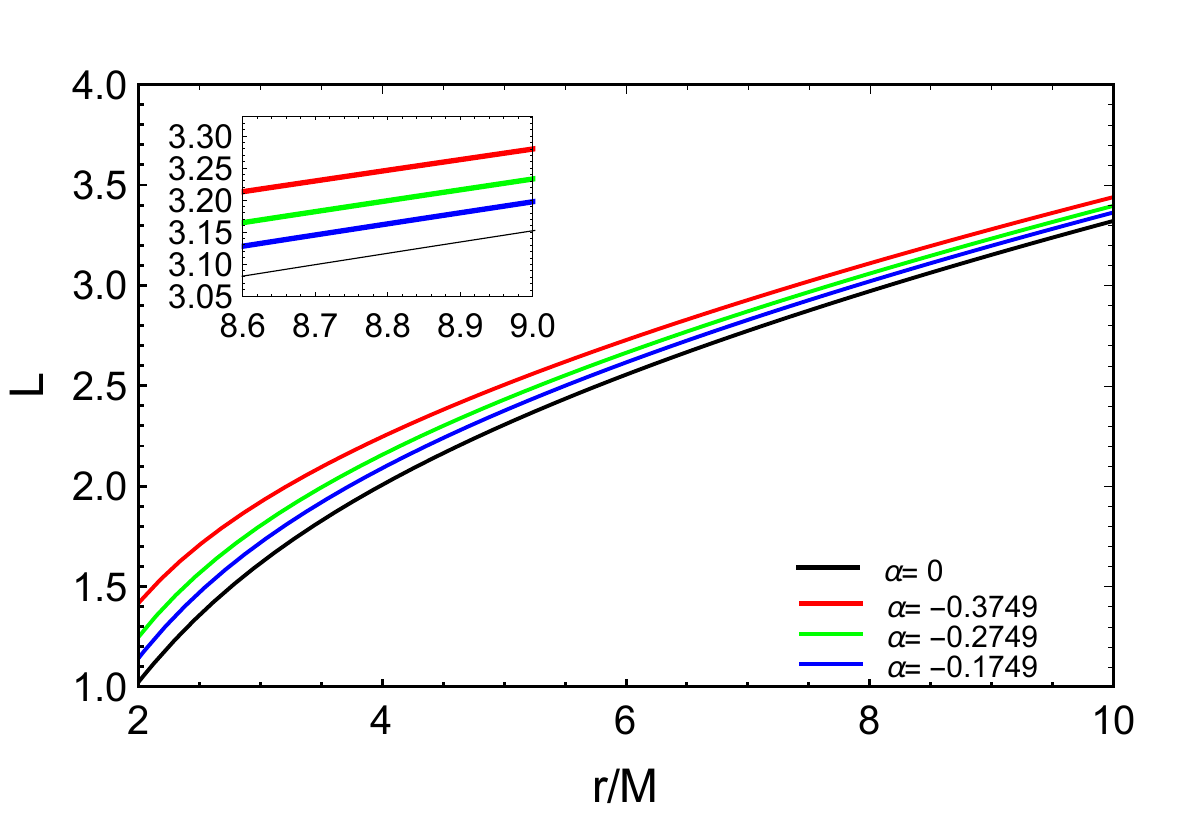}
    }\\[1.5ex]
    \subfloat[$a=1.02,a=1.16, a=1.3, a=1.5, a=1.7, a=2$]{
        \includegraphics[width=0.47\textwidth]{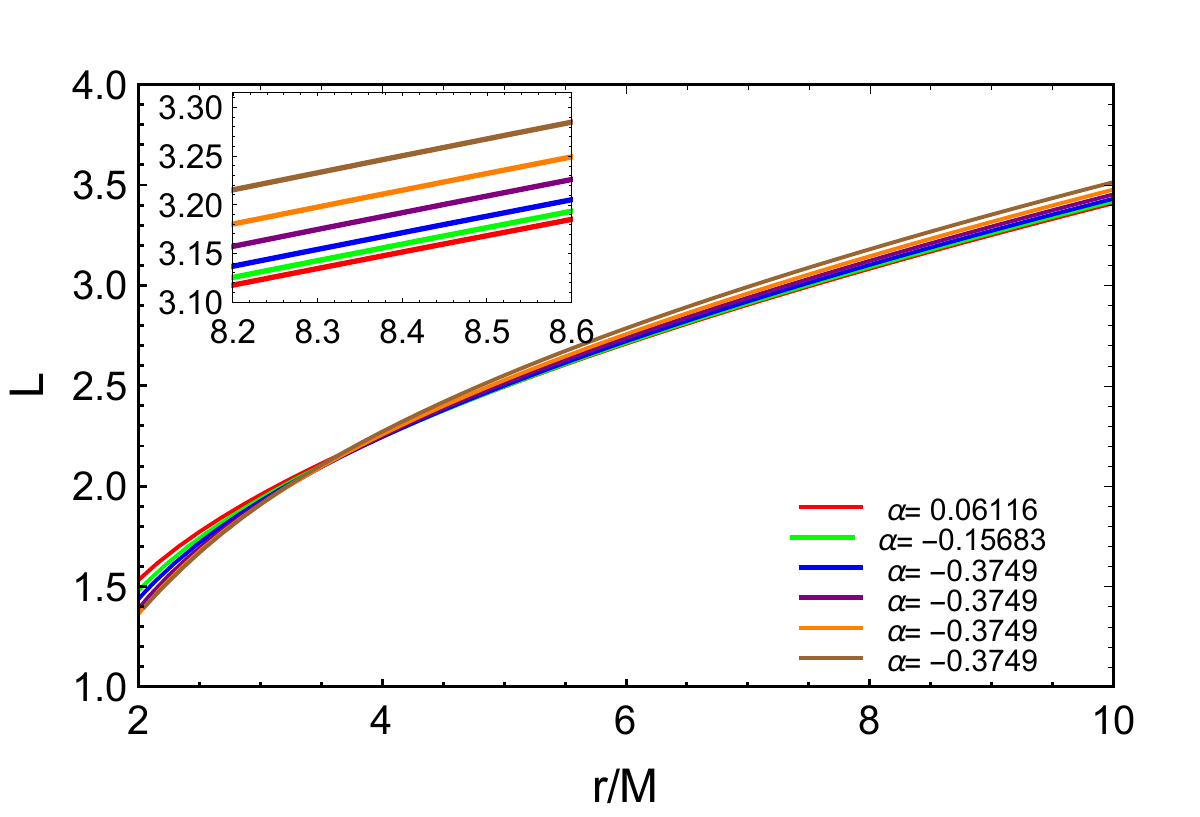}
    }
    \subfloat[$\alpha=-0.028883$]{
        \includegraphics[width=0.47\textwidth]{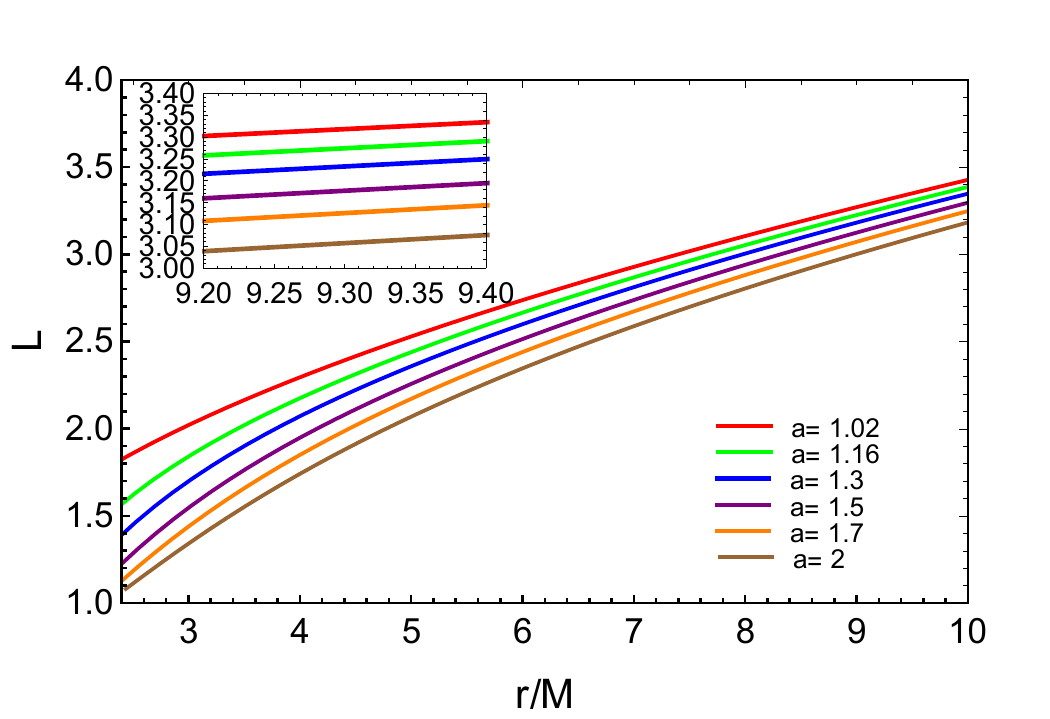}
    }
    \caption{\justifying Angular momentum \( L \) for circular orbits as a function of \( r/{M} \), plotted for different spin parameters \( a \) and MOG parameters \( \alpha \). Panels (a) and (b) show fixed spin with varying \( \alpha \), including the KNS case \( \alpha = 0 \), while panels (c)  vary spin with corresponding  \( \alpha \) values where different colors represent various spin values: red ($a = 1.02$), green ($a = 1.16$), blue ($a = 1.3$), purple ($a = 1.5$), orange ($a = 1.7$), and brown ($a = 2$). In (d) vary spin with fixed $\alpha$.}
    \label{Figmomentum}
\end{figure}
where \( \omega \) is the frame-dragging frequency of the spacetime, defined as
$\omega = -\frac{g_{t\phi}}{g_{\phi\phi}}$. Circular orbits with  \(r> r_{\text{ph}} \), for which the specific energy satisfies \( E < 1 \), are classified as bound orbits. The condition \( E = 1 \) corresponds to the radius \( r_{\text{mb}} \) of the marginally bound orbit, the smallest radius for which a test particle can remain on a bound circular path.
Fig.~\ref{Figenergy} shows that the effect of the MOG parameter $\alpha$ on the specific energy $E$ depends on the spin value  $a$. For small spin ($a=1.01$), increasing $\alpha$ slightly reduces the energy required for circular orbits, indicating that the MOG correction can enhance gravitational attraction in this regime. However, at higher spins, increasing $\alpha $ increases the energy, showing a repulsive effect. In \ref{Figenergy}(a) and \ref{Figenergy}(b), with fixed \( a = 1.01 \) and \( 1.38 \), increasing \( \alpha \) raises the energy required for circular motion, reflecting a repulsive effect of modified gravity. The case \( \alpha = 0 \) corresponds to the KNS baseline. \ref{Figenergy}(c) shows how jointly increasing \( a \) and \( \alpha \) modifies the energy profile, while   \ref{Figenergy}(d) isolates the effect of spin with fixed \( \alpha = -0.02883 \), confirming that higher spin reduces orbital energy via frame-dragging. In all cases, the energy flattens out at large distances, approaching the rest-mass energy as expected in asymptotically flat spacetimes.
Whereas, Fig.~\ref{Figmomentum} illustrates the behavior of the specific angular momentum \( L \) under similar conditions. For lower spin (\( a = 1.01 \)), increasing \( \alpha \) leads to a slight decrease in \( L \), reflecting a stronger effective gravitational binding. In contrast, for higher spin (\( a = 1.38 \)), the angular momentum increases with \( \alpha \), showing that the MOG correction weakens the gravitational attraction and demands higher angular momentum to sustain stable circular orbits.  circular orbits under varying spin \( a \) and MOG parameter \( \alpha \). In ~\ref{Figmomentum}(a) and ~\ref{Figmomentum}(b), increasing \( \alpha \) at fixed spin (\( a = 1.01 \) and \( 1.38 \)) raises the angular momentum required to maintain stable orbits. The \( \alpha = 0 \) case reflects the KNS baseline. ~\ref{Figmomentum}(c) reveals the joint effect of increasing both \( a \) and \( \alpha \), while~\ref{Figmomentum}(d) isolates spin effects at constant \( \alpha =- 0.02883 \). At large distances, all curves converge, indicating that the influence of \( \alpha \) and spin diminishes. 

The flux of radiant energy emitted from the accretion disk is given by the expression \cite{page1974disk}
\begin{equation}
F(r) = -\frac{\dot{M}_0}{4\pi \sqrt{-g}} \, \frac{d\Omega}{dr} \, \frac{1}{({E} - \Omega {L})^2} \int_{r_{\text{ms}}}^{r} ({E} - \Omega{L}) \frac{d{L}}{dr} \, dr,
\label{eq:flux}
\end{equation}
where \( \dot{M}_0 \) represents the mass accretion rate. When the disk is assumed to be in thermal equilibrium, the radiative flux should follow the Stefan–Boltzmann law. Furthermore, using conservation laws, this expression allows for calculating the energy flux across the disk region extending from the ISCO (\( r_{\text{ms}} \)) out to any specified radius \cite{page1974disk,rhie1991global}.
\noindent
 
\begin{figure}[!htb]
    \centering

   \subfloat[]{
        \includegraphics[width=0.47\textwidth]{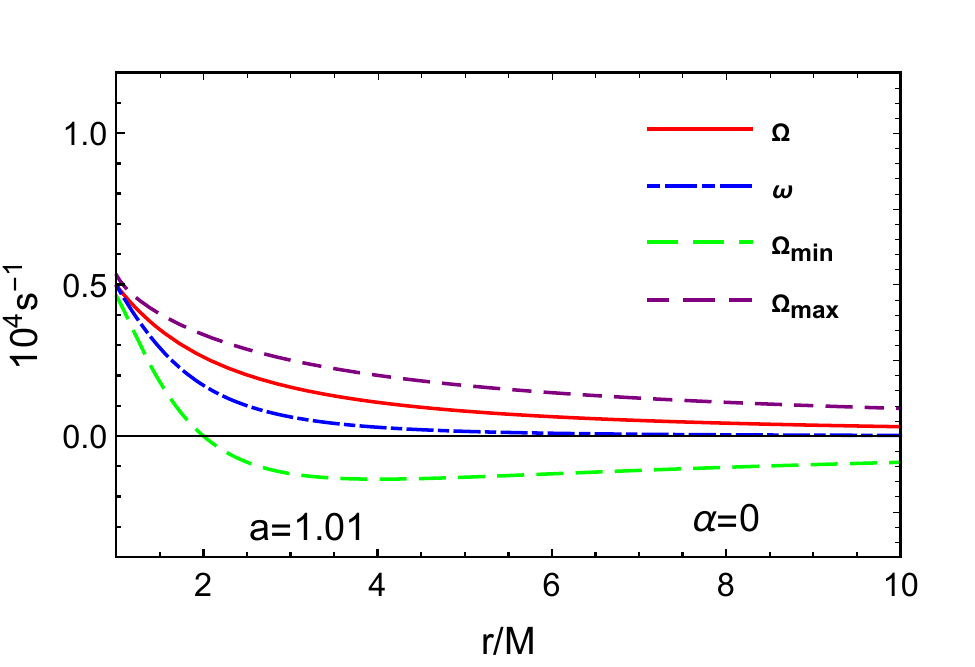}
    } 
    \subfloat[]{
        \includegraphics[width=0.47\textwidth]{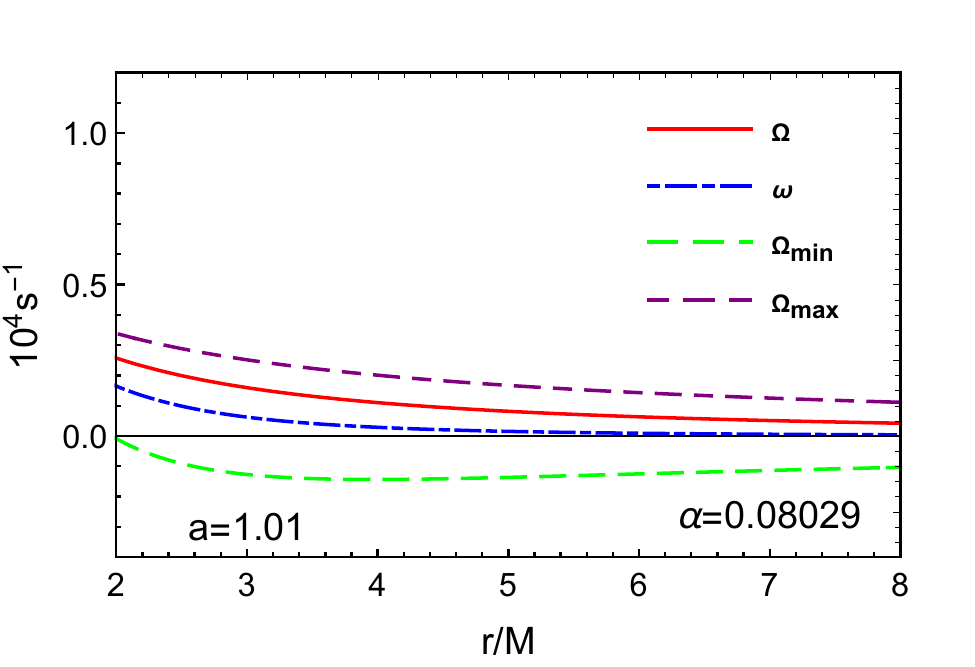}
    }\hfill
    \subfloat[]{
        \includegraphics[width=0.47\textwidth]{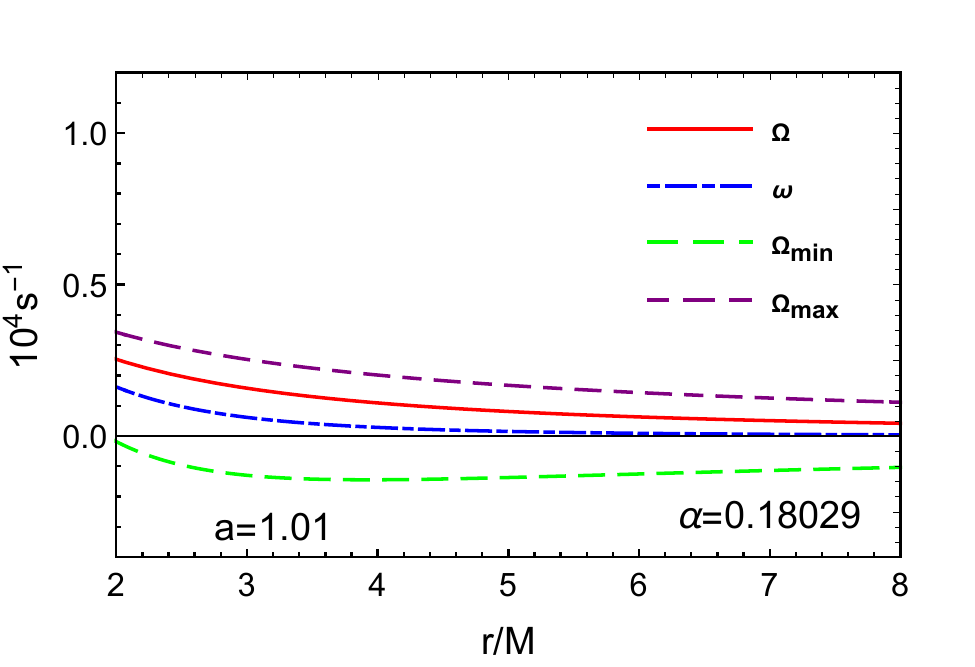}
    }
    \subfloat[]{
        \includegraphics[width=0.47\textwidth]{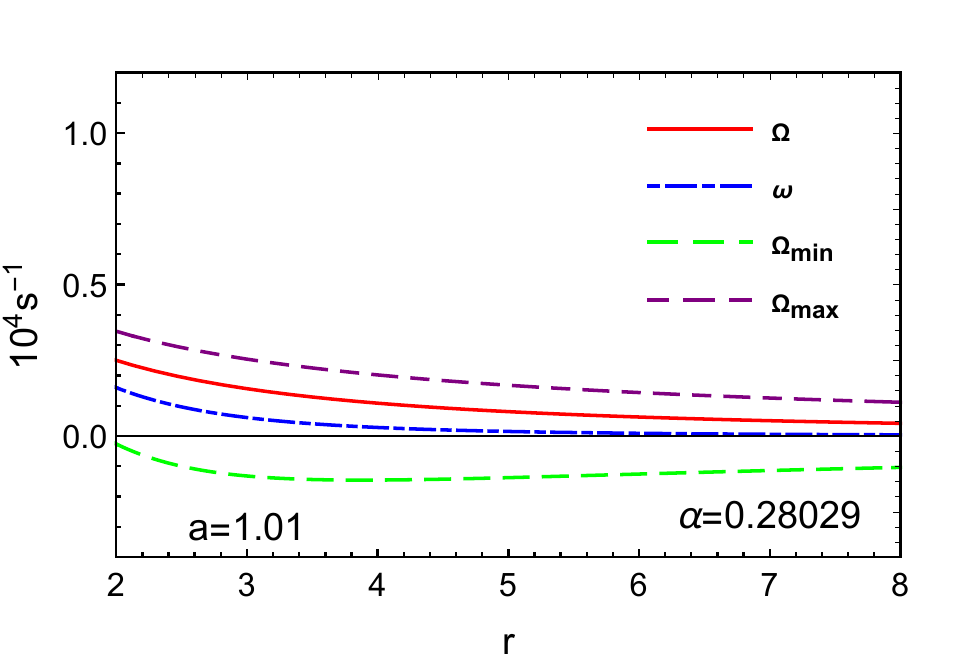}
    }\hfill
    \subfloat[]{
        \includegraphics[width=0.47\textwidth]{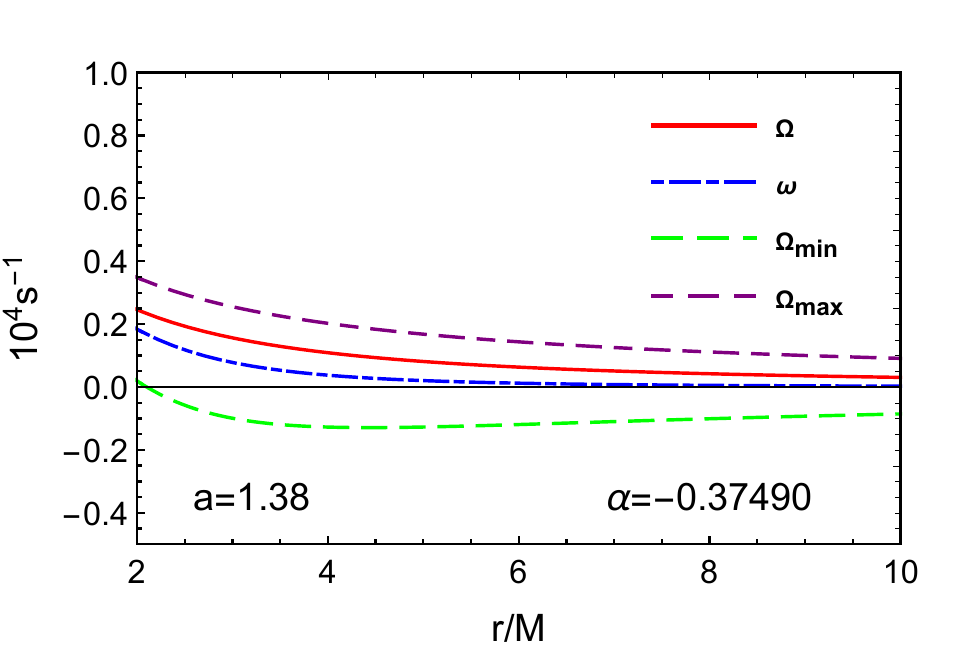}
    }
    \subfloat[]{
        \includegraphics[width=0.47\textwidth]{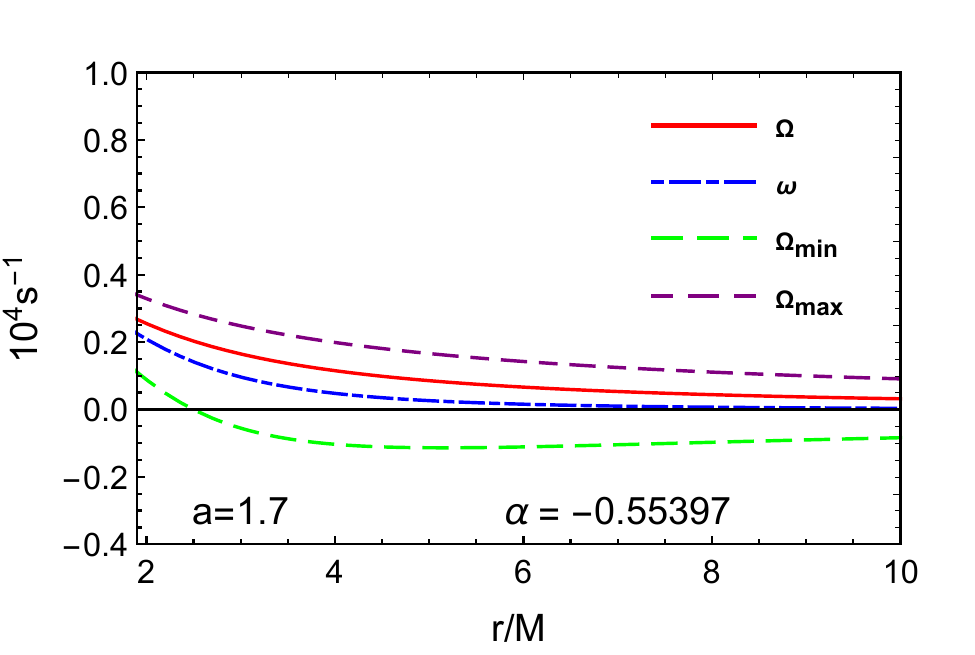}
    }
    \caption{\justifying The radial profiles of $\Omega$, $\Omega_{\text{min}}$, $\Omega_{\text{max}}$, and $\omega$ are shown for  KMNS with different spin parameters and corresponding values of $\alpha$ .
}
    \label{Figangulr1}
\end{figure}

In Fig.~\ref{Figangulr1}, we present the behavior of the angular velocity \( \Omega \), frame-dragging frequency \( \omega \), and the limiting angular velocities \( \Omega_{\min} \) and \( \Omega_{\max} \) for circular motion in the equatorial plane, for various spin values \( a \) and MOG parameter \( \alpha \). In the KNS case (\( \alpha = 0 \)), shown in panel \ref{Figangulr1}(a), frame-dragging is weaker, and the separation between \( \Omega_{\min} \) and \( \Omega_{\max} \) is narrower. As \( \alpha \) increases, \ref{Figangulr1}(b) to \ref{Figangulr1}(d) show that the causal range of allowed angular velocities tightens further, while \( \omega \) shifts upward due to enhanced frame-dragging. For higher spin values in \ref{Figangulr1}(e) and \ref{Figangulr1}(f), this effect becomes more significant, particularly near smaller radii, indicating a stronger rotational influence of the spacetime. As \( r \to \infty \), all curves converge, reflecting the asymptotic flatness of the geometry. At the limiting surface, \( \Omega = \Omega_{\min} = \Omega_{\max} = \omega \), indicating the synchronization of motion near the inner boundary for rapidly rotating configurations.

\begin{figure}[!htb]
    \centering

   \subfloat[$a=1.01$]{
        \includegraphics[width=0.47\textwidth]{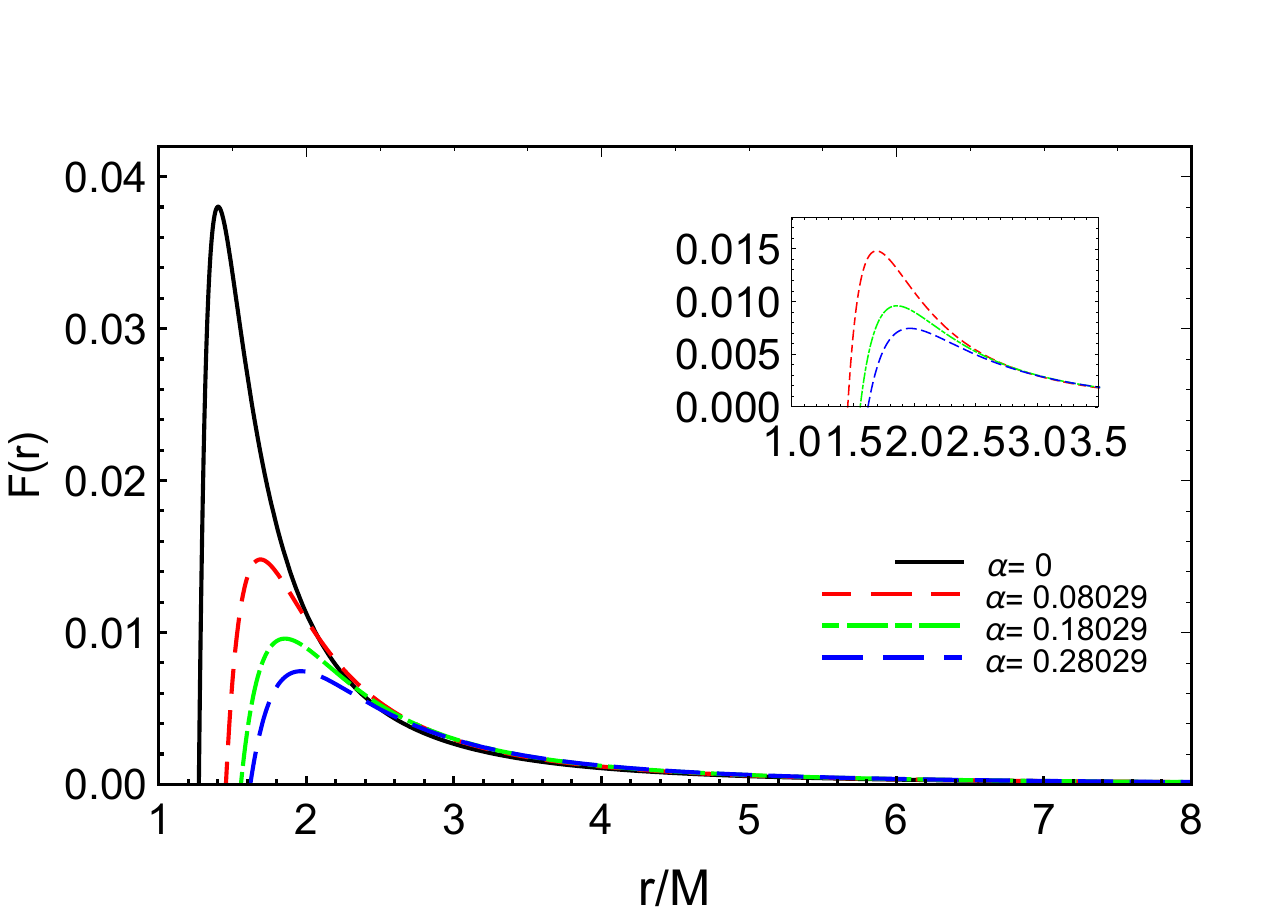}
    } 
    \subfloat[$a=1.38$]{
        \includegraphics[width=0.47\textwidth]{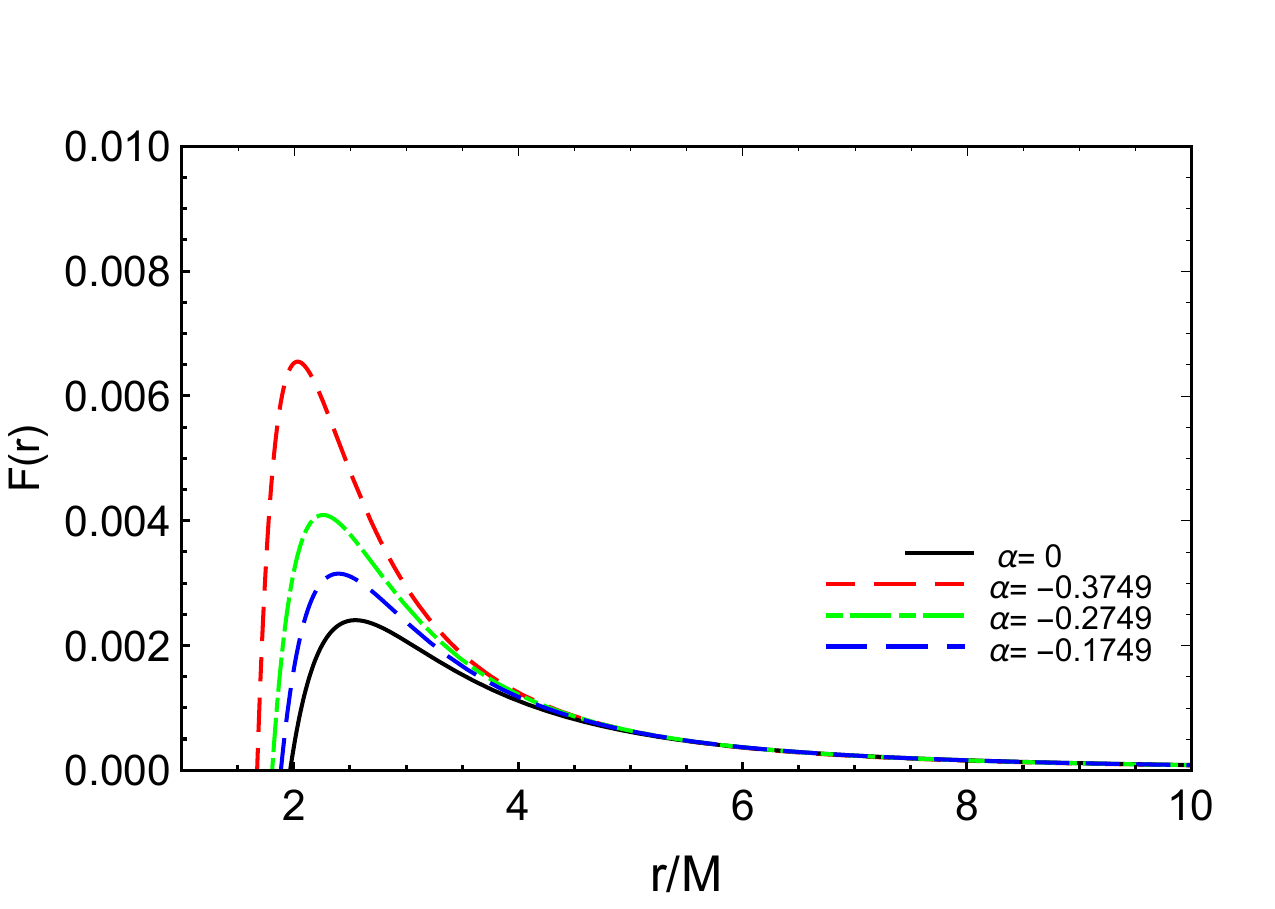}
    }\hfill
    \subfloat[$a=1.5$]{
        \includegraphics[width=0.47\textwidth]{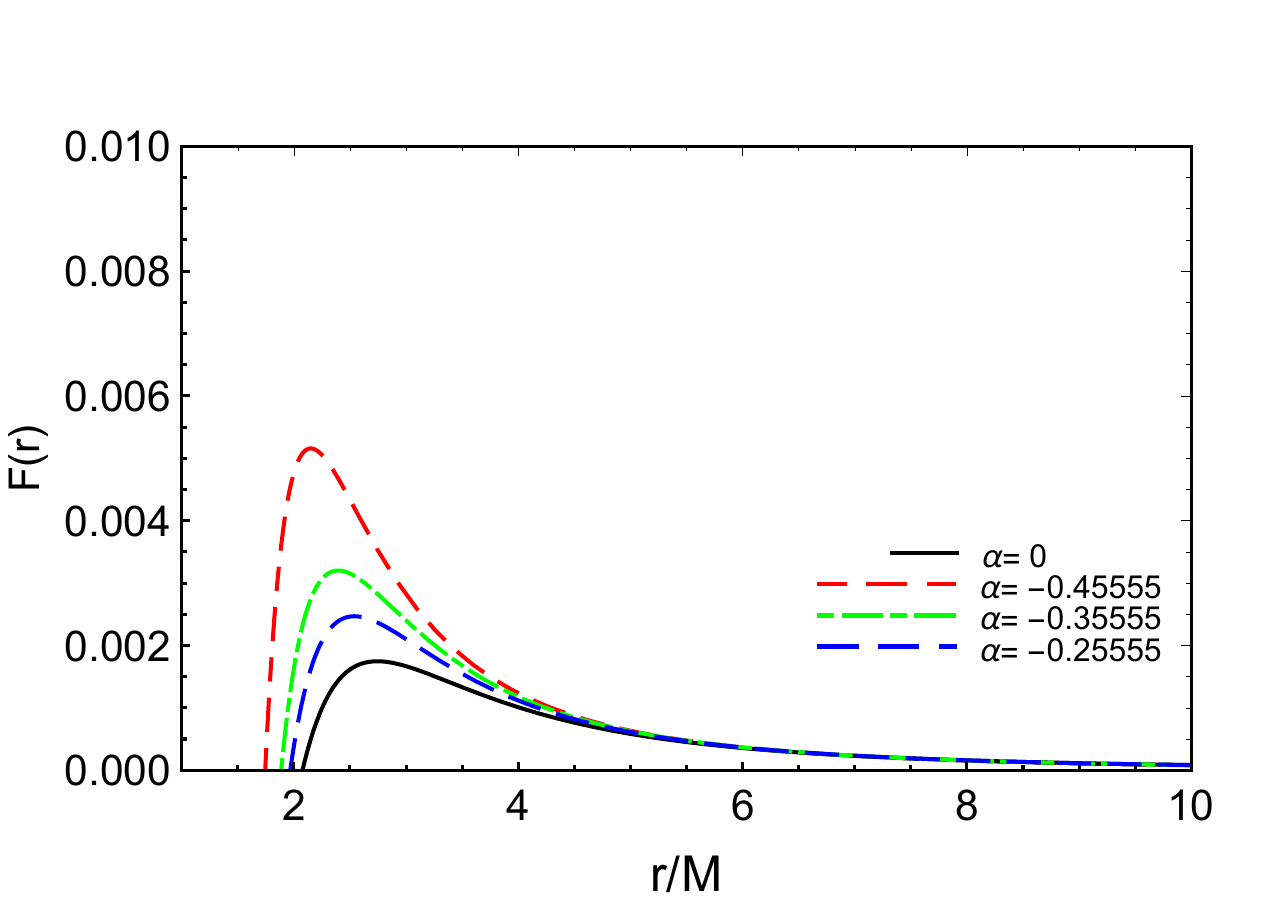}
    }
    \subfloat[$a=1.7$]{
        \includegraphics[width=0.47\textwidth]{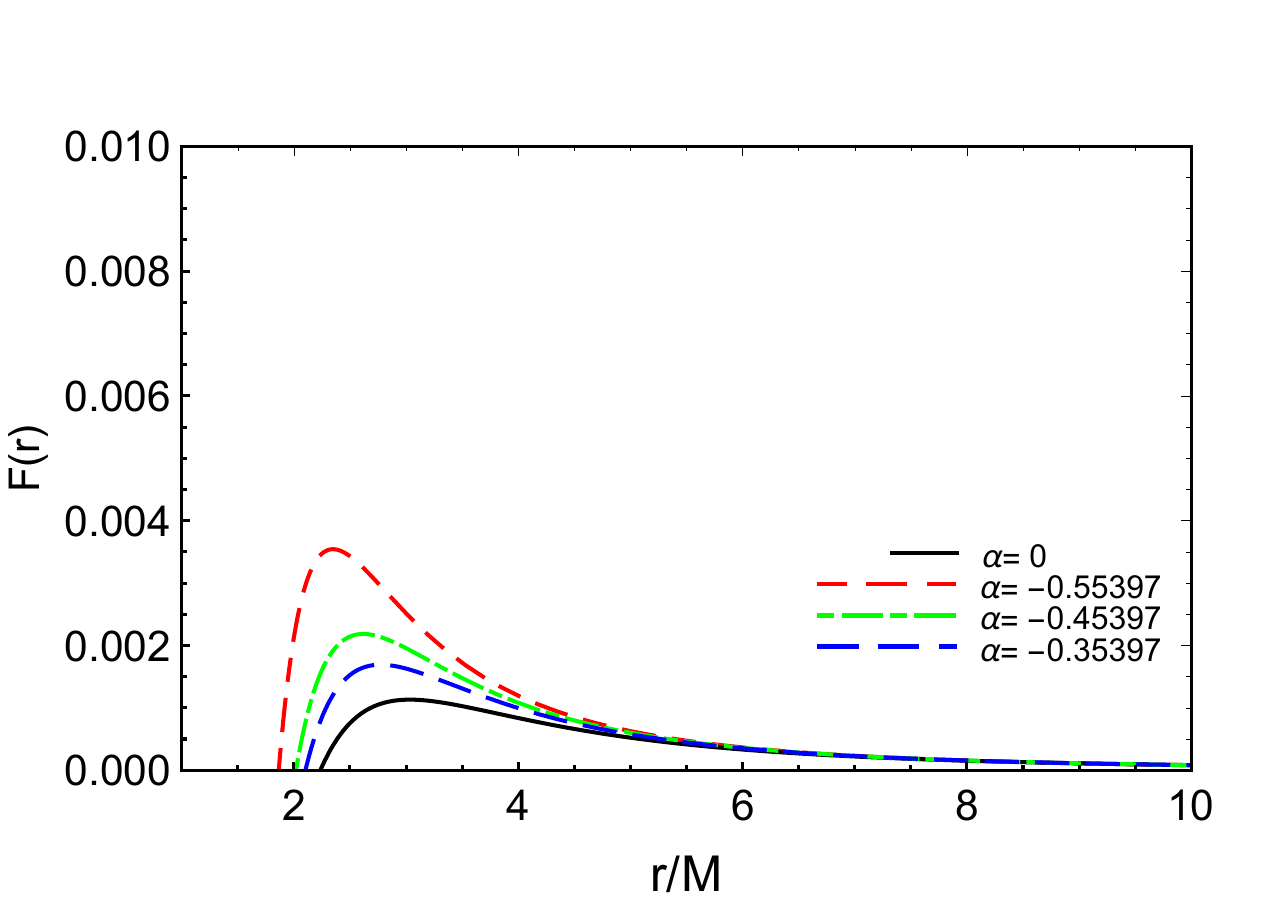}
    }\hfill
    \subfloat[$a=1.02,a=1.16, a=1.3, a=1.5, a=1.7, a=2$]{
        \includegraphics[width=0.47\textwidth]{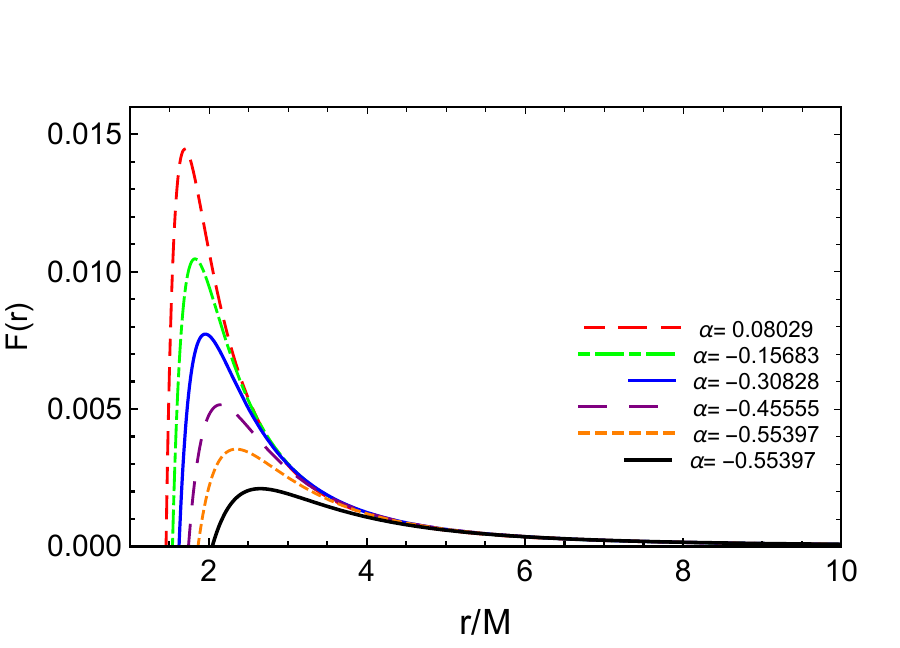}
    }
    \subfloat[$\alpha=-0.02883$]{
        \includegraphics[width=0.47\textwidth]{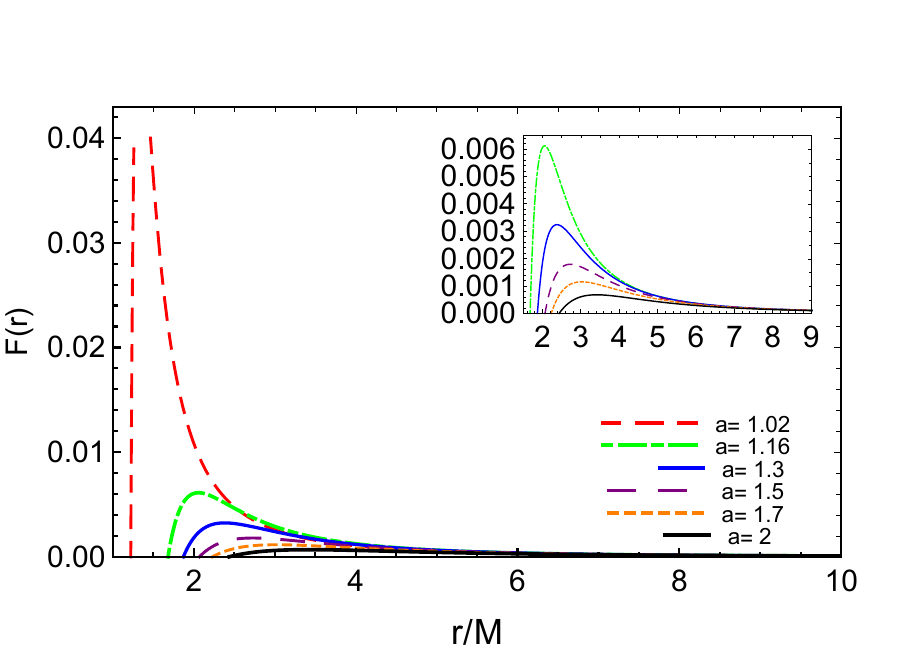}
    }
    \caption{\justifying The energy flux $F(r)$ radiated by the disk around KMNS and KNS ($\alpha = 0$), is computed for different values of the spin parameter $a$ and corresponding $\alpha$. In plot (e), different colors represent various spin values: red ($a = 1.02$), green ($a = 1.16$), blue ($a = 1.3$), purple ($a = 1.5$), orange ($a = 1.7$), and black ($a = 2$).
}
    \label{Figflux}
\end{figure}

The flux profiles in Fig.~\ref{Figflux} examine the effects of varying $\alpha$ and $a$ on the energy emission from thin disks around KMNS. In  \ref{Figflux}(a), for $a = 1.01$, increasing $\alpha$ enhances the flux and slightly shifts the peak inward, reflecting mild gravitational repulsion effects at low spin. In \ref{Figflux}(b), with $a = 1.38$, this enhancement becomes more pronounced, a higher $\alpha$ results in steeper and more compact flux profiles, indicating stronger MOG influence. \ref{Figflux}(c), for $a = 1.5$, shows a significant amplification of the peak flux with increasing $\alpha$, accompanied by a noticeable inward shift, as relativistic and scalar corrections dominate disk structure. In \ref{Figflux}(d), for high spin $a = 1.7$, the increase in flux with $\alpha$ persists but begins to saturate, suggesting a crossover where frame-dragging limits further radiative gain. \ref{Figflux}(e) explores varying $\alpha$ and $a$ simultaneously as both increase, the flux peak becomes sharper and more localized, demonstrating the strong synergy between spin and MOG corrections in enhancing disk luminosity. Lastly, \ref{Figflux}(f) fixes $\alpha$ and varies $a$, revealing that for each $\alpha$, increasing spin boosts the flux and shifts its maximum inward, though at high spin the profiles flatten, indicating saturation effects. Together, these results confirm that both $\alpha$ and $a$ play critical roles in shaping the flux, with higher values leading to brighter and more compact emission zones. The sharp rise in the flux near the inner disk edge reflects the contribution of the ISCO radius approaching the compact object as spin increases.
Across all panels, the flux $F(r)$ decreases rapidly with radius and asymptotically approaches zero, as expected in the outer regions of the disk where local energy dissipation becomes negligible. In some configurations, particularly at high spin and large values of $\alpha$, $F(r)$ develops negative regions near the ISCO. These unphysical values result from a sign change in the integrand of the flux expression, specifically due to the radial derivative of angular velocity ($\Omega_{,r}$). This behavior signals a breakdown of the standard thin-disk approximation, implying that alternative transport mechanisms, such as advection or turbulent convection, may dominate energy transfer in the innermost disk regions.
In the low-spin regime ($a = 1.02$), the KNS case with $\alpha = 0$ exhibits a higher peak flux than its Kerr MOG counterparts. This suggests that, at small spin, the repulsive gravitational correction introduced by a non-zero $\alpha$ suppresses inner-disk efficiency and reduces radiative output. However, as the spin increases, the flux from KMNSs surpasses that of the Kerr case, indicating a crossover. In this high-spin regime, relativistic frame-dragging and scalar field effects dominate the dynamics, enhancing disk luminosity under modified gravity. This crossover reflects the interplay between spin-enhanced energy extraction and $\alpha$-induced gravitational repulsion in shaping the thermal structure of the disk.
\begin{figure}[!htb]
    \centering

   \subfloat[$a=1.01$]{
        \includegraphics[width=0.47\textwidth]{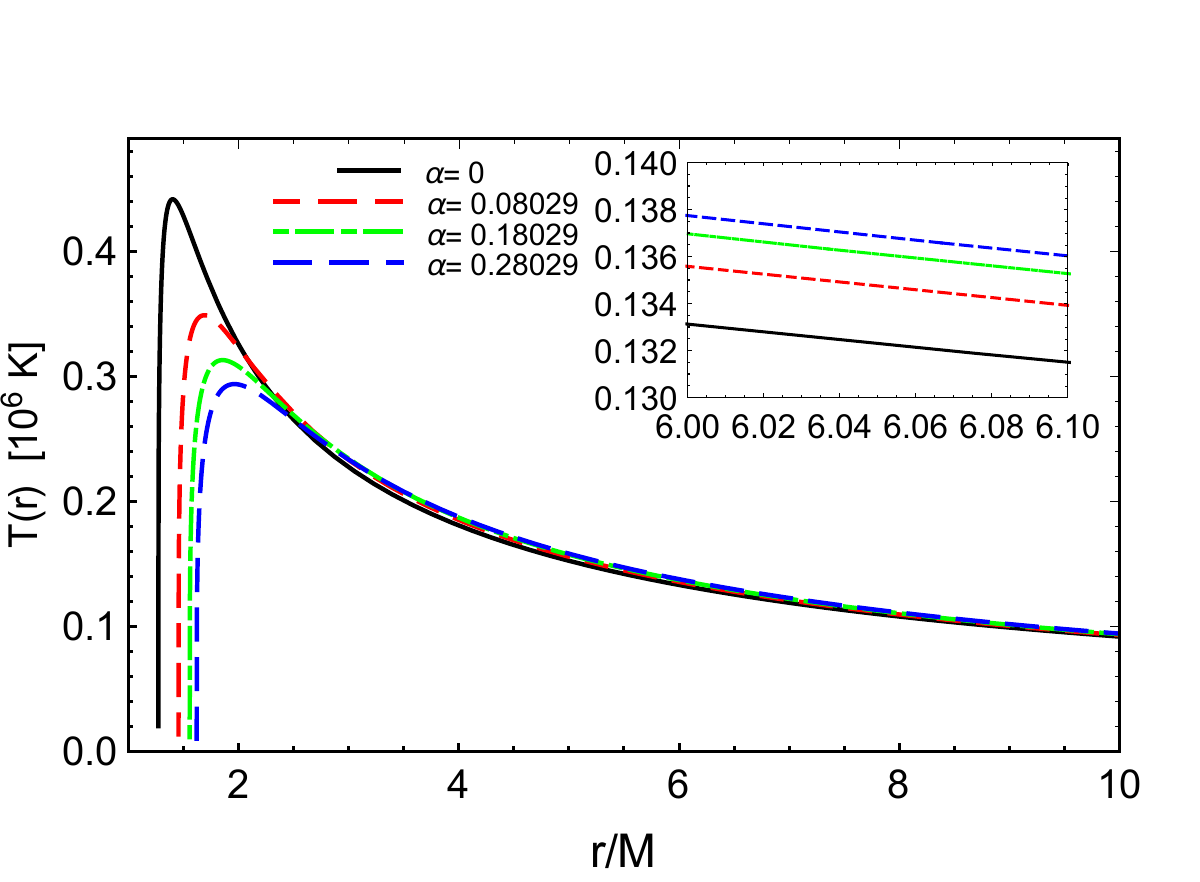}
    } 
    \subfloat[$a=1.38$]{
        \includegraphics[width=0.47\textwidth]{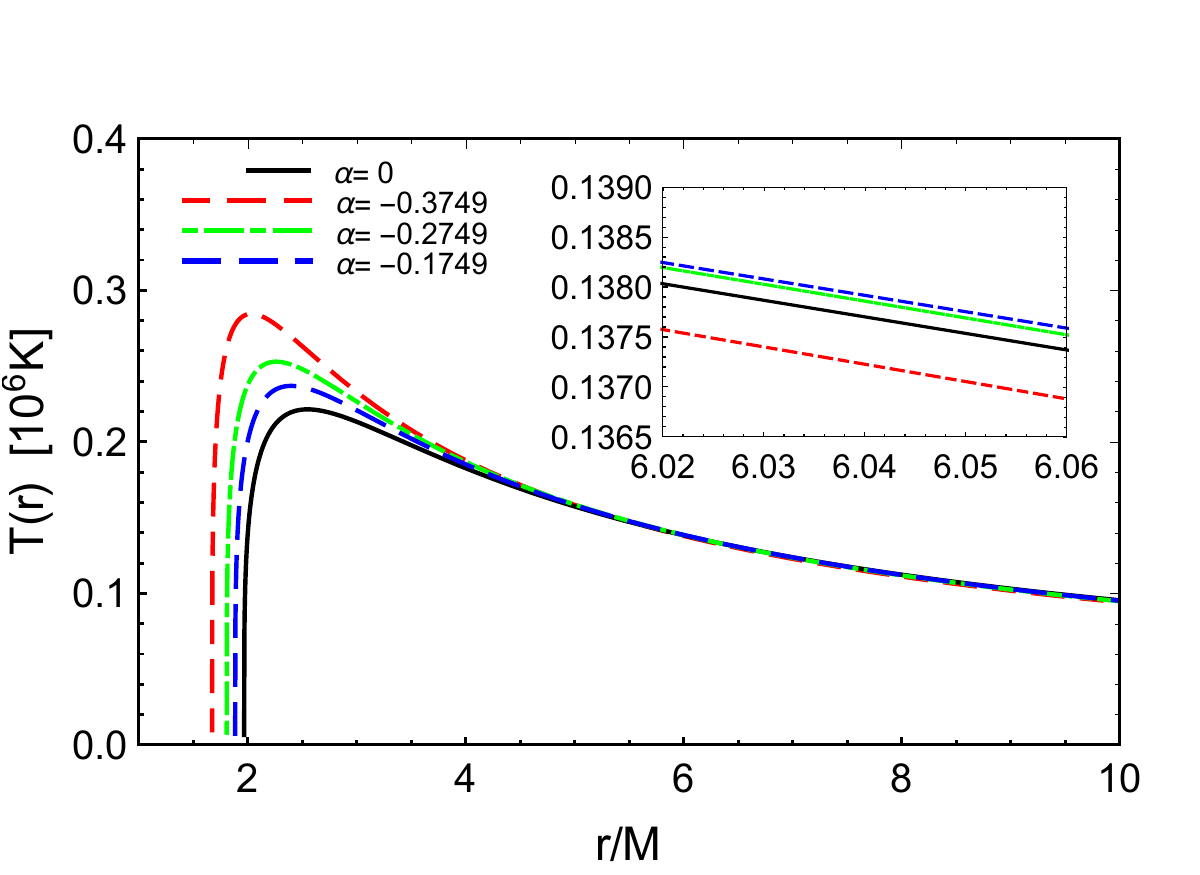}
    }\hfill
    \subfloat[$a=1.5$]{
        \includegraphics[width=0.47\textwidth]{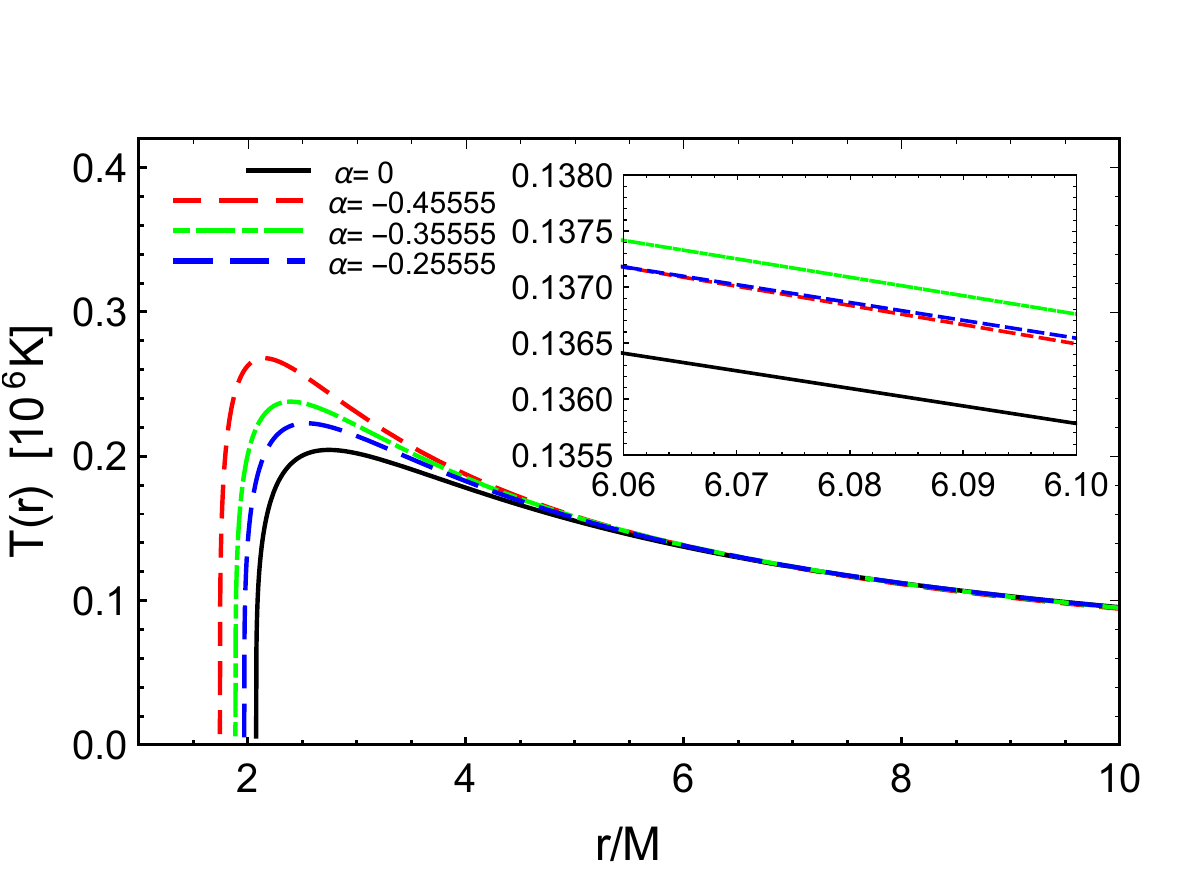}
    }
    \subfloat[$a=1.7$]{
        \includegraphics[width=0.47\textwidth]{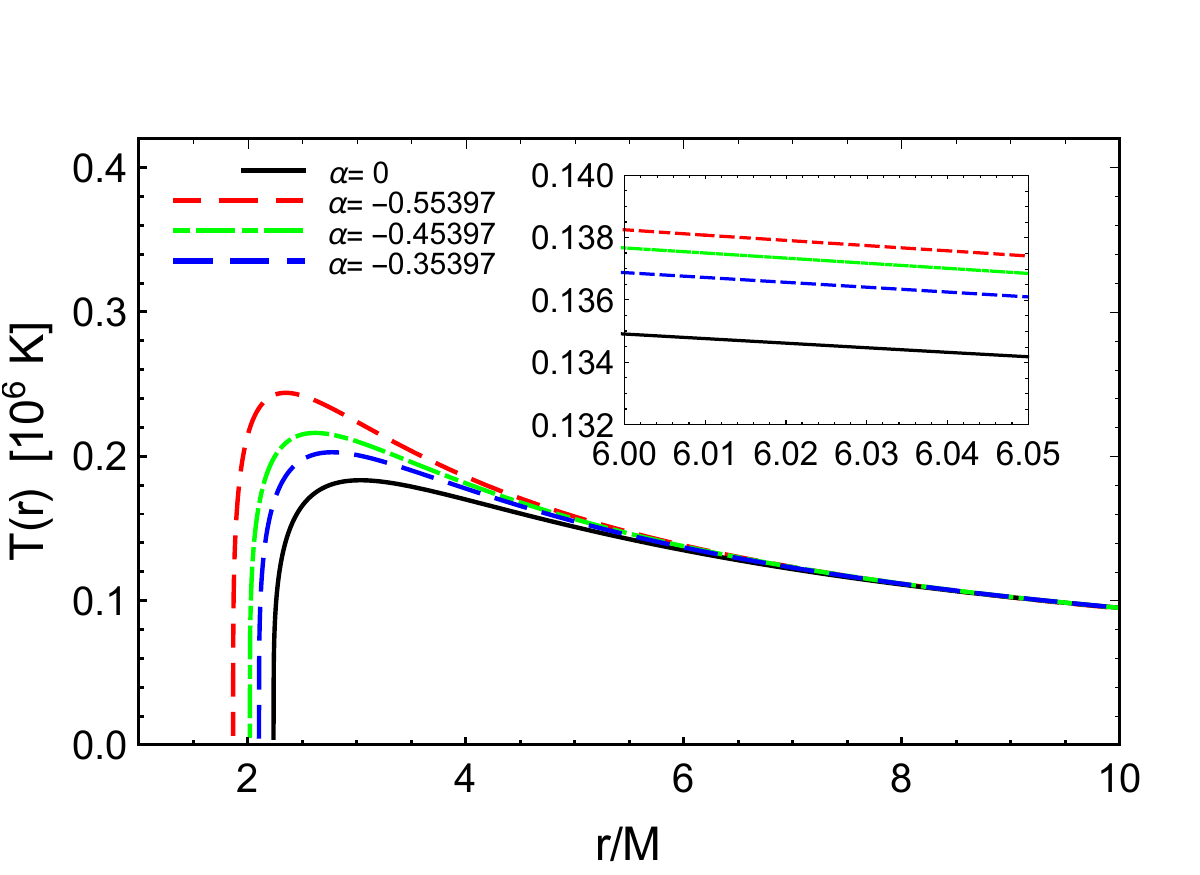}
    }\hfill
    \subfloat[$a=1.02,a=1.16, a=1.3, a=1.5, a=1.7, a=2$]{
        \includegraphics[width=0.47\textwidth]{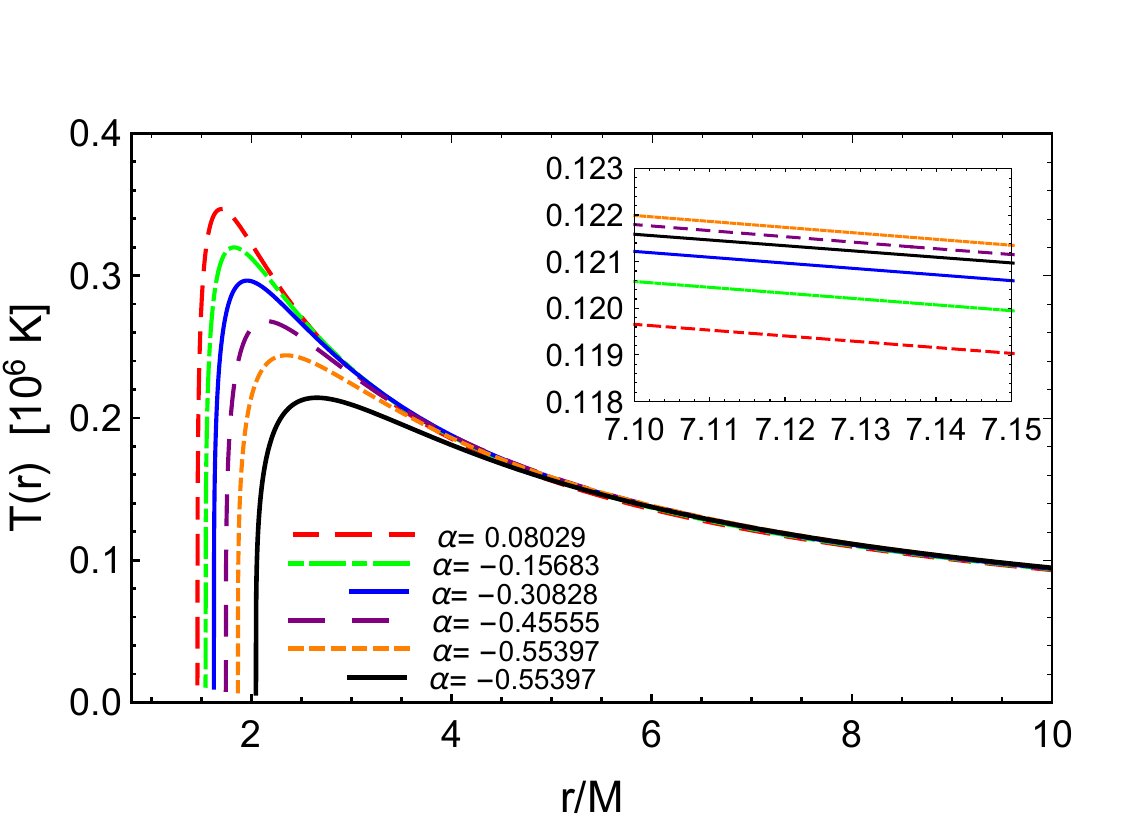}
    }
    \subfloat[$\alpha=-0.02883$]{
        \includegraphics[width=0.47\textwidth]{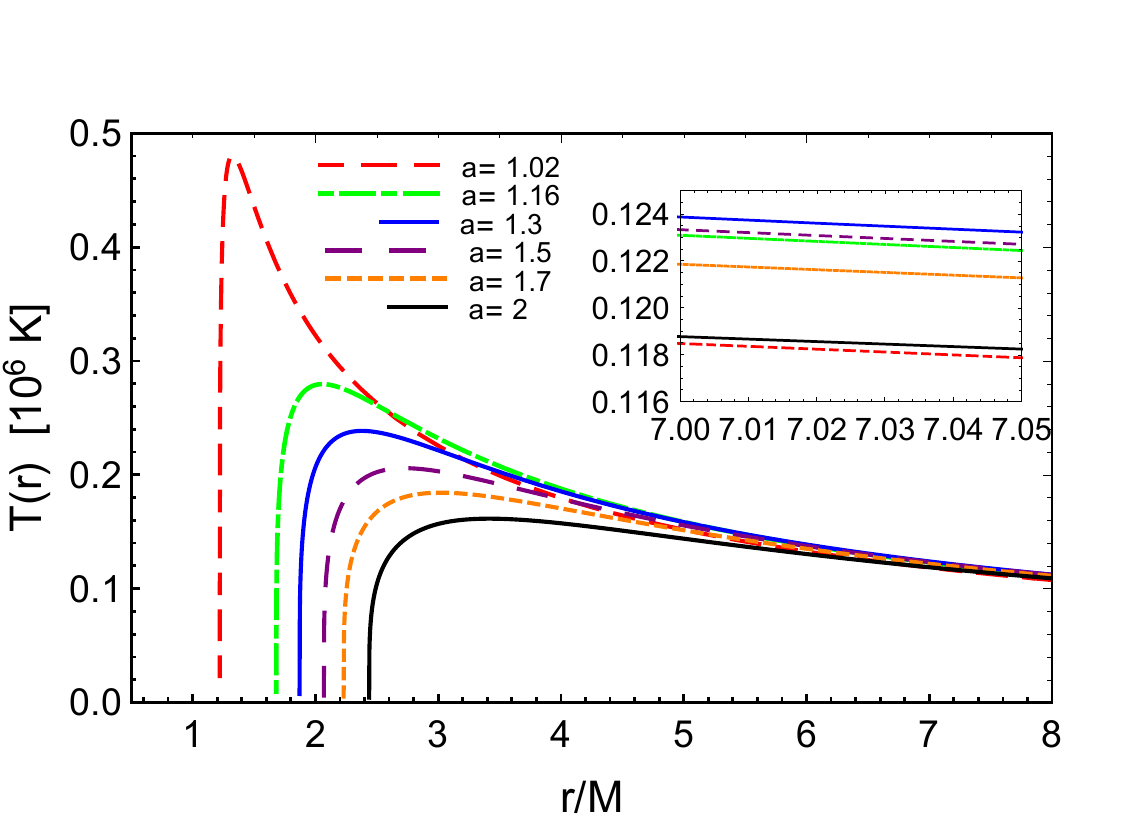}
    }
    \caption{\justifying The energy flux $F(r)$ radiated by the disk around KMNS and KNS ($\alpha = 0$), is computed for different values of the spin parameter $a$ and corresponding $\alpha$. In plot (e), different colors represent various spin values: red ($a = 1.02$), green ($a = 1.16$), blue ($a = 1.3$), purple ($a = 1.5$), orange ($a = 1.7$), and black ($a = 2$).
}
    \label{Figtemp}
\end{figure}
Overall, the flux profiles reveal that both $a$ and $\alpha$ strongly influence the thermal properties of the accretion disk. While $\alpha$ enhances luminosity and shifts the emission region inward by allowing accretion to proceed closer to the singularity, $a$ compresses the flux peak further inward and amplifies its magnitude. These combined effects significantly alter the disk’s radiative characteristics, particularly in high-spin KMNS configurations, and may provide distinct observational signatures relevant for distinguishing between classical and modified gravity scenarios.
It should also be noted that thermodynamic equilibrium is a fundamental assumption in the model describing a steady-state accretion disk. As a result, the radiation emitted from the disk surface is assumed to follow a perfect blackbody spectrum \cite{2017PhRvD..95j4047P}. This assumption allows the thermal emission to be related to the local energy flux through the Stefan–Boltzmann law $F(r) = \sigma_{\text{SB}} T^4(r)$, where $\sigma_{\text{SB}}$ is the Stefan–Boltzmann constant. Therefore, under this framework, one can directly determine the effective temperature profile $T(r)$ of the accretion disk from the energy flux $F(r)$.
 Fig.~\ref{Figtemp} reflects the thermal response of the disk to the flux distributions, with the Stefan–Boltzmann constant set to unity. In \ref{Figtemp}(a), for low spin $a = 1.01$, the KNS case ($\alpha = 0$) yields the highest peak temperature, while increasing $\alpha$ suppresses the peak and shifts the emission zone outward, indicating reduced inner-disk efficiency due to MOG-induced gravitational repulsion. In \ref{Figtemp}(b), at moderate spin $a = 1.38$, the trend reverses, the peak temperature increases with $\alpha$ and moves inward, reflecting enhanced energy extraction facilitated by MOG corrections. 
\begin{figure}[!htb]
    \centering

   \subfloat[$a=1.01$]{
        \includegraphics[width=0.47\textwidth]{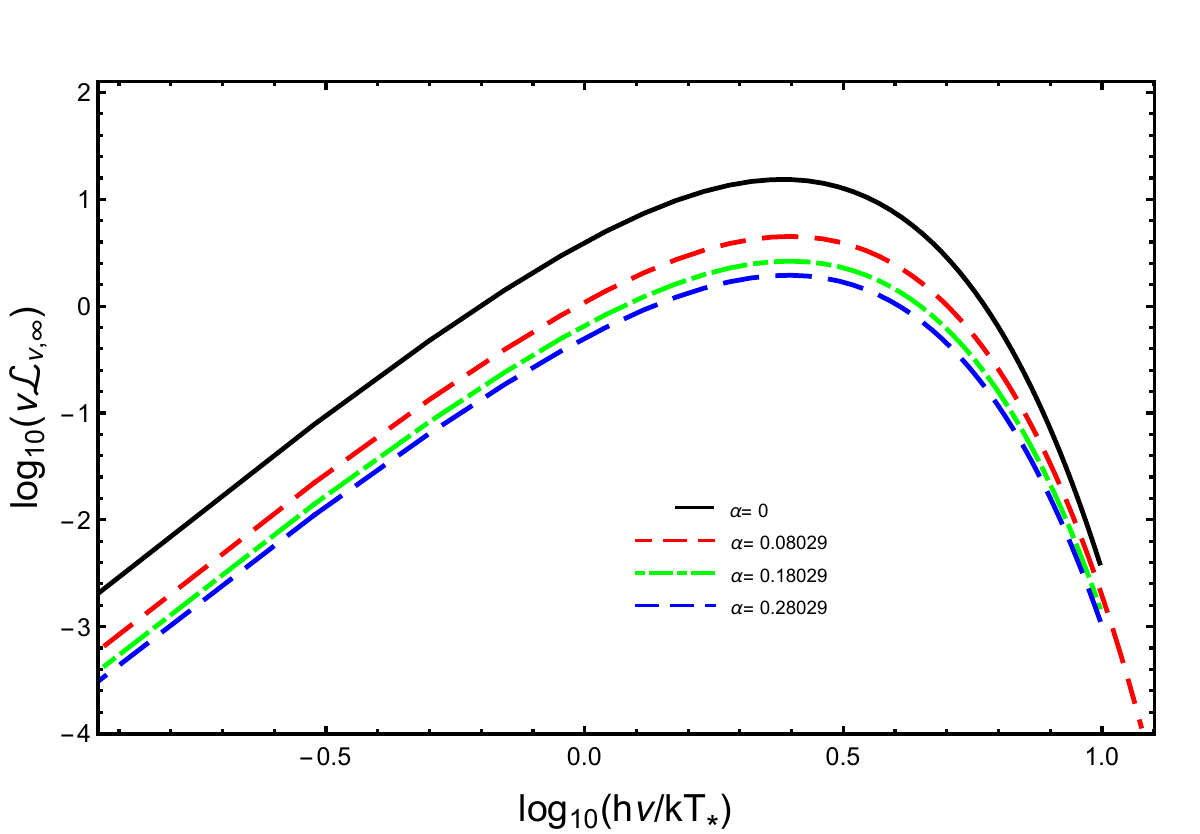}
    } 
    \subfloat[$a=1.38$]{
        \includegraphics[width=0.47\textwidth]{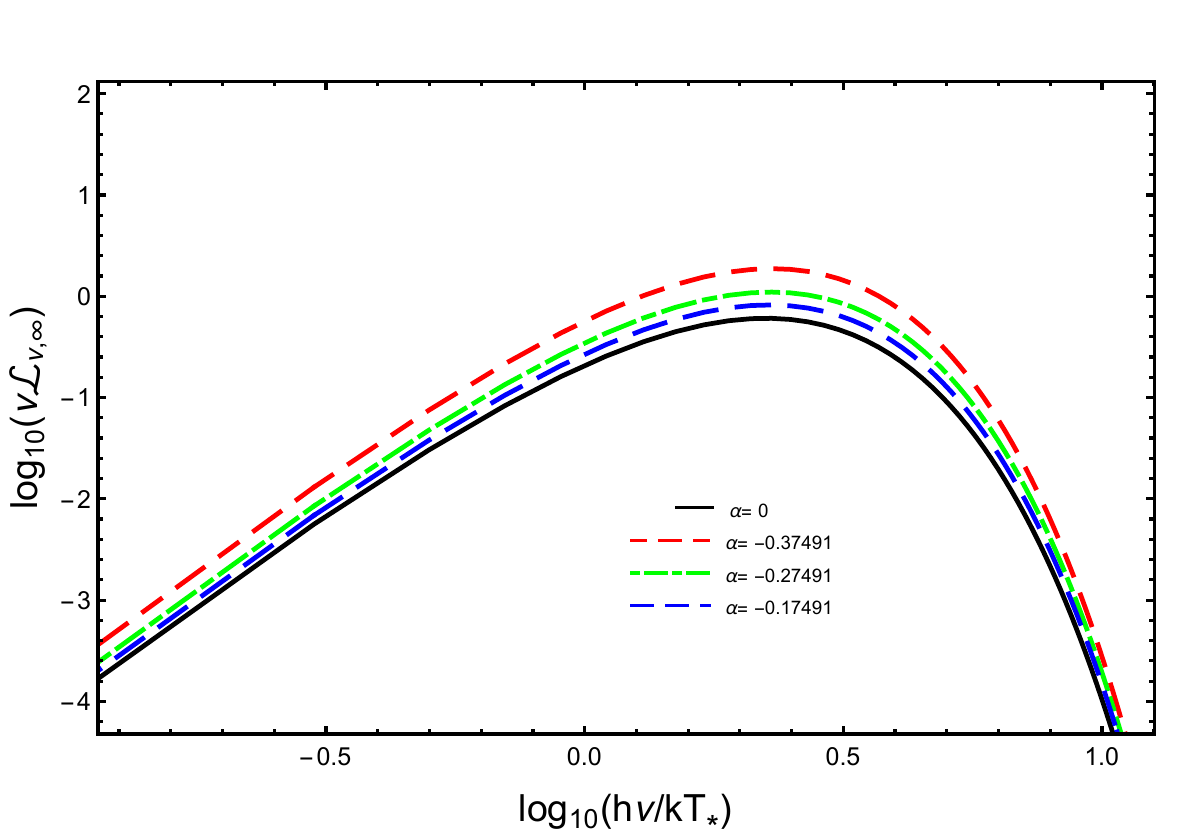}
    }\hfill
    \subfloat[$a=1.5$]{
        \includegraphics[width=0.47\textwidth]{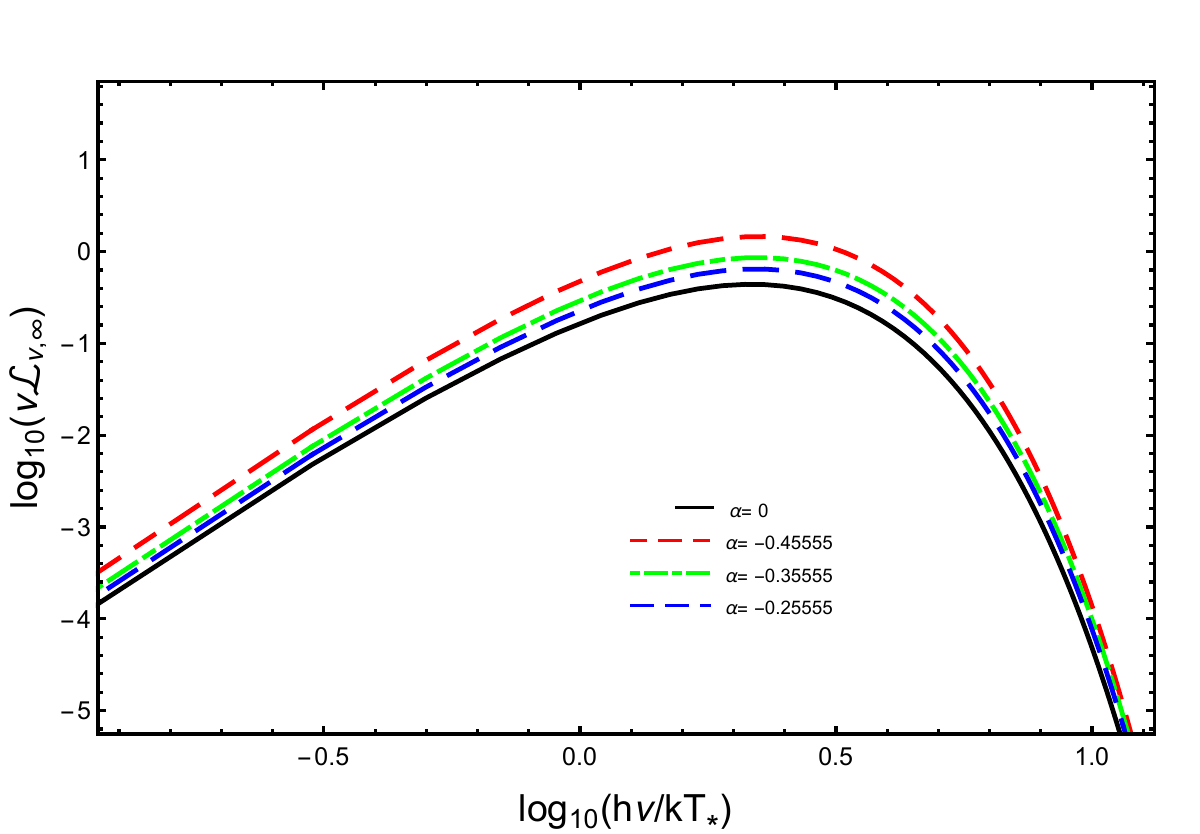}
    }
    \subfloat[$a=1.7$]{
        \includegraphics[width=0.47\textwidth]{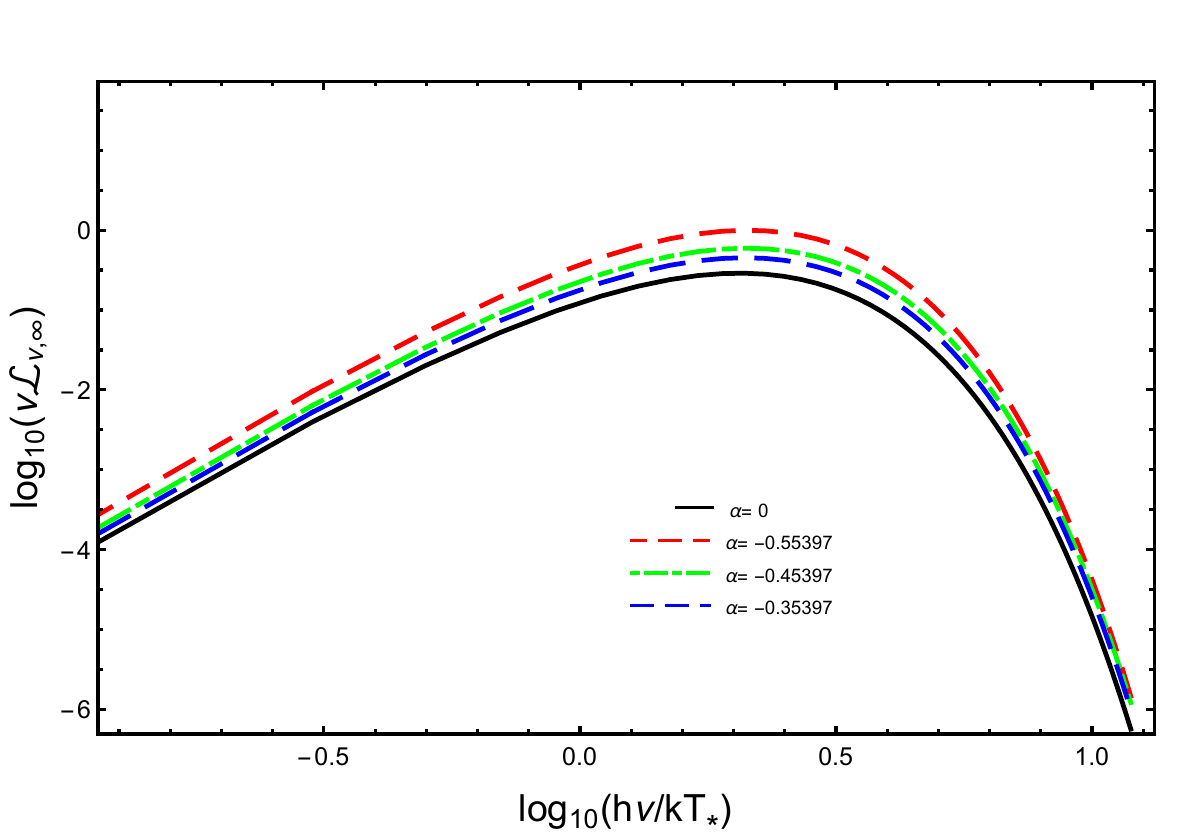}
    }\hfill
    \subfloat[$a=1.02,a=1.16, a=1.3, a=1.5, a=1.7, a=2$]{
        \includegraphics[width=0.47\textwidth]{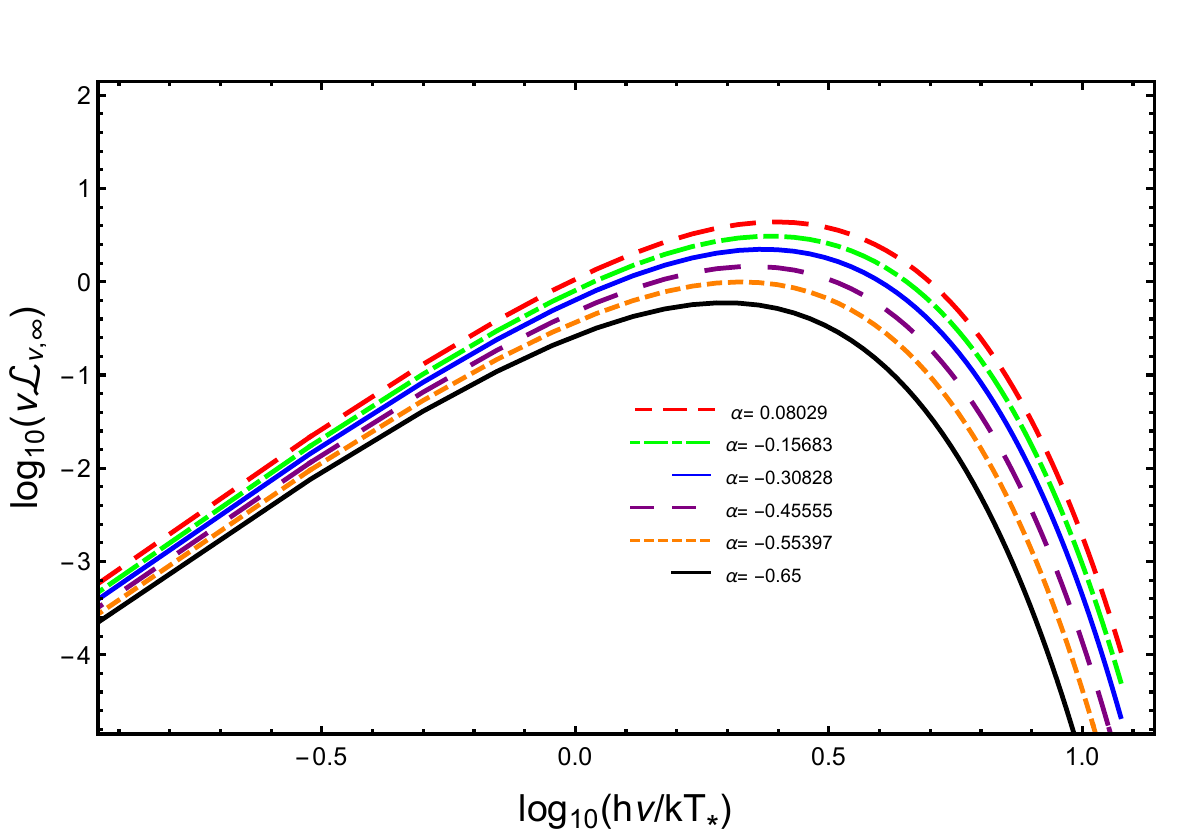}
    }
    \subfloat[$\alpha=-0.02883$]{
        \includegraphics[width=0.47\textwidth]{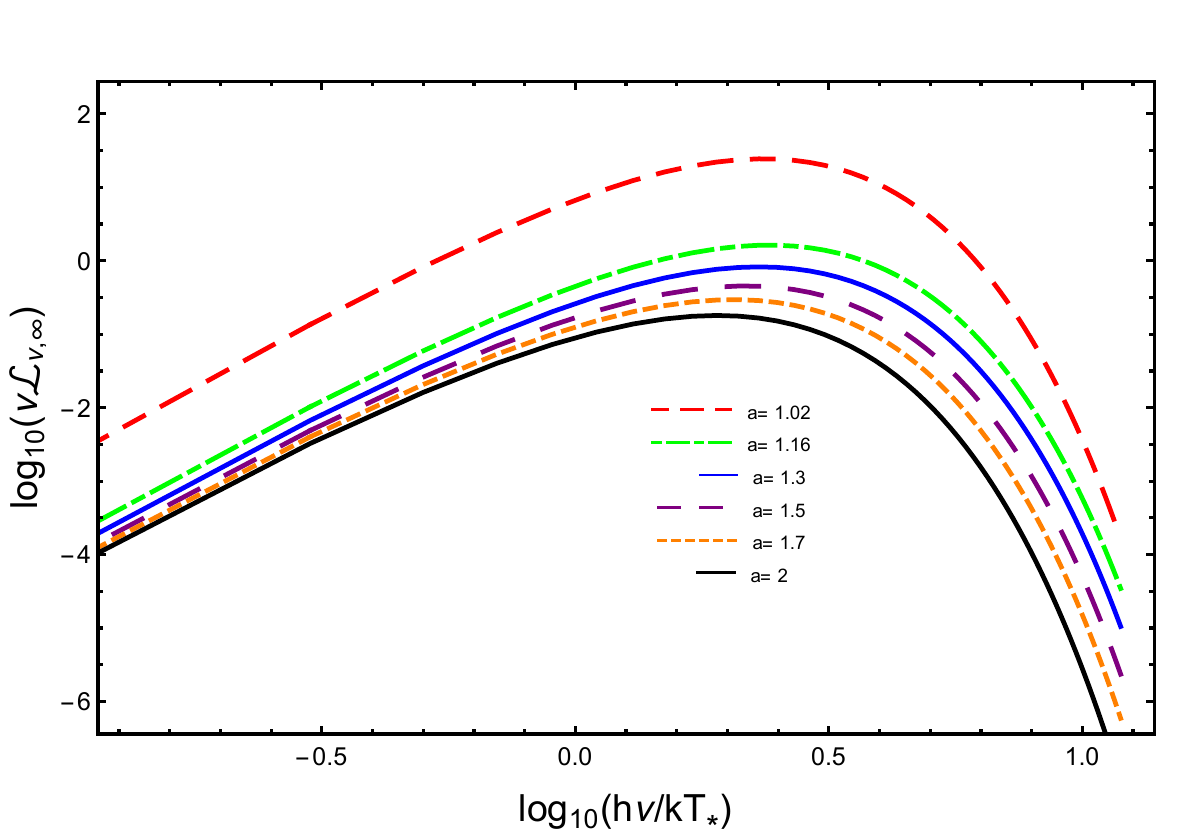}
    }
    \caption{The spectral luminosity as a function of radiation frequency around KMNS and KNS ($\alpha = 0$), is computed for different values of the spin parameter $a$ and corresponding $\alpha$. In plot (e), different colors represent various spin values: red ($a = 1.02$), green ($a = 1.16$), blue ($a = 1.3$), purple ($a = 1.5$), orange ($a = 1.7$), and black ($a = 2$).
}
    \label{Figspect}
\end{figure}
Moreover, in practice, we observe the spectrum of light emitted from the accretion disk as a function of frequency. Therefore, it is useful to consider the spectral luminosity distribution as measured at infinity, denoted by \( L_{\nu,\infty} \).
Assuming black body radiation from the disk, the observed luminosity is given by \cite{Boshkayev2020kle}
\begin{equation}
\nu \mathcal{L}_{\nu,\infty} = \frac{60}{\pi^3} \int_{r_{\text{ISCO}}}^{\infty} \, \frac{ \sqrt{-g} E}{\mathcal{M}^2} \frac{\left(u^t y\right)^4 }{\left[ \exp\left( \frac{u^t y}{F^{1/4}} \right)\right] - 1 } dr,
\label{eq:spectral_luminosity}
\end{equation}
where \( u^t \) is the contravariant time component of the four-velocity, defined as \cite{Kurmanov2025acceration}
\begin{equation}
u^t(r) = \frac{1}{\sqrt{ -g_{tt} - 2\Omega g_{t\phi} - \Omega^2 g_{\phi\phi} }},
\label{eq:ut_definition}
\end{equation}
and \( y = \frac{h \nu}{k T_*} \), with \( h \) being Planck’s constant, \( \nu \) the radiation frequency, \( k \) Boltzmann’s constant, and \( T_* \) the characteristic temperature defined by the Stefan-Boltzmann law \(
\sigma_{SB} T_*^4 = \frac{M_0}{4 \pi \mathcal{M}^2}\) \cite{Kurmanov2025acceration}.
For higher spin values in \ref{Figtemp}(c) and (d), with $a = 1.5$ and $a = 1.7$, increasing $\alpha$ sharply elevates the peak temperature and compresses the radiating region toward the singularity, although saturation effects emerge at very high spin. \ref{Figtemp}(e), where $a$ and $\alpha$ increase simultaneously, shows a pronounced rise in temperature and inward peak shift, highlighting the synergistic role of spin and scalar charge. Finally, \ref{Figtemp}(f), with fixed $\alpha$ and varying spin, confirms that increasing $a$ enhances and localizes the thermal output. Overall, the temperature profiles indicate that both $a$ and $\alpha$ significantly shape the thermal structure of the disk, with high-spin KMNS configurations producing hotter and more compact emission zones than their Kerr counterparts. The spectral luminosity distribution as a function of radiation frequency is presented in Fig.~\ref{Figspect}. It is evident that all key characteristics of the accretion disk are more sensitive to the parameter \( \alpha \) than to the spin parameter \( a \), with the mimicking effect being apparent across all quantities.
\section{Discussion}\label{secconc}

In this work, we have shown that the shadow of KMNSs defined as the projection of unstable spherical photon orbits at infinity, can appear closed, open, or even vanish entirely. Through an analytical study, we identified how the shadow topology changes with the spin parameter \( a \), modified parameter $\alpha$ and observer inclination angle \( i \). 
 We also examined the causality structure of KMNSs and found that the emergence of CTCs is significantly influenced by both the spin parameter and the parameter \( \alpha \). Our analysis shows that higher spin values, in conjunction with non-zero \( \alpha \), lead to more extensive regions where CTCs form, implying that causality violations are more widespread in the KMNS geometry than in the standard KNS case. While both solutions exhibit a shrinking CTC region as \( \alpha \) decreases, the KMNS consistently supports a broader domain of causality violation. This distinct difference underscores the profound impact of modified gravity on spacetime structure and further highlights how observational and theoretical studies of such features could serve as critical tests for alternative gravity theories.

 We further analyzed the influence of the \( \alpha \) on photon motion in Kerr MOG spacetime, keeping the spin fixed. Our findings show that as \( \alpha \) increases, the region forbidden to photon trajectories contracts in both radial and angular directions. This contraction implies a reduction in the effective potential barrier, allowing a wider range of photon paths to escape the vicinity of the singularity. As a result, the parameter \( \alpha \) significantly alters the photon orbit structure, directly impacting the morphology of the shadow seen by a distant observer. 
We have also explored the properties of accretion disks that may form around naked singularities, hypothetical objects predicted by general relativity that lack an event horizon.  This geometry smoothly reduces to the standard Kerr metric as the parameter \( \alpha \) approaches zero. 

 We have also investigated the behavior of angular velocities and energy flux profiles associated with circular motion and thin disk accretion in KMNS spacetimes. Our analysis shows that increasing the MOG parameter \( \alpha \) enhances frame-dragging effects and narrows the causal range of allowed angular velocities, particularly at higher spin values and smaller radii. This reflects a stronger rotational influence of the spacetime near the compact object, while all dynamical quantities asymptotically converge at large distances, consistent with flat spacetime behavior. In parallel, the energy flux from accretion disks is found to increase with both \( \alpha \) and spin \( a \), with higher values leading to more compact and intense emission profiles. At moderate spin, the increase is gradual, but at higher spin and deformation, the flux grows sharply and peaks closer to the singularity. This behavior signals a strong interplay between relativistic rotation and scalar field corrections, with the flux eventually saturating at extreme spins, suggesting a limit to radiative efficiency. These findings highlight the critical roles of both modified gravity and rotational dynamics in shaping observable properties of accretion disks around naked singularities.

\noindent The thermal properties of accretion disks are also affected by the spin parameter \( a \) and the MOG parameter \( \alpha \). For low-spin configurations, the maximum disk temperature is highest when \( \alpha = 0 \) (KNS case), while increasing \( \alpha \) reduces the peak temperature and shifts the emission zone outward. This behavior suggests a decrease in inner-disk efficiency due to gravitational repulsion introduced by modified gravity. At moderate spin, the pattern reverses, the peak temperature increases with larger \( \alpha \), and the radiating region moves inward, indicating more efficient energy extraction enabled by MOG effects. In high-spin regimes, both the temperature and localization of the emission region increase significantly with \( \alpha \), although saturation effects begin to limit further enhancement at extreme spins. When both spin and scalar charge increase simultaneously, the thermal output becomes highly concentrated near the central object. Even with fixed \( \alpha \), a rise in spin enhances the thermal response and compacts the emission zone. These findings indicate that KMNS produces hotter and more compact thermal disks than their Kerr counterparts, offering potentially observable differences in high-resolution spectral imaging.

The thermodynamic and electromagnetic properties of thin accretion disks, such as energy flux, temperature distribution, and equilibrium radiation spectrum, exhibit significant differences between KMNSs and KNSs. These distinctions lead to clear observational signatures that may be used to differentiate between the two compact objects. For objects with comparable energy flux, the disk surface can reach notably higher peak values in KMNS, particularly for specific combinations of spin \( a \) and \( \alpha \). All key thermodynamic quantities are found to be highly sensitive to variations in both parameters. These differences are consistently reflected in the corresponding disk temperatures and radiation spectra. Furthermore, within certain ranges of spin and $\alpha$, the conversion efficiency of accreting mass into radiation is substantially higher in the KMNS case, indicating a more effective radiative mechanism compared to BHs. Collectively, these observational features provide a strong foundation for distinguishing KMNSs in modified gravity from classical KNS through future high-resolution astrophysical measurements.

\bibliographystyle{ieeetr}
\bibliography{reference}

@article{pradhan2020distinguishing,
  title={Distinguishing black hole and naked singularity in {MOG} via inertial frame dragging effect},
  author={Pradhan, P.},
  journal={ arXiv:2007.01347},
  year={(2020)}
}

@article{moffat2015black,
  title={Black holes in modified gravity {(MOG)}},
  author={Moffat, J. W.},
  journal={The European Physical Journal C},
  volume={75},
  number={},
  pages={},
  year={175(2015)},
  publisher={Springer}
}

@article{Joshi2011zm,
    author = "Joshi, P. S. and Malafarina, D. and Narayan, R.",
    title = "{Equilibrium configurations from gravitational collapse}",
    journal = "Class. Quant. Grav.",
    volume = "28",
    pages = "",
    year = "235018(2011)"
}

@BOOK{1995lwetbookV,
       author = {{Visser}, Matt},
        title = "{Lorentzian wormholes. From Einstein to Hawking}",
         year = 1996,
       publisher={American Institute of Physics}
}

@ARTICLE{1968AmSci..56....1W,
       author = {{Wheeler}, J.~A.},
        title = "{Our universe: the known and the unknown.}",
      journal = {American Scientist},
         year = 1968,
        month = jan,
       volume = {56},
       number = {1},
        pages = {1-20},
       adsurl = {https://ui.adsabs.harvard.edu/abs/1968AmSci..56....1W}
}

@INPROCEEDINGS{1974IAUS...64..132B,
       author = {{Bardeen}, J.~M.},
        title = "{Properties of Black Holes Relevant to Their Observation (invited Paper)}",
    booktitle = {Gravitational Radiation and Gravitational Collapse},
         year = 1974,
       editor = {{Dewitt-Morette}, C.},
       series = {IAU Symposium},
       volume = {64},
        month = jan,
        pages = {132},
       adsurl = {https://ui.adsabs.harvard.edu/abs/1974IAUS...64..132B}
}

@article{lee2017innermost,
  title={Innermost stable circular orbit of {K}err-{MOG} black hole},
  author={Lee, H. C. and Han, Y. J.},
  journal={The European Physical Journal C},
  volume={77},
  pages={},
  year={9(2017)},
  publisher={Springer}
}

@article{nguyen2023shadow,
  title={Shadow geometry of {K}err naked singularities},
  author={Nguyen, B. and Christian, P. and Chan, C.-K.},
  journal={The Astrophysical Journal},
  volume={954},
  number={},
  pages={},
  year={78(2023)},
  publisher={IOP Publishing}
}

@article{perlick2022calculating,
  title={Calculating black hole shadows: review of analytical studies},
  author={Perlick, V. and Tsupko, O. Y.},
  journal={Physics Reports},
  volume={947},
  pages={},
  year={39(2022)},
  publisher={Elsevier}
}

@article{patel2022light,
  title={Light trajectory and shadow shape in the rotating naked singularity},
  author={Patel, V. and Tahelyani, D. and Joshi, A. B. and Dey, D. and Joshi, P. S.},
  journal={The European Physical Journal C},
  volume={82},
  number={},
  pages={},
  year={798(2022)},
  publisher={Springer}
}

@article{wang2024ring,
  title={The ring-shaped shadow of a rotating naked singularity with a complete photon sphere},
  author={Wang, M. and Guo, G. and Yan, P. and Chen, S. and Jing, J.},
  journal={Chinese Physics C},
  volume={48},
  number={},
  pages={},
  year={105103(2024)},
  publisher={IOP Publishing}
}

@article{azreg2014generating,
  title={Generating rotating regular black hole solutions without complexification},
  author={Azreg-A{\"\i}nou, M.},
  journal={Physical Review D},
  volume={90},
  number={},
  pages={},
  year={064041(2014)},
  publisher={APS}
}

@article{gott1991closed,
  title={Closed timelike curves produced by pairs of moving cosmic strings: Exact solutions},
  author={Gott III, J. R.},
  journal={Physical Review Letters},
  volume={66},
  number={},
  pages={},
  year={1126(1991)},
  publisher={APS}
}

@article{penrose1965gravitational,
  title={Gravitational collapse and space-time singularities},
  author={Penrose, R.},
  journal={Physical Review Letters},
  volume={14},
  number={},
  pages={},
  year={57(1965)},
  publisher={APS}
}

@article{akiyama2019first,
  title={First {M87*} {E}vent {H}orizon {T}elescope results. {I}. {T}he shadow of the supermassive black hole},
  author={Akiyama, K. and others},
  journal={The Astrophysical Journal Letters},
  volume={875},
   number={},
  year={6(2019)},
  publisher={IOP Publishing}
}

@article{akiyama2022first,
  title={First {S}agittarius {A*} {E}vent {H}orizon {T}elescope results. {I}. {T}he shadow of the supermassive black hole in the center of the {M}ilky {W}ay},
  author={Akiyama, K. and others},
  journal={The Astrophysical Journal Letters},
  volume={930},
  number={},
  pages={},
  year={12(2022)},
  publisher={IOP Publishing}
}

@article{joshi2020shadow,
  title={Shadow of a naked singularity without photon sphere},
  author={Joshi, A. B. and Dey, D. and Joshi, P. S. and Bambhaniya, P.},
  journal={Physical Review D},
  volume={102},
  number={},
  pages={},
  year={024022(2020)},
  publisher={APS}
}

@article{janis1968reality,
  title={{R}eality of the {S}chwarzschild singularity},
  author={Janis, A. I. and Newman, E. T. and Winicour, J.},
  journal={Physical Review Letters},
  volume={20},
  number={},
  pages={},
  year={878(1968)},
  publisher={APS}
}

@article{yasmin2025shadow,
  title={Shadow cast by the {K}err {MOG} black hole under the influence of plasma and constraints from {EHT} observations},
  author={Yasmin, S. and Jafarzade, K. and Jamil, M.},
  journal={Chinese Physics  C},
  volume={49},
  number={},
  pages={},
  year={065107(2025)},
  publisher={IOP Publishing}
}

@misc{novikov1973black,
  title={Black Holes ed {C}. {D}eWitt and {B}. {D}eWitt},
  author={Novikov, ID and Thorne, KS},
  year={1974},
  publisher={New York: Gordon and Breach}
}

@article{luminet2019image,
  title={ Image of a spherical black hole with thin
 accretion disk},
  author={Luminet, J.-P.},
  journal={Astronomy and Astrophysics},
volume={75},
 year={ 228 (1979)},
}

@article{cunningham1973optical,
  title={The optical appearance of a star orbiting an extreme {K}err black hole},
  author={Cunningham, C. T. and Bardeen, J. M.},
  journal={Astrophysical Journal},
  volume={183},
  pages={},
  year={264(1973)}
}

@article{Gyulchev2019imageo,
  title={ Image of the {J}anis-{N}ewman-{W}inicour naked singularity with a thin accretion disk},
  author={Gyulchev, G. and  Nedkova, P. and Vetsov, T. and Yazadjiev, S.},
  journal={Physical Review D},
  volume={100},
  pages={},
  year={024055(2019)}
}

@article{moffat2006scalar,
  title={Scalar tensor vector gravity theory},
  author={Moffat, J. W.},
  journal={Journal of Cosmology and Astroparticle Physics},
volume={2006},
  pages={},
  year={004(2006)}
}

@article{brownstein2007bullet,
  title={The bullet cluster 1{E}0657-558 evidence shows modified gravity in the absence of dark matter},
  author={Brownstein, J. R. and Moffat, J. W.},
  journal={Monthly Notices of the Royal Astronomical Society},
  volume={382},
  number={},
  pages={},
  year={47(2007)},
  publisher={Blackwell Publishing Ltd Oxford, UK}
}

@article{hawking1992chronolgy,
  title={Chronology protection conjecture},
  author={Brownstein, J. R. and Moffat, J. W.},
  journal={Physical Review D},
  volume={46},
  number={},
  pages={},
  year={603(1992)},
}

@article{vaishak2019chronolgy,
  title={Chronology protection problem in modified {K}err-{N}ewman spacetimes},
  author={Vaishak, P. and Rahul, S. and Sashideep, G.

},
  journal={Physical Review D},
  volume={99},
  number={},
  pages={},
  year={024023 (2019)},
}

@article{bajowald2005black,
  title={Black hole mass threshold from nonsingular quantum gravitational collapse},
  author={Bojowald, M. and Goswami, R. and Maartens, R. and Singh, P.

},
  journal={Physical Review D},
  volume={95},
  number={},
  pages={},
  year={091302(2005)},
}

@article{hossenfelder2010model,
  title={Model for nonsingular black hole collapse and evaporation},
  author={Hossenfelder, S. and Modesto, L. and Pr{\'e}mont, S. I.},
  journal={Physical Review D},
  volume={81},
  number={},
  pages={},
  year={044036(2010)},
  publisher={APS}
}

@article{germain2020reversible,
  title={Reversible dynamics with closed time-like curves and freedom of choice},
  author={Germain, T. and Fabio, C.},
  journal={Classical and Quantum Gravity},
  volume={37},
  number={},
  pages={},
  year={205011(2020)},
  publisher={IOP Science}
}

@article{ikeda2021black,
  title={Black-hole microstate spectroscopy: Ringdown, quasinormal modes, and echoes},
  author={Ikeda, T. and  Bianchi, M. and  Consoli, D. and others},
  journal={Physical Review D},
  volume={104},
  number={},
  pages={},
  year={066021(2021)},
  publisher={APS}
}

@article{lum2021closed,
  title={Closed timelike curves, singularities and causality: A survey from {G}ödel to chronological protection Universe},
  author={Luminet, J.P.},
  journal={Universe},
  volume={7},
  number={},
  pages={},
  year={12(2021)},
  publisher={}
}

@article{calvani1978time,
  title={Time machine and geodesic motion in {K}err metric},
  author={Calvani, M. and De Felice, F. and Muchotrzeb, B. and Salmistraro, F.},
  journal={General Relativity and Gravitation},
  volume={9},
  number={},
  pages={},
  year={163(1978)},
  publisher={}
}

@article{bcarter1973equi,
  title={Black hole equilibrium states, in Black Holes
 (Les Astres Occlus), edited by C. Dewitt and B. S. Dewitt},
  author={B. Carter},
  journal={Elsevier, Taiwan},
  volume={},
  number={},
  pages={},
  year={214( 1973)},
  publisher={}
}

@article{GIMON2009astro,
title = {Astrophysical violations of the {K}err bound as a possible signature of string theory},
author={Eric G. G. and Petr, H. },
journal = {Physics Letters B},
volume = {672},
number = {},
pages = {},
year = {302(2009)},
publisher={}
}

@article{kazempour2022analysis,
  title={Analysis of accretion disk around a black hole in d{RGT} massive gravity},
  author={Kazempour, S. and Zou, Y. C. and Akbarieh, A. R.},
  journal={The European Physical Journal C},
  volume={82},
  number={},
  pages={},
  year={190(2022)},
  publisher={Springer}
}

@article{page1974disk,
  title={Disk-accretion onto a black hole. {T}ime-averaged structure of accretion disk},
  author={Page, D. N. and Thorne, K. S.},
  journal={Astrophysical Journal},
  volume={191},
  pages={},
  year={506(1974)}
}

@article{rhie1991global,
  title={Global monopoles do not ‘‘collapse’’},
  author={Rhie, S. H. and Bennett, D. P.},
  journal={Physical Review Letters},
  volume={67},
  number={},
  pages={},
  year={1173(1991)},
  publisher={APS}
}

@ARTICLE{2017PhRvD..95j4047P,
       author = {{P{\'e}rez}, D. and {Armengol}, F. G. L. and {Romero}, G. E.},
        title = "{Accretion disks around black holes in scalar-tensor-vector gravity}",
      journal = {\prd},
         year = {104047(2017)},
       volume = {95},
       number = {},
        pages = {},
         

}

@article{dutta2024role,
  title={On the role of closed timelike curves and confinement structure around {K}err--{N}ewman singularity},
  author={Dutta, A. and Roy, D. and Chakraborty, S.},
  journal={International Journal of Modern Physics D},
  volume={33},
  number={},
  pages={},
  year={2450034(2024)},
  publisher={World Scientific}
}

@inproceedings{shakura1973black,
  title={Black Holes in Binary Systems: Observational Appearances},
  author={Shakura, N.I. and Sunyaev, R.A.},
  booktitle={Symposium-International Astronomical Union},
  volume={55},
  pages={},
  year={164(1973)},
  organization={Cambridge University Press}
}

@article{thorne1974disk,
  title={Disk-accretion onto a black hole. {II}. {E}volution of the hole},
  author={Thorne, K. S.},
  journal={Astrophysical Journal},
  volume={191},
  pages={},
  year={520(1974)}
}

@article{harko2011thin,
  title={Thin accretion disk signatures of slowly rotating black holes in {H}o{\v{r}}ava gravity},
  author={Harko, T. and Kov{\'a}cs, Z. and Lobo, F. SN},
  journal={Classical and Quantum Gravity},
  volume={28},
  number={},
  pages={},
  year={165001(2011)},
  publisher={IOP Publishing}
}

@article{Boshkayev2020kle,
    author = "Boshkayev, K. and Idrissov, A. and Luongo, O. and Malafarina, D.",
    title = "{Accretion disc luminosity for black holes surrounded by dark matter}",
    journal = "Mon. Not. Roy. Astron. Soc.",
    volume = "496",
    number = "",
    pages = "",
    year = "1123(2020)"
}

@ARTICLE{Kurmanov2025acceration,
       author = {{Kurmanov}, Y. and {Boshkayev}, K. and {Konysbayev}, T. and {Muccino}, M. and {Luongo}, O. and {Urazalina}, A. and {Dalelkhankyzy}, A. and {Belissarova}, F. and {Alimkulova}, M.},
        title = "{Accretion disk luminosity around rotating naked singularities}",
      journal = {Physics of the Dark Universe},
         year =" 101917(2025)",
       volume = {48},
        pages = {},

}

@article{PhysRevD.92.043008,
  title = {Timelike geodesics of a modified gravity black hole immersed in an axially symmetric magnetic field},
  author = {Hussain, S. and Jamil, M.},
  journal = {Physical Review D},
  volume = {92},
  issue = {},
  pages = {},
  numpages = {},
  year = {043008(2015)},
  
  publisher = {American Physical Society},
}

@article{Rehman:2025rxw,
    author = {Rehman, N. and Chan, Z. C. S. and Jamil, M. and Azreg-A{\"\i}nou, M.},
    title = "{Gravitational lensing by a dark compact object in modified gravity and observational constraints from {E}instein rings}",
    
    journal = "European Physical Journal C",
    volume = "85",
    number = "",
    pages = "",
    year = "233(2025)",
}

@article{Shabbir:2025kqh,
    author = {Shabbir, O. and Jamil, M. and Azreg-A{\"\i}nou, M.},
    title = "{Periodic orbits and their gravitational wave radiations around the {S}chwarzschild-{MOG} black hole}",
    journal = "Physics of the Dark Universe",
    volume = "47",
    pages = "",
    year = "101816(2025)",
}

@article{Babar2016dyna,
author = {Babar, G. Z. and Jamil, M. and Lim, Yen. K.},
title = {Dynamics of a charged particle around a weakly magnetized naked singularity},
journal = {International Journal of Modern Physics D},
volume = {25},
number = {},
pages = {},
year = {1650024(2016)},
}

@article{Jafarzade2025xcn,
    author = "Jafarzade, K. and Shaymatov, S. and Jamil, M.",
    title = "{Shadows and optical appearances of black holes in ${R}^2$ gravity}",
   
    journal = "Astroparticle Physics",
    volume = "168",
    pages = "",
    year = "103100(2025)",
}

@article{khaijde2024zqq,
  author  = {Jafarzade, K. and Bazyar, Z. and Jamil, M.},
  title   = {A study of black holes in $F(R)$-{M}od{M}ax gravity: {G}ravitational lensing and constraints from {EHT} observations},
  journal = {Physics Letters B},
  volume  = {864},
  pages   = {},
  year    = {137903(2025)}
}

@article{universe8020102,
  author  = {Jusufi, K. and Azreg-Aïnou, M. and Jamil, M. and Saridakis, E. N.},
  title   = {Constraints on {B}arrow entropy from {M}87* and {S}2 Star observations},
  journal = {Universe},
  volume  = {8},
  number  = {},
  pages   = {},
  year    = {102(2022)}
}

@article{PhysRevD.101.044035,
  author  = {Jusufi, K. and Jamil, M. and Chakrabarty, H. and Wu, Q. and Bambi, C. and Wang, A.},
  title   = {Rotating regular black holes in conformal massive gravity},
  journal = {Physical Review D},
  volume  = {101},
  number  = {},
  pages   = {},
  year    = {044035(2020)},
}

@article{PhysRevD.100.sh,
  author  = {Zhu, T. and Wu, Q. and Jamil, M. and Jusufi, K.},
  title   = {Shadows and deflection angle of charged and slowly rotating black holes in {E}instein-\AE{}ther theory},
  journal = {Physical Review D},
  volume  = {100},
  number  = {4},
  pages   = {},
  year    = {044055(2019)},
}

@article{Wang2018prk,
    author = "Wang, H. M. and Xu, Y. M. and Wei, S.W.",
    title = "{Shadows of {K}err-like black holes in a modified gravity theory}",

    journal = "JCAP",
    volume = "03",
    pages = "",
    year = "046(2019)"
}

@article{zheng2025shadows,
  title={Shadows and accretion disk images of charged rotating black hole in modified gravity theory},
  author={Zheng, H. B. and Wu, M. Q. and Li, G. P. and Jiang, Q. Q.},
  journal={The European Physical Journal C},
  volume={85},
  number={},
  pages={46},
  year={46(2025)},
  publisher={Springer}
}

@article{Jafarzade2025nbe,
    author = "Jafarzade, K. and Yasmin, S. and Jamil, M.",
    title = "{Shadow of {F}({R})-{E}uler{\textendash}{H}eisenberg black hole and constraints from {EHT} observations}",
    doi = "",
    journal = "Annals Physics",
    volume = "482",
    pages = "",
    year = "170198(2025)"
}
\end{document}